*Heaven's light is our guide*

# Rajshahi University of Engineering &Technology, Bangladesh

## Department of Mechanical Engineering

## A Project & Thesis Report on
## Development and analysis of a secured VoIP system for surveillance activities

**Supervised by**

Dr. Mhia Md. Zaglul Shahadat
Professor
Department of Mechanical Engineering
Rajshahi University of Engineering & Technology

**Submitted by**

M. Matsive Ali
Roll No: 1602023
Mohammad Nashik Islam
Roll No: 1602042

*Heaven's light is our guide*

# Rajshahi University of Engineering &Technology, Bangladesh

## Department of Mechanical Engineering

A Project & Thesis Report on

**Development and analysis of a secured VoIP system for surveillance activities**

This report is submitted in partial fulfillment of the requirements for the degree of Bachelor of Science in Mechanical Engineering at Rajshahi University of Engineering & Technology, Rajshahi, Bangladesh.

**<u>Supervised by</u>**

Dr. Mhia Md. Zaglul Shahadat
Professor
Department of Mechanical Engineering
Rajshahi University of Engineering & Technology

**<u>Submitted by</u>**

M. Matsive Ali
Roll No: 1602023
Mohammad Nashik Islam
Roll No: 1602042

*Heaven's light is our guide*

# Rajshahi University of Engineering &Technology, Bangladesh

## Department of Mechanical Engineering

## Certificate

This is to certify that the thesis entitled "**Development and analysis of a secured VoIP system for surveillance activities**" has been carried out by M. Matsive Ali, Roll: 1602023, and Mohammad Nashik Islam, Roll: 1602042 under my supervision in the Department of Mechanical Engineering at Rajshahi University of Engineering & Technology, Rajshahi-6204, Bangladesh.

**Signature of the Supervisor**

…………………………...

Dr. Mhia Md. Zaglul Shahadat
Professor
Department of Mechanical Engineering
Rajshahi University of Engineering & Technology

**Signature of the students**

………………………….

M. Matsive Ali
Roll No.: 1602023

………………………….

Mohammad Nashik Islam
Roll No.: 1602042

**Countersigned**

…………………………...

Dr. Md Rabiul Islam Sarker
Professor and Head
Department of Mechanical Engineering
Rajshahi University of Engineering & Technology

**Signature of the External**

…………………………...

Dr. Md. Rokunuzzaman
Professor
Department of Mechanical Engineering
Rajshahi University of Engineering & Technology

# ACKNOWLEDGEMENT

This thesis work has been finally completed, by the grace of Almighty Allah. The authors would like to express their heartiest gratitude to their thesis supervisor, **Dr. Mhia Md. Zaglul Shahadat**, Professor, Department of Mechanical Engineering, Rajshahi University of Engineering & Technology for his constant guidance, suggestion, support, and inspiration in performing and completing the thesis works.

We would like to express our gratitude towards honorable teachers of the Mechanical Engineering department at RUET for their valuable suggestions during the research work

Special thanks go to Md. Saiful Islam, Asst. Chief Technical Officer, Metrology Lab, Department of Mechanical Engineering, Rajshahi University of Engineering & Technology, for his support related to thesis work.

Also, the authors are also thankful to their elder brother Avijit Mallik, Department of Mechanical Engineering, Rajshahi University of Engineering & Technology, for his effective help.

RUET, Rajshahi

November, 2022

M. Matsive Ali

Roll No: 1602023

Mohammad Nashik Islam

Roll No: 1602042

# ABSTRACT

Since 1990s, telephone has been the primary mode of communication. However, Voice over Internet Protocol (VoIP) which is a highly straightforward and affordable form of data transfer, is now becoming an important part of daily communication. VoIP is the technology that makes it possible to send speech and multimedia data packets across either a public or private IP network. However, a cyberattack known as "man-in-the-middle" is a serious concern issue in transferring data through any network these days. Thereby, the authors have designed a system that sends voice over the internet within the range of a router using encrypted data transfer. Thus an embedded system comprising of Electret Microphone, Embedded C++, Node.JS & Particle Photon Microcontroller along with 'Internet of Things (IoT)' is fabricated. Due to its compact size, this type of device may be incorporated into automobiles, spy bugs, and surveillance systems. The VoIP system gathers sound signals using MAX9814 microphone while particle photon microcontroller securely transmits the data. Devices with accessibility can download data from VoIP system Transmission Control Protocol (TCP) server. The accessed device stores audio locally as well as upload the corresponding data to Google Drive. This VoIP system provides a secured way of communication conserving the original signal.

# List of Contents

# List of Figures

# List of Tables

# CHAPTER 1

# INTRODUCTION

Radio technology, transmission, and detection of communication signals consisting of electromagnetic waves that travel through the air in a straight line or by reflection from the ionosphere or a communications satellite.

Radio is the technology of signaling and communicating using radio waves. Radio waves are electromagnetic waves of frequency between 30 hertz (Hz) and 300 gigahertz (GHz). They are generated by an electronic device called a transmitter connected to an antenna that radiates the waves and is received by another antenna connected to a radio receiver. Radio is very widely used in modern technology, in radio communication, radar, radio navigation, remote control, remote sensing, and other applications.[1-3]

## 1.1. Walkie-talkie Network & Vulnerability

A walkie-talkie is a hand-held portable two-way radio transceiver. These devices normally work in a frequency range of 136 MHz to 900 MHz.[4]

In the below figure, we can see how a Walkie-Talkie Transmits and another Walkie=Talkie Receives. The signal range of the transmitter is dependent on the power of the transmitter. Normally a 0.25Watt transmitter can send data across 0.15k m to 0.3 Km.

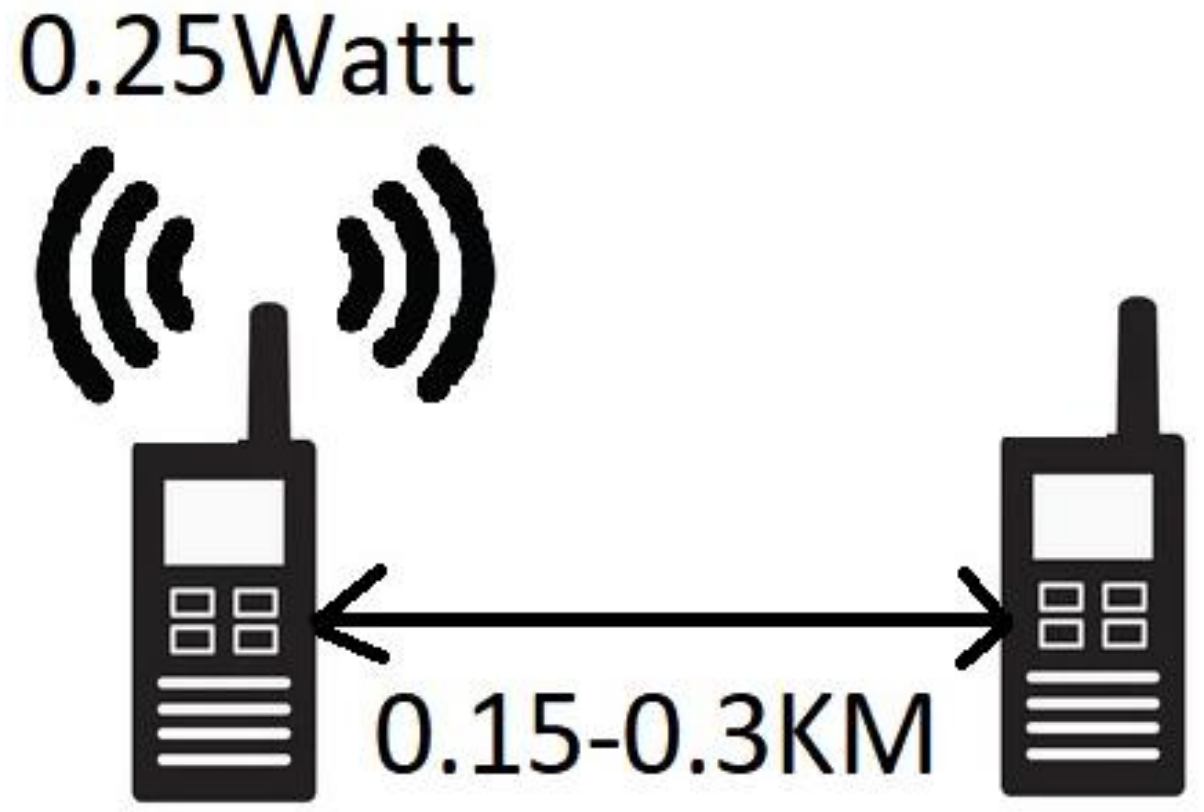

Figure 1 : Walkie-talkie Communication.

But as the transmitter transmit the data in a radius of 0.15-0.3 km other unwanted receivers are also able to receive the data of the transmitter which makes the system very vulnerable.

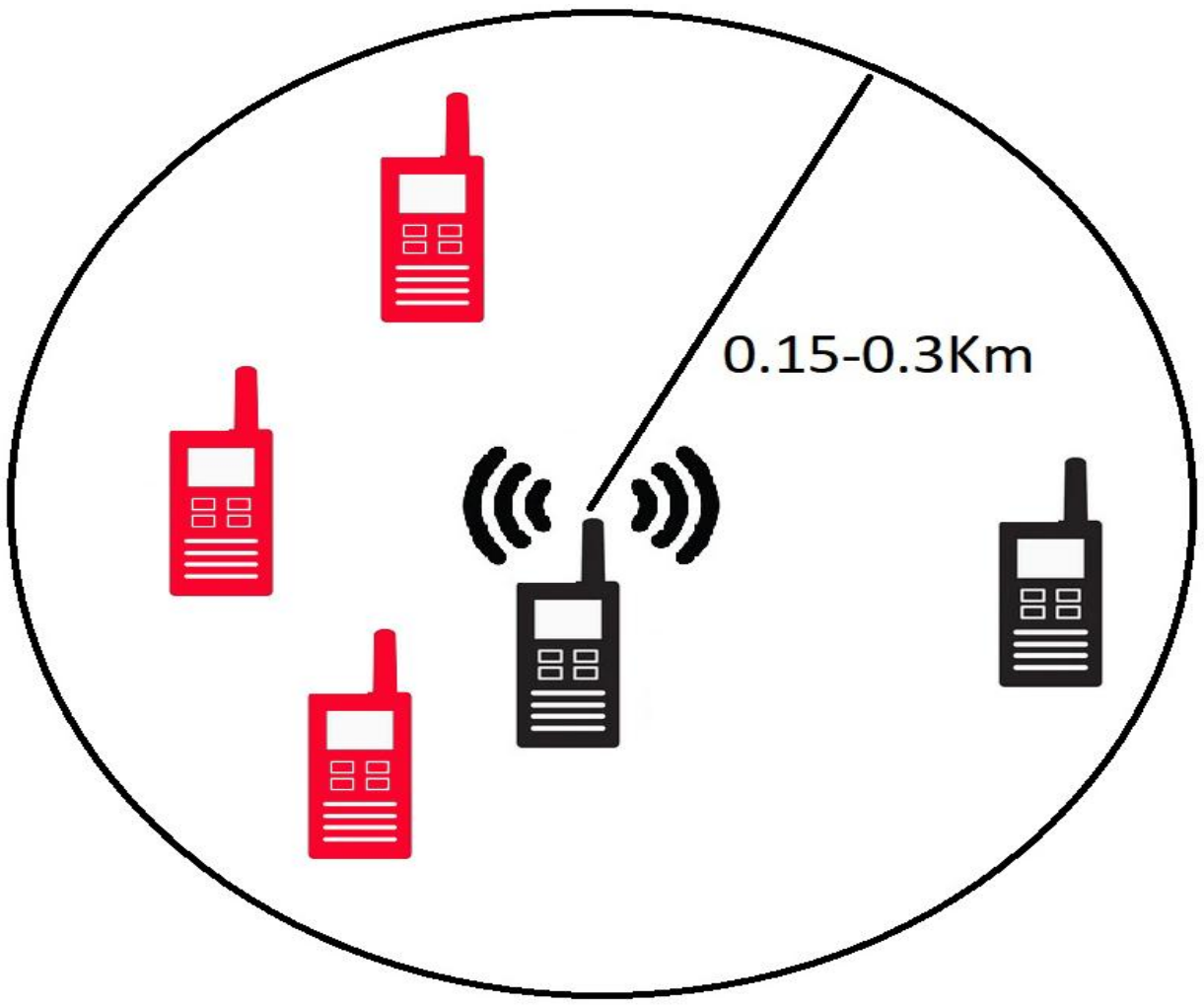

Figure 2 : Walkie-talkie Communication Abuse.

## 1.2. WiFi's Radio Technology & Vulnerability

Wi-Fi is a wireless networking technology that allows devices such as computers, mobile devices, and other equipment to interface with the Internet. The radios used in WiFi are not so different from the radios used in walkie talkies. WiFi normally works in a frequency range of 900 MHz to 5 GHz.[5]

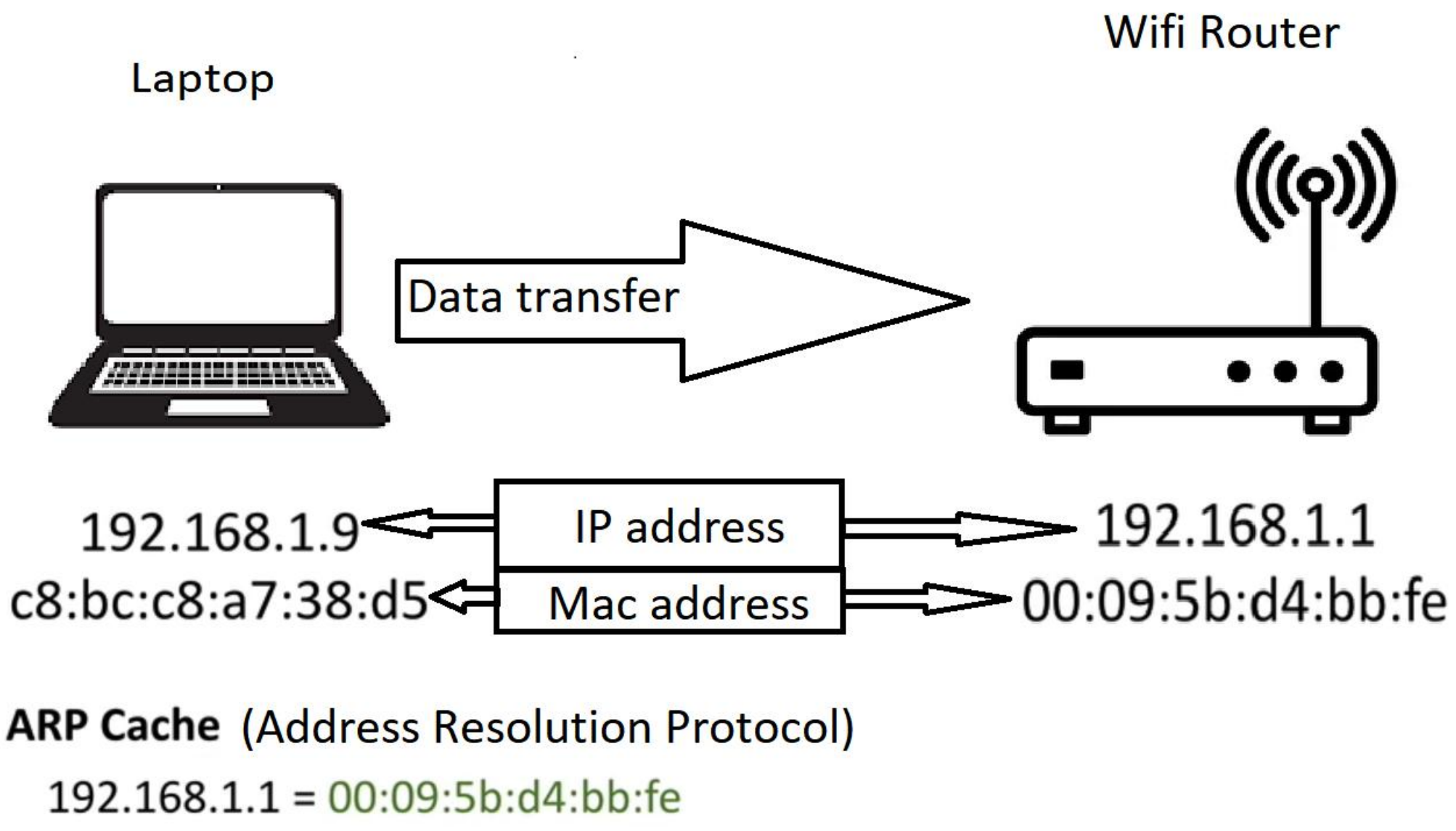

Figure 3 : Wifi Technology.

In the procedure of connecting the laptop to the WiFi, the laptop first searches for the MAC address of the specified IP of the router which is 192.168.1.1. The router with the IP 192.168.1.1 sends the MAC address to the laptop which the laptop receives and saves to its ARP Cache (Address Resolution Protocol Cache). Next, all the data that the laptop wants to send to the internet goes through the MAC address of the Router then to the internet, and vice versa.

However, in WiFi Technology the most common vulnerability is the man-in-the-middle attack. A man-in-the-middle attack is a type of eavesdropping attack, where attackers interrupt an existing conversation or data transfer. After inserting themselves in the "middle" of the transfer, the attackers pretend to be both legitimate participants. This enables an attacker to intercept information and data from either party while also sending malicious links or other information to both legitimate participants in a way that might not be detected until it is too late. [6]

The most common man-in-the-middle attack is ARP Cache poisoning. In which a third-party laptop is connected to the same WiFi router the knowledge of the Use Laptop changes the ARP Cache MAC address to its own MAC address. Thus the data transfer occurs as follows, data is sent from the user's laptop to the third-party laptop then it is sent to the router. Thus the user is unaware that his data is being taken or monitored as his laptop is working normally.

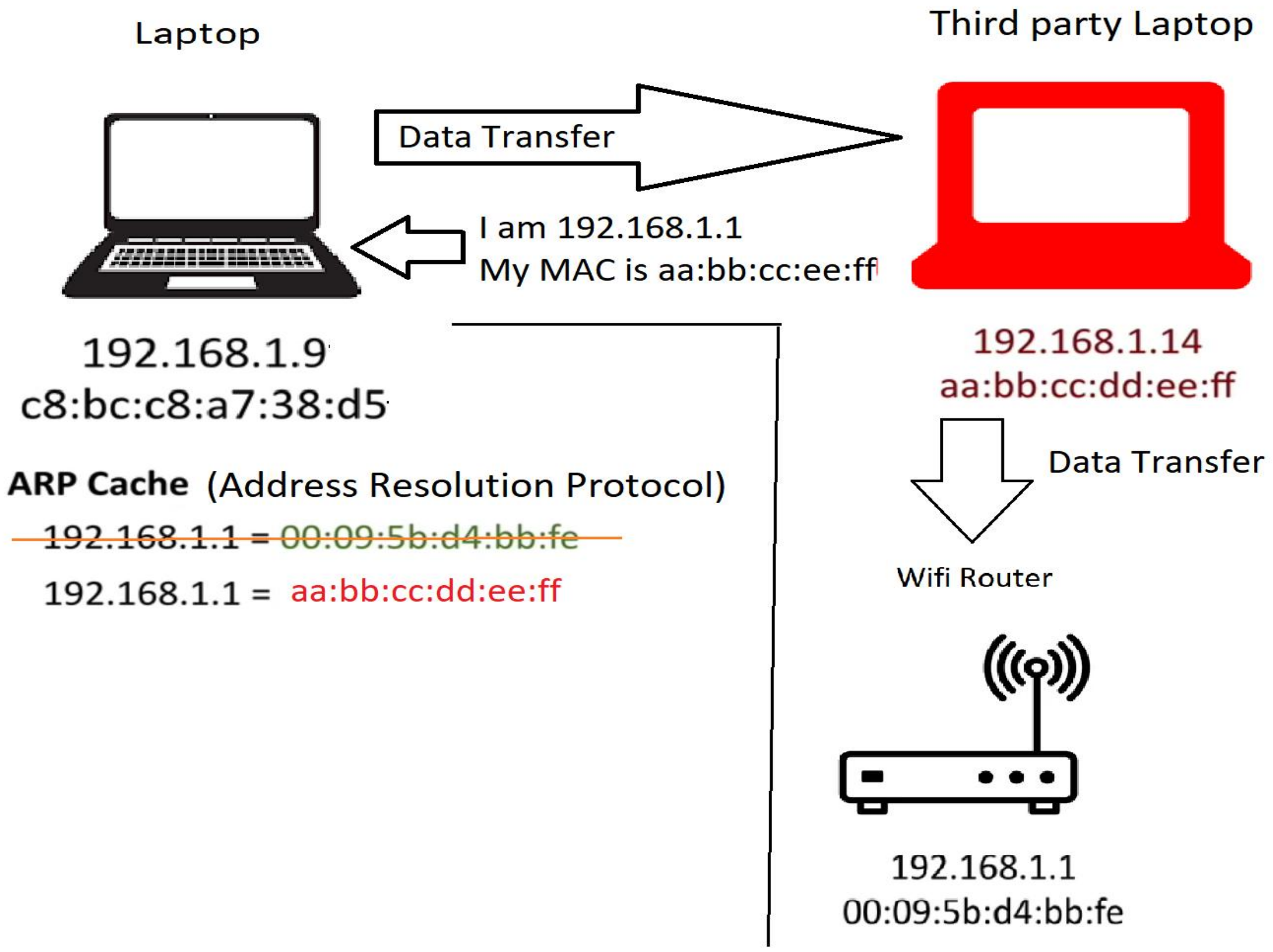

Figure 4 : ARP Cache Poisoning.

## 1.3. Internet of Things (IoT)

IoT has emerged as a new network paradigm, which allows various physical entities in the world to connect with each other. The observed or generated information of these entities has a great potential to provide useful knowledge across different service domains, such as building management, energy-saving systems, surveillance services, smart homes, smart cities, etc. IoT was first proposed in 1999 by Kevin Ashton, who is the co-founder of Auto-ID center at the Massachusetts Institute of Technology (MIT) [7].

One foundational technology of IoT is Radio-Frequency Identification (RFID) technology, which allows microchips to transmit the identification number of the objects to a reader through wireless communication. Through RFID technology, physical objects can be identified, tracked, and monitored automatically. Nowadays, RFID technology has been widely adopted in logistics, pharmaceutical production, retailing, and supply chain management [8, 9-11].

Another foundational technology of IoT is Wireless Sensor Networks (WSNs), which adopt interconnected intelligent sensors to periodically sense the monitored environment and send the information to the sink (or base station), at which the gathered/collected information can be further processed for end-user queries [12]. The applications include disaster control, environment and habitat monitoring, battlefield surveillance, traffic control, and health care applications. Additionally, many other technologies and devices such as Near Field Communication (NFC) [13], short-range wireless communication (i.e., ZigBee [14] and Bluetooth [15]), universal mobile accessibility (i.e., Wi-Fi hotspots [16], and cellular networks [17]), social networking [18] and cloud computing [19] support internet of things to compose an extensive network infrastructure.

## 1.4. Internet Protocol (IP)

The Internet Protocol (IP) is the primary communications protocol in the Internet Protocol Suite for the transmission of data graphics across network boundaries. Its routing function allows you to connect to the Internet and configure the Internet in bulk. [20]

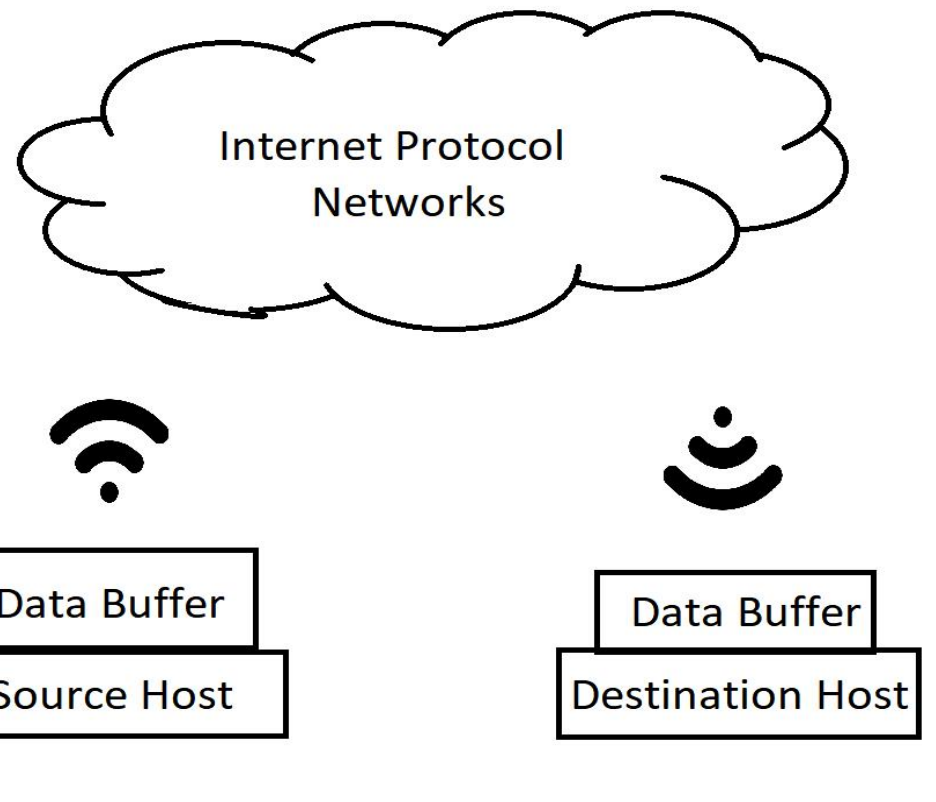

Figure 5 : Data transmission through IP.

IP is responsible for delivering packets from the source host to the destination server based only on the IP address in the packet header. To this end, IP defines the packet structures which encapsulate the data to be delivered. It also defines the addressing methods used to label datagrams with source and destination information.

Internet Protocol is responsible for resolving host interfaces, encapsulating data in data schemas (including fragmentation and reassembly), and routing data schemas from source host interfaces to the destination host interface on one or more IP networks.[21] For these purposes, the Internet Protocol defines the format of packets and provides an addressing scheme. Each datagram has two components: a header and a payload. The IP header includes the source IP address, destination IP address, and other metadata needed to route and deliver the datagram. The payload is the data being transported. The method of nesting the data payload in a header packet is called encapsulation. IP addressing involves assigning IP addresses and related parameters to host interfaces. The address space is divided into subnets, regarding the designation of network prefixes. IP routing is performed by all hosts, as well as routers, whose main function is to carry packets across network boundaries. Routers communicate with each other using specially designed routing protocols, which can be either an internal gateway protocol or an external gateway protocol, depending on the needs of the network topology.[22]

## 1.5. Voice Over Internet Protocol (VoIP)

Voice over Internet Protocol (VoIP), also known as IP telephony, is a method and group of technologies for providing voice communications and multimedia sessions over an Internet Protocol (IP) network, such as the Internet. Telephone calls are similar to traditional digital telephones and include signaling, channeling, analog voice digitization, and encoding. Instead of being transmitted over a switched network, digital information is punctuated and transmission occurs as IP packets over a packet-switched network. They carry media streams using special media streaming protocols that encode audio and video with audio and video codecs. VoIP is also available on many personal computers and access devices. Other Internet. SMS calls and text messages can be sent over Wi-Fi or your operator's mobile data network [23].

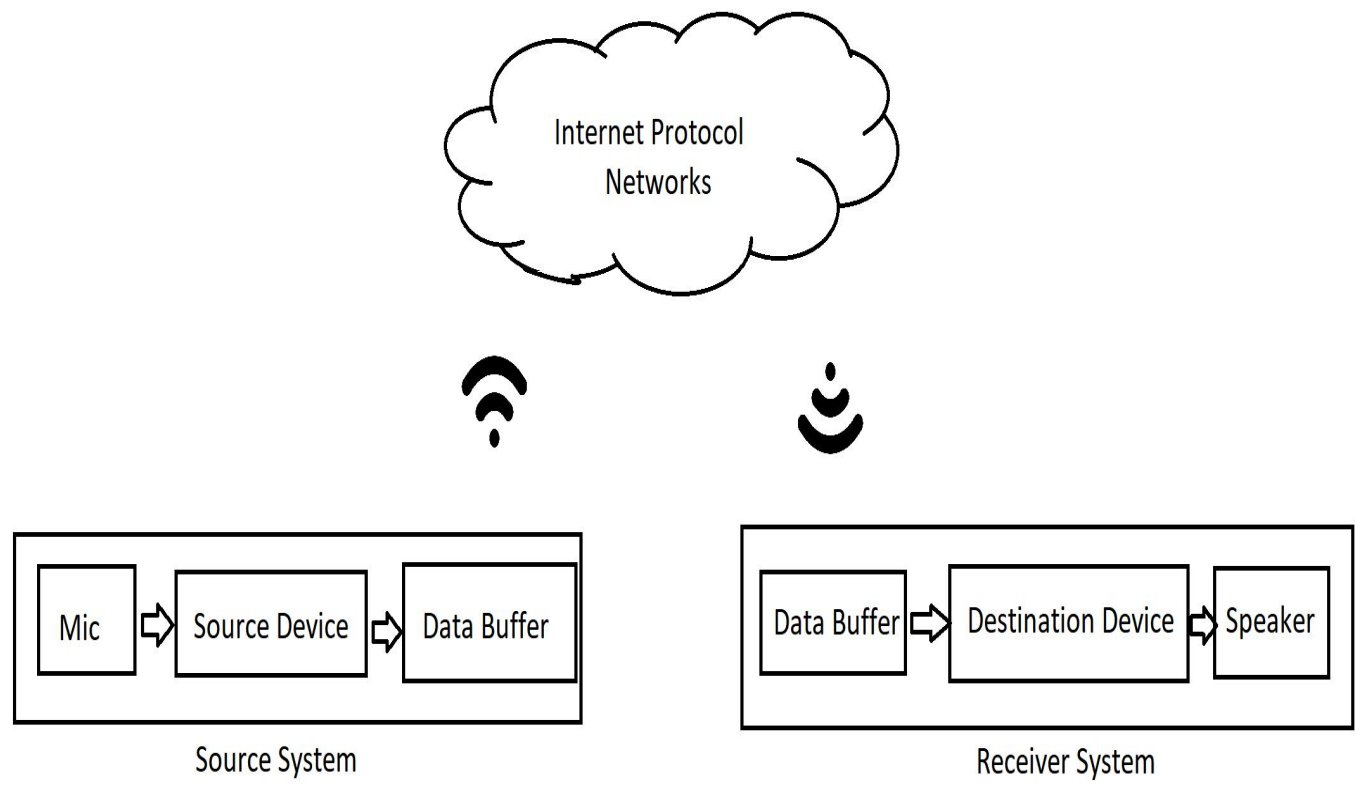

Figure 6 : Voice Transmission Through IP.

VoIP provides a framework to unify all modern communication technologies using a single unified communication system. The security concerns of VoIP phone systems are similar to those of other devices connected to the Internet. This means that hackers with knowledge of VoIP vulnerabilities can perform denial of service attacks, collect customer data, record chats, and compromise voicemail. Compromised session credentials or VoIP user accounts could allow attackers to incur significant charges from third-party services, such as long signals or international calls to be carried out than traditional telephone circuits. The result of the lack of encryption is that it is relatively easy to eavesdrop on VoIP calls when the data network is accessible.

## 1.6. Comparison of Conventional Telephony and VoIP

International calls are phone calls that are made from one country to another. There are numerous international telephone call gateway exchanges or switches, that facilitate these phone calls. Calling cards usually come equipped with PIN for user protection and security. Most companies require user to enter the PIN before granting access to the calling card's funds. PINs often are printed on a piece of paper found inside the calling card's packaging. Once the users make their first call, some companies offer the option of eliminating the PIN to speed up the calling process. Companies that sell virtual calling cards online typically PIN via email. The calling card provides you with a set of numeric prefixes that enable you to make the call.[24]

It costs a lot to make calls normally or with international calling cards as the whole system may include a combination of wires, towers, satellites, etc to transmit the audio from one place to another.

VoIP is the set of standardized technologies that deliver voice calls via the internet or a private IP-based network. Here VoIP system consists of a combination of router, optic fiber, submarine cable, etc. As it is very cheap this method of voice calls is becoming popular.

Table 1: Difference between Telephony and VoIP.[25]

| VoIP | Telephone |
|---|---|
| Its cost is lower. | Its cost is high |
| It is portable. | It has to be connected to telephone line. |
| It transfers data over the air. | It transfer data through telephone lines. |
| It is dependent on high bandwidth. | It work one low bandwidth, |
| It required multiple power sources. | It only require one source of energy. |

## 1.7. Data Transmission Security

Data transmission is the transfer of information across a point-to-point or point-to-multipoint communication channel as a digital bitstream or digitized analog signal.[26]

In a wireless network, security challenges grow more complicated. It is important to make sure that network access is restricted and that data is not at risk of being accessed while being transmitted across the network. Data can be encrypted using a variety of technologies, helping to protect its privacy and integrity.[27] They ensure that by doing the following:

- Data remains confidential.
- Data cannot be modified.
- Data cannot be replayed.
- Lost packets can be detected.

Secure data transmission is the transfer of data, such as private or confidential information, over a secure channel, making the data transmission hard to intercept and tamper with.

Some secure transmission methods are given below :

Wi-Fi Protected Access (WPA):

A security standard for computing devices using wireless internet connections is called WPA. The key that computers utilize is dynamically changed by the temporal key integrity protocol (TKIP), which is used by WPA. As a result, attackers are unable to produce an encryption key that is identical to the one used by the secure network.[28]

Encryption:

Encryption scrambles the original representation of information, known as plain text, into an unreadable format known as cipher text via various procedures.

There are numerous encryption methods,[29]such as:

1. Symmetric Encryption
   - DES (Data Encryption Standard)
   - 3DES (Triple DES)
   - Advanced Encryption Standard (AES)
2. Asymmetric Encryption
   - RSA (Rivest–Shamir–Adleman)
   - ECC (Elliptic-curve cryptography)
3. Hybrid Encryption
4. PGP (Pretty Good Privacy) or GPG (Gnu Privacy Guard)
5. Public key infrastructure (PKI)

Tokenization:

When sensitive piece of data is replaced with a non-sensitive equivalent referred to as a token that has no exploitable value as part of the tokenization process. A tokenization system uses the reference (also known as an identifier) of the token to map back to the sensitive data.[30]

Communication protocol:

A set of guidelines known as a communication protocol enables information to be transmitted between two or more entities in a communications system.[31] There are different methods of communication protocol, such as:

1) Transmission Control Protocol (TCP)
2) Internet Protocol (IP)
3) User Datagram Protocol (UDP)
4) Post office Protocol (POP)
5) Simple mail transport Protocol (SMTP)
6) Secure File Transfer Protocol (SFTP)
7) Hyper Text Transfer Protocol Secure (HTTPS)
8) Telnet
9) Gopher

Cryptographic protocol:

Secure connections are made possible by cryptographic protocols, allowing two parties to interact while maintaining their privacy and data integrity. [32]

- Off-the-Record Messaging (OTR)
- SSL (Secure Sockets Layer)
- Transport Layer Security (TLS)

## 1.8. Motivation of work

There are many advantages to using a VoIP system for audio transmission. One of the most important advantages is the possibility of saving money. As normal phone calls from one country to another cost a lot of money, while VoIP only takes the amount of data needed to upload and download the given audio. In addition, the equipment used by VoIP is mobile, so one can move from one place to another during a call session. VoIP uses bandwidth efficiently because 50% of audio conversations are silent, so users do not receive bandwidth when they are not talking. . This VoIP system can be created in such a way that it can be used for monitoring purposes, for example, if the source system with the microphone is hidden in the room and people are talking to each other, the system will automatically listen, record and download their conversation to the cloud from where the conversation can be broadcast over and over again. This system was designed to create a modern replica of a mysterious machine. How dolphins in the sea communicate with each other.

### 1.9. Application

The VoIP system can be implimented into automobile, ships, submarine, surveillance devices as the system requires very small space. However, the specific application of this VoIP system is as a covert listening device, also known as a bug or a wire. It is a common technique in surveillance, espionage and police investigations where a confidential conversation needs to be investigated.

### 1.10. Objectives of the Thesis

1. To design VOIP systems for secured transmission of data.
2. To construct a VOIP system according to Design.
3. To analyze sound wave of the voice module used in system.
4. To analyze sound received from the VoIP system.

## 1.11. Scope and Limitations

This is a huge sector of work according to research gaps. But, we are focusing mainly on IoT-based microcontroller VoIP system which increases security, helps store the data and as development board is small they can be compacted into very small.

Microcontrollers are small computers that are able to do a lot of tasks. Some microcontrollers are able to connect to the internet and transfer data through it. So for that reason, it is very easy to transfer packets of data or data of voice through the internet to desired location (device or website).

The VoIP system made with microcontrollers is very cheap and secure. These systems can be used in different applications (Home security, Gathering information, etc.). These systems can be used widely. Also as it is programmed to send data to a specified location third-party hackers will have a problem getting data from the system and if they are able to do so they will receive only encoded data which can not be converted to sound signal easily.

# CHAPTER 2

# LITERATURE REVIEW

Numerous kinds of research on VoIP are ongoing because it can provide a simple, secure, cheap, and fast way of communication between two devices.

M. Zubair Rafique, M. Ali Akbar, and Mudassar Farooq (2009) published “Evaluating DoS Attacks against Sip-Based VoIP Systems”. This paper deals with the launching of several Denial of Service (DoS) attacks on a SIP-based VoIP infrastructure that can severely compromise its reliability. Here it is shown that the robustness and reliability of generic SIP servers are more inadequate than commonly perceived. The measurements are shown in the paper show that a standard SIP server can be easily overloaded by sending simple call requests.[33]

Vinod Namboodiri and Lixin Gao (2010) published a research paper “Energy-Efficient VoIP over Wireless LANs” in which the Greenall algorithm was proposed to derive sleep/wake-up schedules for the WLAN radio to save energy during VoIP calls while ensuring that application quality is preserved within acceptable levels of users. Greenall was evaluated on commodity hardware and studied its performance over diverse network paths and describe our experiences in the process. It further extensively investigated the effect of different application parameters on possible energy savings through trace-based simulations. It was shown that, despite the interactive, real-time nature of voice, energy consumption during calls can be reduced by close to 80 percent in most instances.[34]

In “Voice-over-IP Security: Research and Practice”, Here authors discuss the rapidly adopting Voice-over-IP technologies, which offer higher flexibility and more features than traditional telephony infrastructures. The author surveyed vulnerabilities in the Common Vulnerabilities Exposures (CVE) database and two IETF RFC Internet drafts. These issues ranged from relatively straightforward problems that can lead to server or equipment crashes to more serious problems.[35]

Sung-Min Oh, and Jae-Hyun Kim (2010) in their paper “Application-Aware Design to Enhance System Efficiency for VoIP Services in BWA Networks” a cross-layer framework for voice over Internet protocol (VoIP) services in IEEE 802.16 systems. It uses the application session information of the session description protocol to generate the quality of service parameters in IEEE 802.16 systems. This feature allows the system to efficiently allocate the radio resource because it can exactly estimate the properties of VoIP services such as packet size and packet-generation interval. In other words, the cross-layer framework is expected to achieve a novel resource request scheme for a VoIP service that dynamically assigns the radio resource. This paper has analyzed the maximum number of supportable VoIP users for the resource request schemes in terms of the packet generation interval in the silent period, the duration of the silent period, and the major VoIP speech codec. The numerical results show that the proposed scheme can efficiently support VoIP services for various communication.[36]

E. Jammeh, I. Mkwawa, and A. Khan (2010) present a research “Quality of experience (QoE) driven adaptation scheme for voice/video over IP ” a QoE-driven adaptation scheme for voice and video over IP to solve the optimization problem to provide optimal QoE for networked voice and video applications. The adaptive VoIP architecture was implemented and tested both in NS2 and in an Open IMS Core network to allow extensive simulation and test-bed evaluation. Results show that the scheme was optimally responsive to available network bandwidth and congestion for both voice and video and optimized delivered QoE for different network conditions, and is friendly to TCP traffic.[37]

Young Feng Huang, Shanyu Tang, Jian Yuan (2011) “Stenography in Inactive Frames of VoIP Streams Encoded by Source Codec” describes a novel high-capacity steganography algorithm for embedding data in the inactive frames of low bit rate audio streams encoded by G.723.1 source codec, which is used extensively in Voice over Internet Protocol (VoIP). This study reveals that, contrary to existing thought, the inactive frames of VoIP streams are more suitable for data embedding than the active frames of the streams; that is, steganography in the inactive audio frames attains a larger data embedding capacity than that in the active audio frames under the same imperceptibility. Performance evaluation shows embedding data in various speech parameters led to different levels of concealment.[38]

The paper “A comprehensive review on VoIP over Wireless LAN networks” demonstrates a survey on VOIP over WLAN, its advantages and challenges, and also VOIP capacity over WLAN and the number of calls for different voice codecs and intervals based on IEEE 802.11b standard.[39]

M. A. Mohamed (2012) “Performance Analysis of VoIP Codecs over WiMAX Networks” Here authors evaluate the performance of different VOIP codecs over the WiMAX network. Simulation results show that G.723 is better than codecs G.711, G.726, G.728, and G.729 because it has a lower delay and higher MOS, traffic received, and throughput.[40]

This paper “Performance Evaluation of Audio Codecs using VoIP Traffic in Wireless LAN using RSVP” deals with VOIP over VPN implementation between open-source applications and hardware devices in a campus environment. In this, the evaluation of the performance of CPU Utilization, Jitter, Delay, Memory Usage, and MOS(mean opinion score).[41]

Elechi Onyekachi O., and Eze Elias C. in their paper “Investigating the QoS of Voice over IP using WiMAX Access Networks in a Campus Network” attempted to identify some of the network performance parameters that service providers will focus on to develop a VOIP over WIMAX communication tool that will serve as a voice communication broadband replacement technology to old circuit switch voice communication.[42]

The paper titled “Speech Quality Evaluation Based Codec for VOIP Over 802.11P” provides a quick and technical overview of the concept, standard, technology, and architecture for IEEE 802.16 WiMAX.[43]

Ramesh, D. (2011) in his paper “Techniques to Improve Performance of VoIP over 802.11e WLAN.” presented a Media Access Control Protocol that provides the quality of service for VOIP over WLAN. In this, the characteristics of our proposed protocol are No hardware modification of VOIP STA. Backward compatibility to minimize the cost of development no modification of access points.[44]

In the research article “Predictive Modeling in a VoIP System,” the authors look at how the user profiles, peak hours, or call patterns are shaped for a real system and, in a second step, at constructing a model that is capable of predicting trends. An important problem one needs to deal with in a Voice over IP system is server overload. One way to prevent such problems is to rely on prediction techniques for the incoming traffic, namely to proactively scale the available resources. Anticipating the computational load induced on processors by incoming requests can be used to optimize load distribution and resource allocation.[45]

Preetinder Singh, and Ravneet Kaur's review paper “VOIP Over Wimax: A Comprehensive Review” focused on the quality of service scheduling services and performance-related metrics such as jitter, packet end-to-end delay, and MOS(mean opinion score).[46]

Liping Zhang, Shanyu Tang, and Zhihua Cai (2013) in the paper “Efficient and flexible password-authenticated key agreement for Voice over Internet Protocol Session Initiation Protocol using the smart card” presents an efficient and flexible password-authenticated key agreement protocol for session initiation protocol associated with Voice over Internet Protocol. The proposed protocol has many unique properties, such as session key agreement, mutual authentication, and password updating function and the server does not need to maintain a password or verification table, and so on. In addition, its protocol is secure against the replay attack, the impersonation attack, the stolen-verifier attack, the man-in-the-middle attack, the Denning–Sacco attack, and the offline dictionary attack with or without the smart card.[47]

Again Liping Zhang, Shanyu Tang, and Shaohui Zhu (2016) in the study “A lightweight privacy-preserving authenticated key agreement protocol for SIP-based VoIP” biometrics is employed to design a lightweight privacy-preserving authentication protocol for SIP based on symmetric encryption, achieving a delicate balance between performance and security. In addition, the proposed authentication protocol can fully protect the privacy of biometric characteristics and data identity, which has not been considered in previous work. The completeness of the proposed protocol is demonstrated by Gong, Needham, and Yahalom (GNY) logic. Performance analysis shows that the proposed protocol increases efficiency significantly in comparison with other related protocols.[48]

The review paper “A Review Paper: Voice over Internet Protocol” deals with VOIP to a level that allows discussion of security purposes and concerns. In this, VOIP components will include network components, gateway, end-user equipment, call processors, and two common architectures.[49]

Jonathan Lee, Kyumin Cho, ChangYong Lee, Seungjoo Kim (2014) “VoIP-aware network attack detection based on statistics and behavior of SIP traffic” a VoIP-aware attack-detection scheme is proposed. The proposed scheme can detect VoIP network attacks including VoIP DoS and SPAM. It can detect VoIP DoS attacks with low false negatives using a statistics-based detection algorithm and can recognize SPAM with low false positives using a caller behavior-based detection algorithm. We have included experimental results to confirm the proposed scheme.[50]

De Pessemier, Toon & Stevens, Isabelle & Marez, Lieven & Martens, Luc & Joseph, Wout. (2014).“Analysis of the quality of experience of a commercial voice-over-IP service.” demonstrates the influence of technical parameters (such as the audio codec, type of data network, and handovers during the call), device characteristics (such as the platform, manufacturer, model, and operating system), and application aspects (such as the software version and configuration) on the subjective quality of a commercial VoIP service. The relative influence of all these parameters is determined and a decision tree combines these results to assess the subjective quality.[51]

Jalendry, Sheetal & Verma, Shradha (2015), “A Detail Review on Voice over Internet Protocol (VoIP)” gives a brief introduction to VoIP technology: the network structure, protocols, echo and delay, jitter, and packet loss in VoIP network. Finally, the survey concludes with a discussion on the feasibility of providing VoIP over challenging satellite links.[52]

Allawi, Khamis & Al-Aqrabi, Hussain. (2015) “Quality of service evaluation of VoIP over wireless networks” demonstrates the performance of Voice over IP (VoIP) in 802.11 wireless networks and elaborates on the evaluation of voice packet end-to-end delay and throughput. Employing literature reviews and an experimental model created on OPNET that is simulated to assess the quality of service (QoS) of VoIP in 802.11g legacy and 802.11e wireless network; shows the enhancement of 802.11 reflects as enhancement in the quality of the VoIP service. The simulation results have indicated that the quality of VoIP service is influenced by the quality of the carrier which is the IEEE 802.11 network. Therefore, the voice service over a wireless network can be improved significantly by developing a quality of service policy that prioritizes the packet transmission based on the controlled access mechanisms.[53]

Tian, Hui & Qin, Jie & Guo, Shutting & Huang, Yongfeng & Liu, Jin & Wang, Tian & Chen, Yonghong & Cai, Yiqiao. (2015), “Improved adaptive partial-matching steganography for Voice over IP.” Adaptive Partial-matching Stenography (APMS) was tried to improve from three aspects. First, when the current partial similarity value (PSV) falls between the two given substitution thresholds, the embedding process is guided with unequal probabilities increased with the PSVs instead of identical probabilities, which enhances the embedding chances of covering parts with larger PSVs. Second, for cover parts considered unsuitable for information hiding in the original strategies, the matrix encoding strategy is employed to conceal secret messages with no more than one bit changed, which improves the cover usage rate while maintaining high embedding efficiency. Third, in the APMS strategy with two flag bits, one more encrypted form of the secret message is added to the comprehensive evaluation of partial matching, which makes the embedding process more efficient. The proposed strategies are evaluated with ITU-T G.711 (A-law) as the codec of the cover speech and compared with the original ones and some existing approaches. The experimental results demonstrate that the improved strategies can strike excellent balances between steganographic transparency and bandwidth, and significantly outperform both the former ones and other previous approaches. Furthermore, the results of detection tests show they have an outstanding performance of steganalysis resistance.[54]

K. Mohamed, O. Mohamed, M. Hamoudi, and M. Masmoudi, (2015) "QoS evaluation in VoIP software with and without Blowfish encryption module," deals with considerable cost savings in comparison with traditional phone lines. In this paper, a proposal was given to use the Blowfish encryption algorithm to encrypt the audio communication between VoIP clients as the used algorithm was previously compared to the Advanced Encryption Standards (AES) and it was shown that it is faster and offers better throughput.[55]

Emily J Ricketts, Amy R Goetz, Matthew R Capriotti (2015) ”A randomized waitlist-controlled pilot trial of voice over Internet protocol-delivered behavior therapy for youth with chronic tic disorders” Comprehensive Behavioral Intervention for Tics (CBIT) is efficacious for chronic tic disorders (CTDs), but utilization is limited by a lack of treatment providers and perceived financial and time burden of commuting to treatment. A promising alternative to in-person delivery is the voice-over-Internet protocol (VoIP), allowing for remote, real-time treatment delivery to patients' homes.[56]

Sonali Golhar, V. S. Dhamdhere (2016), "Voice over Internet Protocol (VoIP) Based IP PBX System Design" explains and analyzes the feasibleness of the results of the VoIP system designed if enforced during a field network like national capital University telephone network. So the convergence strategy of an IP PABX system with the present wire telephone network has been mentioned. Whereas taking this into thought, the upgrading technique of the prevailing system using trendy network appliances has been delineated and at a similar time, a new configuration connecting this science network with the telephone network has been designed. This study entails the simulation and implementation of a VoIP telecom system mistreatment of an IP PBX resolution.[57]

Sha, Mohemmed & Thankappan, Manesh & Abd El-Atty, Saied. (2016). The "VoIP Forensic Analyzer" paper presents an idea of a framework for the forensic analysis of VoIP traffic over the network. This forensic activity includes spotting and scrutinizing the network patterns of the VoIP-SIP stream, which is used to initiate a session for communication, and regenerating the content from the VoIP-RTP stream, which is employed to convey the data. The proposed network forensic investigation framework also accentuates developing an efficient packet restructuring algorithm for tracing the depraved users involved in a conversation.[58]

Hynes, N. and Elwell, A.D. (2016), "The role of inter-organizational networks in enabling or delaying disruptive innovation: a case study of mVoIP", the purpose of this paper is to investigate the role of inter-organizational networks in the emergence of disruptive technology by taking a snapshot view of the UK market for mobile voice over internet protocol (mVoIP) technologies. By delineating the technologies required for a VoIP call to occur, the role of incumbents, technology, and inter-organizational networks is explored. The paper is based on secondary data and the research results may lack generalizability to other technologies.[59]

Lange, Mirjam & Saric, Amela. (2016) "Substitution Between Fixed, Mobile, and Voice over IP Telephony – Evidence from the European Union" explores the access substitution between fixed lines, mobiles, and managed VoIP in a unified EU cross-country framework. We employ a half-yearly dataset for 20 EU countries for the 2008–2011 period and apply dynamic panel data methods. The analysis demonstrates strong access substitution between fixed lines and mobiles and provides indicative evidence of the substitution between fixed lines and VoIP.[60]

Dr Mahdi H. Miraz (2017). "Simulation and Analysis of Quality of Service (QoS) Parameters of Voice over IP (VoIP) Traffic through Heterogeneous Networks" deals with identifying those causes and parameters that affect the Quality of Service (QoS) of Voice-over-Internet Protocol (VoIP) through heterogeneous networks such as WiFi, WiMAX and between them are carried out using the OPNET simulation tool. Optimization of the network for both intra- and intersystem traffic to mitigate the deterioration of the QoS is discussed. The consequences of this optimization, leading to a solution, which can ameliorate the QoS over these networks are analyzed and offered as the conclusion of this research.[61]

Meeran, Mohammad & Annus, Paul & Alam, Muhammad & Le Moullec, Yannick. (2017), "Evaluation of VoIP QoS Performance in Wireless Mesh Networks." focus of this paper is the evaluation of selected voice over Internet protocol (VoIP) solutions in wireless mesh network (WMN) scenarios. While WMNs have self-healing, self-forming, and dynamic topology features, they still pose challenges for the implementation of multimedia applications such as voice in various scenarios. Therefore, various solutions to make WMN more suitable for VoIP applications have been proposed in the scientific literature. In this work, a set of application scenarios were extensively explored by conducting experiments using a network simulator. [62]

Ganesan Vennila, M.S.K. Manikandan, M.N. Suresh (2018), Dynamic voice spammers detection using Hidden Markov Model for Voice over Internet Protocol network, In this paper, a Dynamic Voice Spammer Detection Model (DVSDM) based on the Hidden Markov Model (HMM) is proposed. This model estimates voice spammer's states by using various behavior variables and detects the voice spammers before reaching the victim. The performance of this detection model is experimentally evaluated with two scenarios (mild and heavy distribution of voice spam calls).[63]

Liang, Chen & yu-an, Tan & Zhang, Xiaosong & Wang, Xianmin & Zheng, Jun & Zhang, Quanxin. (2018) in "Building packet length covert channel over mobile VoIP traffics" proposed constructing a covert channel via partitioning the length distribution of legitimate packet traffics and mapping packet length partitions to data symbols. The reliable data transmission of a covert channel is guaranteed by precoding the hidden data into an HDLC-like frame structure and the channel performance can be raised by selective parameter settings. A testbed is a setup to simulate covert communication over natural mobile VoIP traffic. The Undetectability, reliability, and throughput of the scheme are evaluated and experiment results show the potential of the proposed approach. Moreover, the proposed scheme can also be applied to other communication scenarios without or with little modifications.[64]

Chen Liang, Xianming Wang, and Xiaosong Zhang (2018) in "A payload-dependent packet rearranging covert channel for mobile VoIP traffic" described a covert channel scheme that communicates by rearranging the packet sending sequences while keeping the undetectability and robustness. Unlike the existing packet rearranging covert channels which may concentrate on packet rearranging according to the existing packet features such as packet id and packet length, the proposed scheme focuses on building packet rearranging covert channels whose function is regardless of the variation on legitimate traffic. [65]

Waleed Nazih, Yasser Hifny, Wail Elkilani (2019) "Efficient Detection of Attacks in SIP Based VoIP Networks Using Linear I1-SVM Classifier" a system was proposed to detect the Denial of Service (DoS) attacks and Spam over Internet Telephony (SPIT) attack in the SIP Based VoIP networks using a linear support vector machine with I1 regularization classifier. The experimental results show that the proposed system detects malformed messages, and invites flooding and SPIT attacks with high accuracy. In addition, the proposed system outperformed other systems significantly in the detection speed.[66]

Mallik, Avijit & Ahsan, Abid & Shahadat, Mhia & Tsou, Jia-Chi (2019) "Man-in-the-middle-attack: Understanding in simple words" here the authors discuss how a third party can collect data or information from a VoIP session through different types of attacks and describes the security level of VoIP at whole.[67]

From these literature reviews, it was decided to concentrate on IoT-based microcontrollers VoIP systems which increased with security and data transmission.

## 2.1. Research Gaps

A lot of advancements have been done in this sector already but throughout studying various papers and topics, there are still a few things that can be improved. Such as,

1. Security in connectivity between devices.
2. Non-detectable voice telecommunication.
3. Secured data storing in private cloud.
4. Compacting system to be used as non-detectable surveillance device.

## 2.2. Research Methodology

The following research methodology was used in this work:

1. Perform extensive literature research on sound signal processing, conversion, storing.
2. Perform extensive literature research on voice over internet protocol techniques.
3. Identify weaknesses in current voice communication techniques and find ways to make a better VoIP system.
4. Explore different IoT implementation approaches such as Arduino, Raspberry Pi, Particle Photon and choose the optimal platform to fabricate the system.
5. Explore and choose different components.
6. Choose the most optimal technique and implement it.
7. Assemble the system, solder the components, connect all sensors and program them.
8. Write JavaScript which will access the VoIP system server and store the file locally and in online cloud.
9. Evaluate audio file clarity and other characteristics.

# CHAPTER 3

# METHODOLOGY

## 3.1. Algorithm Development

This project deals with the recording of audio of an area and storing the audio to a storage device or a server for secured easy access. For this purpose, the sound waves of an area is converted into electrical analog data by the MAX9814 microphone and the data is sent to particle photon. The particle photon creates a TCP (Transmission Control Protocol) server that continuously uploads encrypted packet of data. If any device wants to listen or download the audio it needs to access the TCP server, next download the encrypted packet of data, then decrypt the packet of data and combine the packets of data to generate a audio file to store locally and online. The audio can no be easily accessed by third party personnel because of the following reasons :

- They need to be connected to the same WiFi network as particle photon.
- They would need the correct IP address and Port to access the TCP server and download the packets data.
- They would need to decrypt the packet of data.
- They would lastly have to convert the data to binary wave file to make it into a audio file.

Figure 7 and figure 8 shows the block diagram of the VoIP system with particle photon and flow chart of the experimental VoIP system respectively.

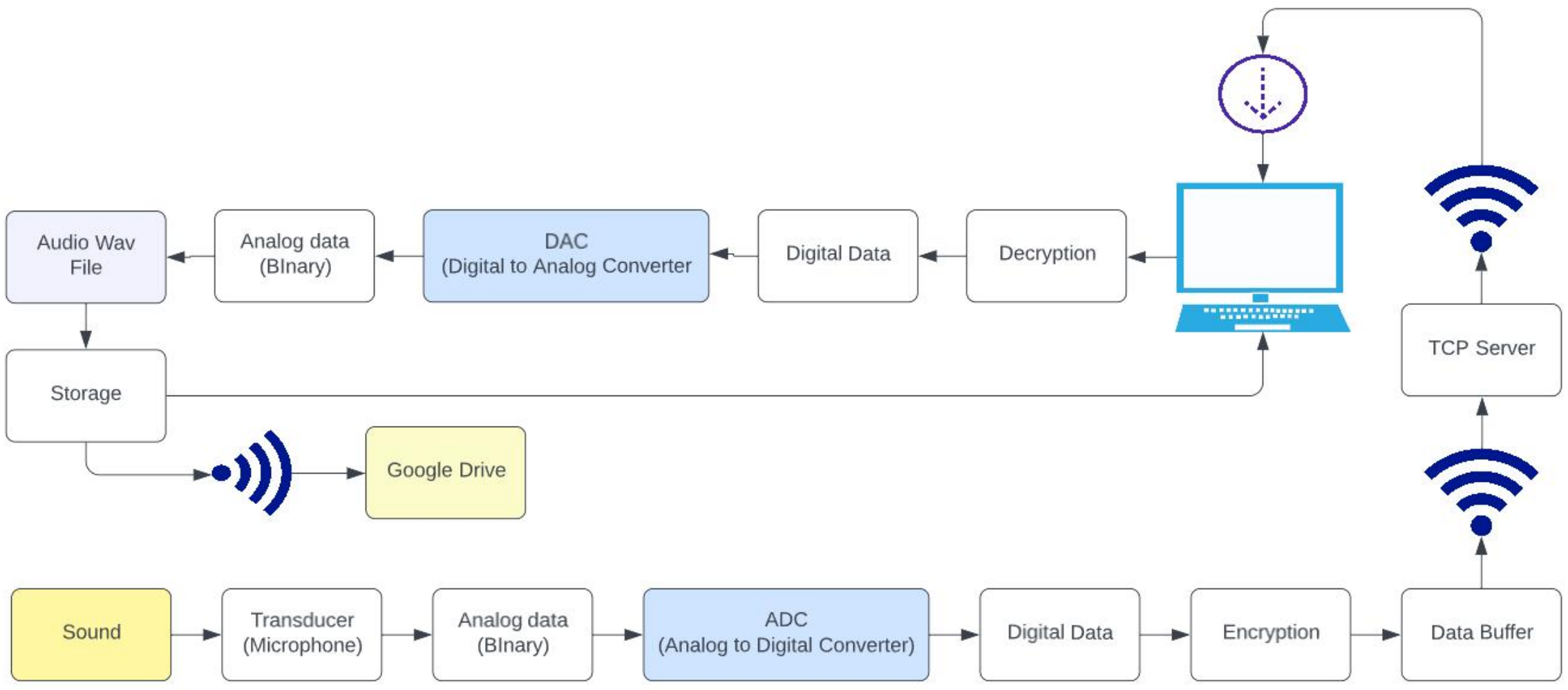

Figure 7 : Block diagram of VOIP system with Particle photon.

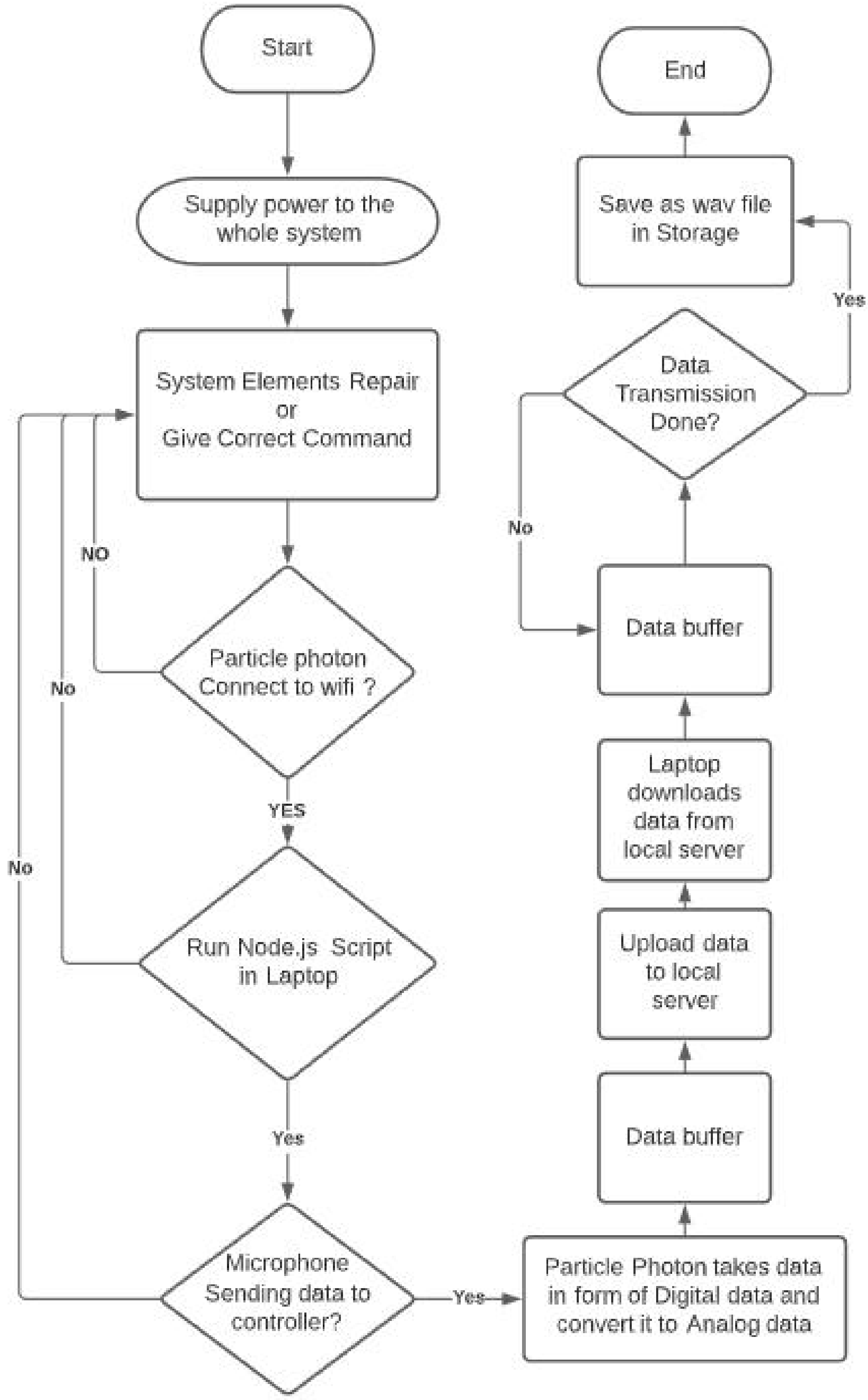

Figure 8 : Flow chart of VoIP system with Particle Photon.

## 3.2. Experimental Setup

Figure 9 shows the schematic diagram of the experimental setup of the VoIP system.

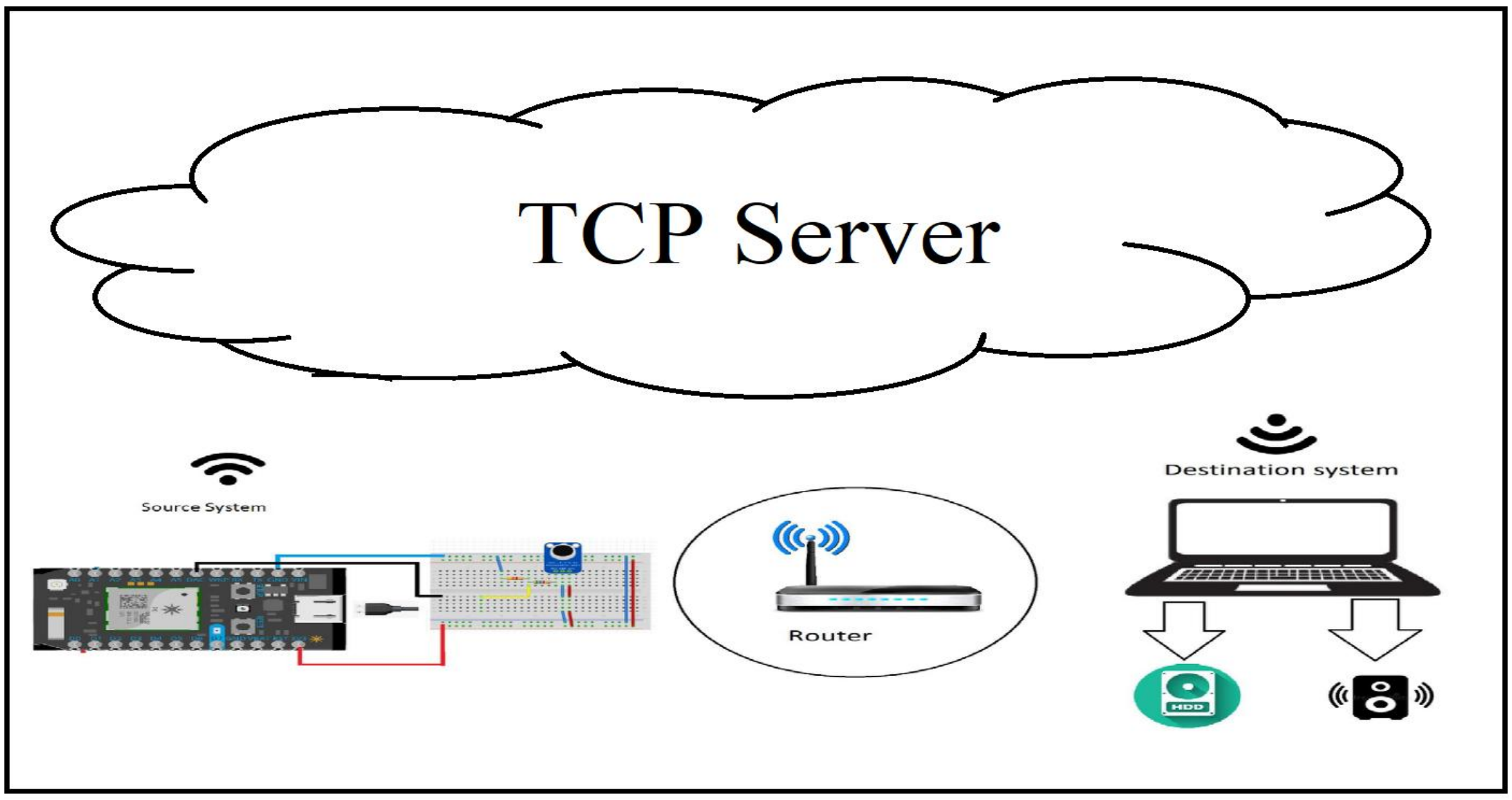

Figure 9 : VoIP system schematic diagram.

The source system circuit diagram for VoIP system with Particle Photon is given below,

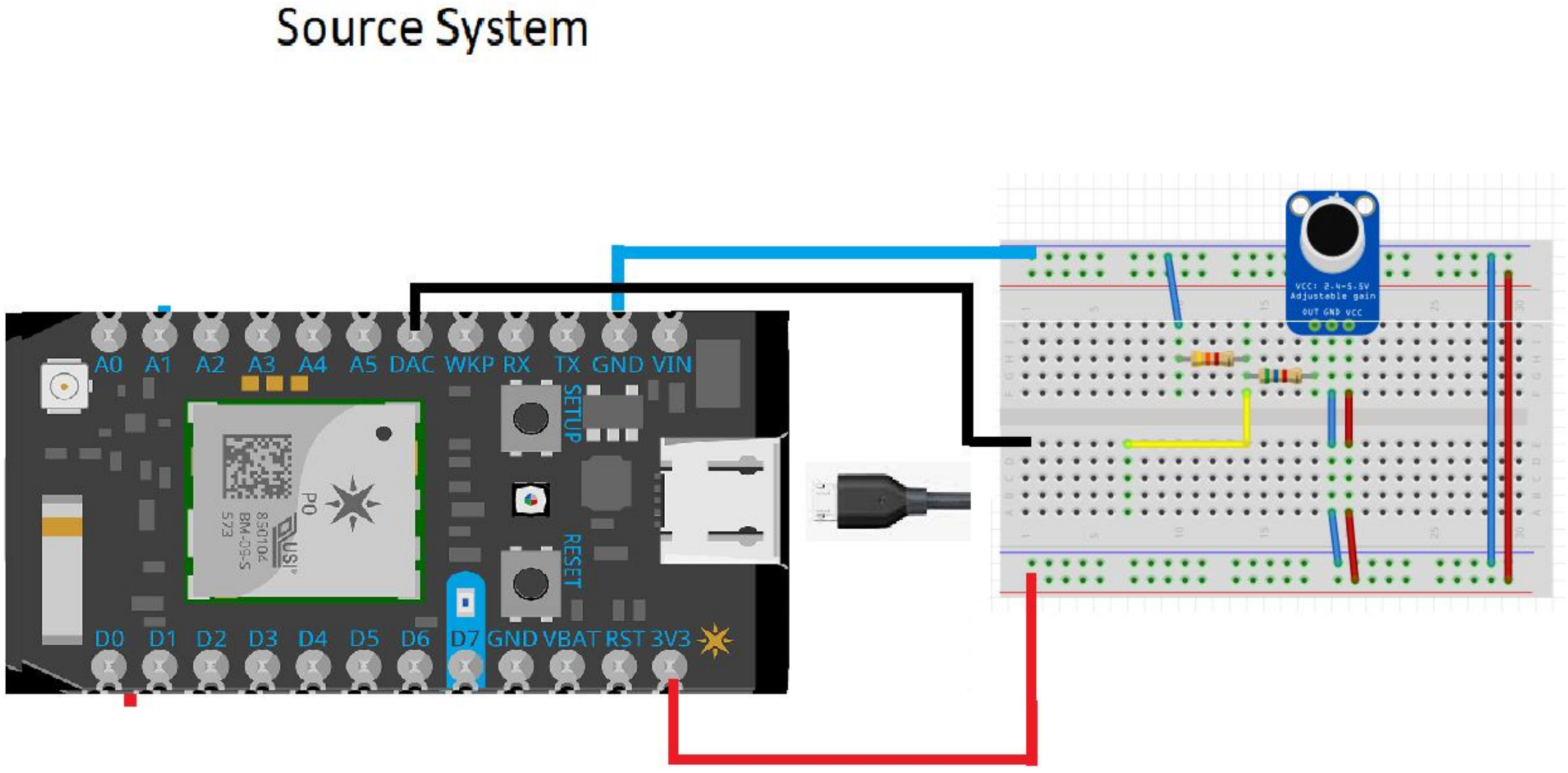

Figure 10 : Particle Photon source system circuit diagram.

The VoIP system with Particle photon was configured according to circuit diagram. The Particle Photon is powered by a USB cable. The positive and negative rails of the model are supplied respectively by the Particle Photons 3v3 pin and the GND pin. The VCC pin of the voice module is connected to the positive bar of the breadboard and the GND pin is connected to the negative terminal of the breadboard. The OUT pin of the Voice Module is connected to the DAC pin of the Particle Photon.

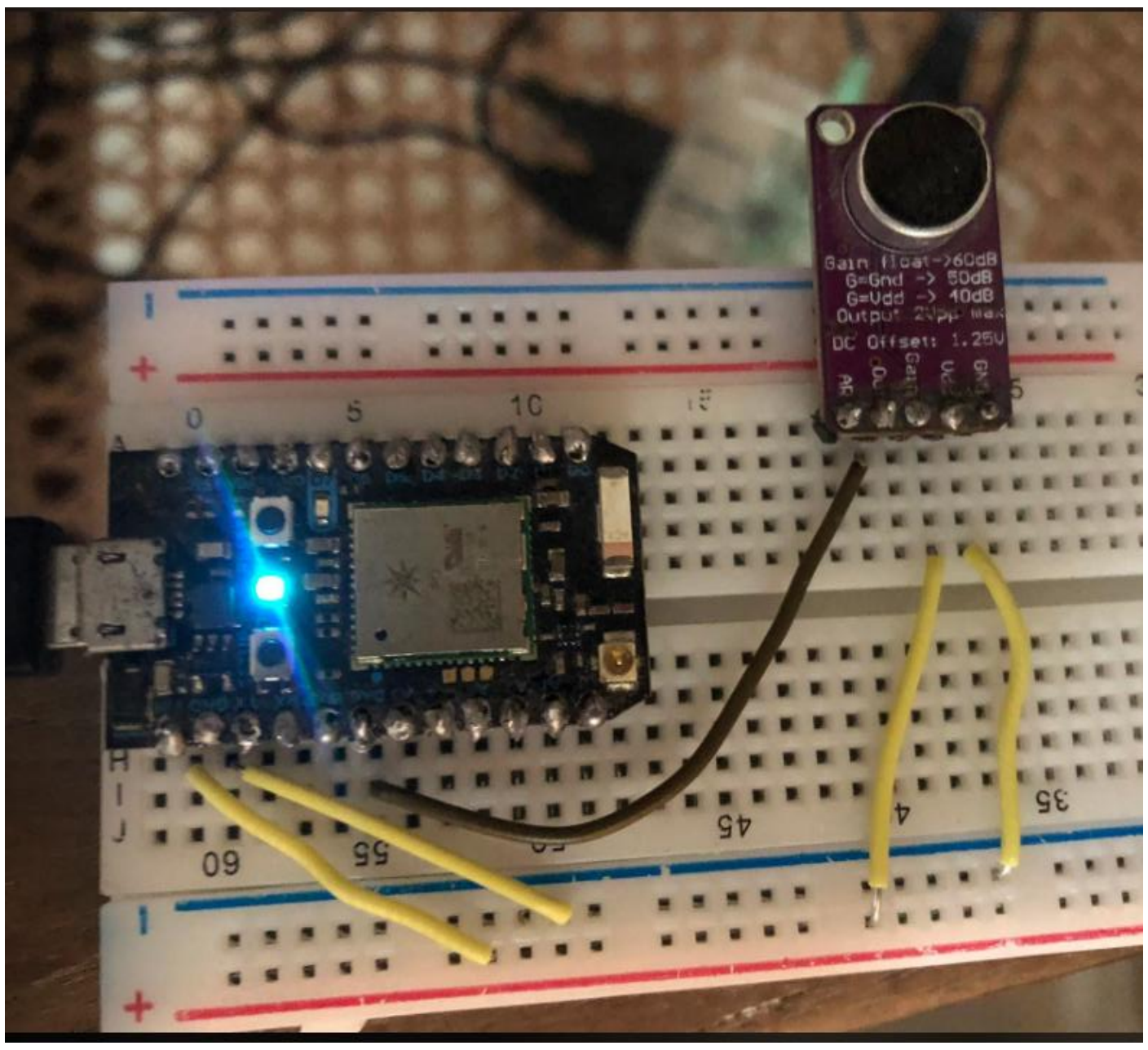

Figure 11 : Experimental Setup of Source system of the VoIP system.

The destination system includes a laptop that can run Node.js script.The Particle Photon microcontroller uses the internet to upload buffer data to its own server,  from which the destination system can download the data for storing, streaming, or future use through using the JavaScript.

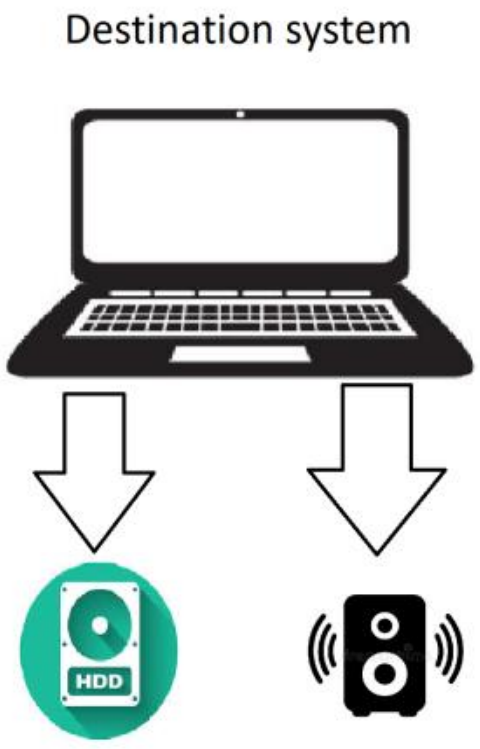

Figure 12 : Laptop Destination System.

## 3.3. Hardware Components

The hardware components consists of the following parts :

- Particle Photon Microcontroller
- MAX9814 Electret Microphone
- Capacitor
- Power Supply 5V
- Router
- Jumper Wires
- Laptop and etc.

## 3.4. Hardware Configurations

### 3.4.1. Particle Photon Microcontroller

Particle Photon is a complete IoT (Internet of Things) hardware development kit that provides everything needed to build a connected product. It has a powerful 120MHz ARM Cortex M3 microcontroller with a Broadcom WiFi chip in a tiny miniature module.[68] Technical specifications are given below:

- Processor: STM32F205 120Mhz ARM Cortex M3
- Memory: 1MB flash, 128KB RAM
- On-board RGB status LED.
- 18 Mixed-signal GPIO and advanced peripherals
- Open source design
- Real-time operating system (Free RTOS)
- Soft AP setup
- On-board Wi-Fi module
- Broadcom BCM43362 Wi-Fi chip
- 802.11b/g/n Wi-Fi

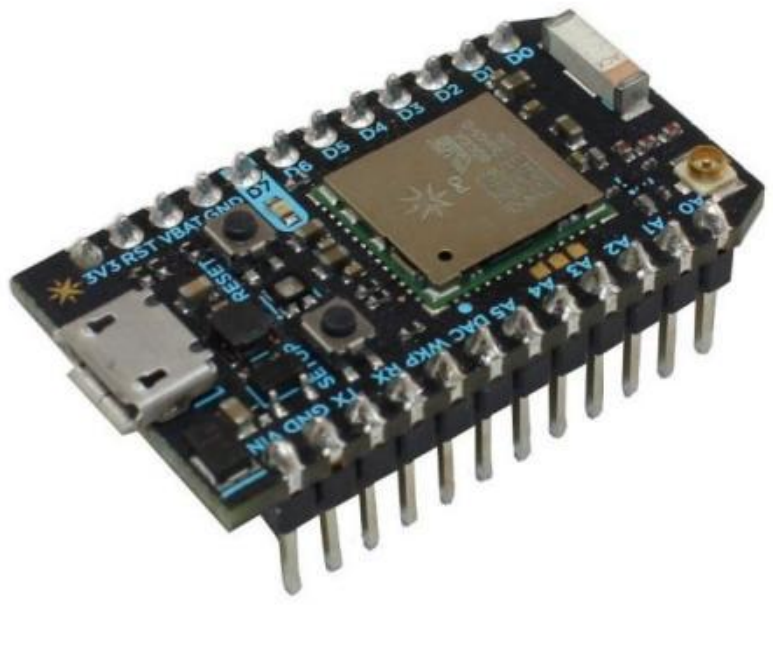

Figure 13 : Particle Phonton Microcontroller.

Particle provides access to a free cloud service of the particle cloud. The Particle Cloud has some great features for building connected projects, including over-the-air firmware updates, an easy-to-use REST API, and firmware development supported by Web and local IDEs.[68]

The Photon has ton of capability in a small footprint, with analog, digital and communication interfaces.

Table 2: Particle Photon Pin Settings.[68]

| Peripheral Type | Qty | Input(I) / Output(O) | 5V / 3V3 |
|---|---|---|---|
| Digital | 18 | I/O | 5V/3V3 |
| Analog (ADC) | 8 | I | 3V3 |
| Analog (DAC) | 2 | O | 3V3 |
| SPI | 2 | I/O | 3V3 |
| I2S | 1 | I/O | 3V3 |
| I2C | 1 | I/O | 5V |
| CAN | 1 | I/O | 3V3 |
| USB | 1 | I/O | 3V3 |
| PWM | 9 | O | 3V3 |

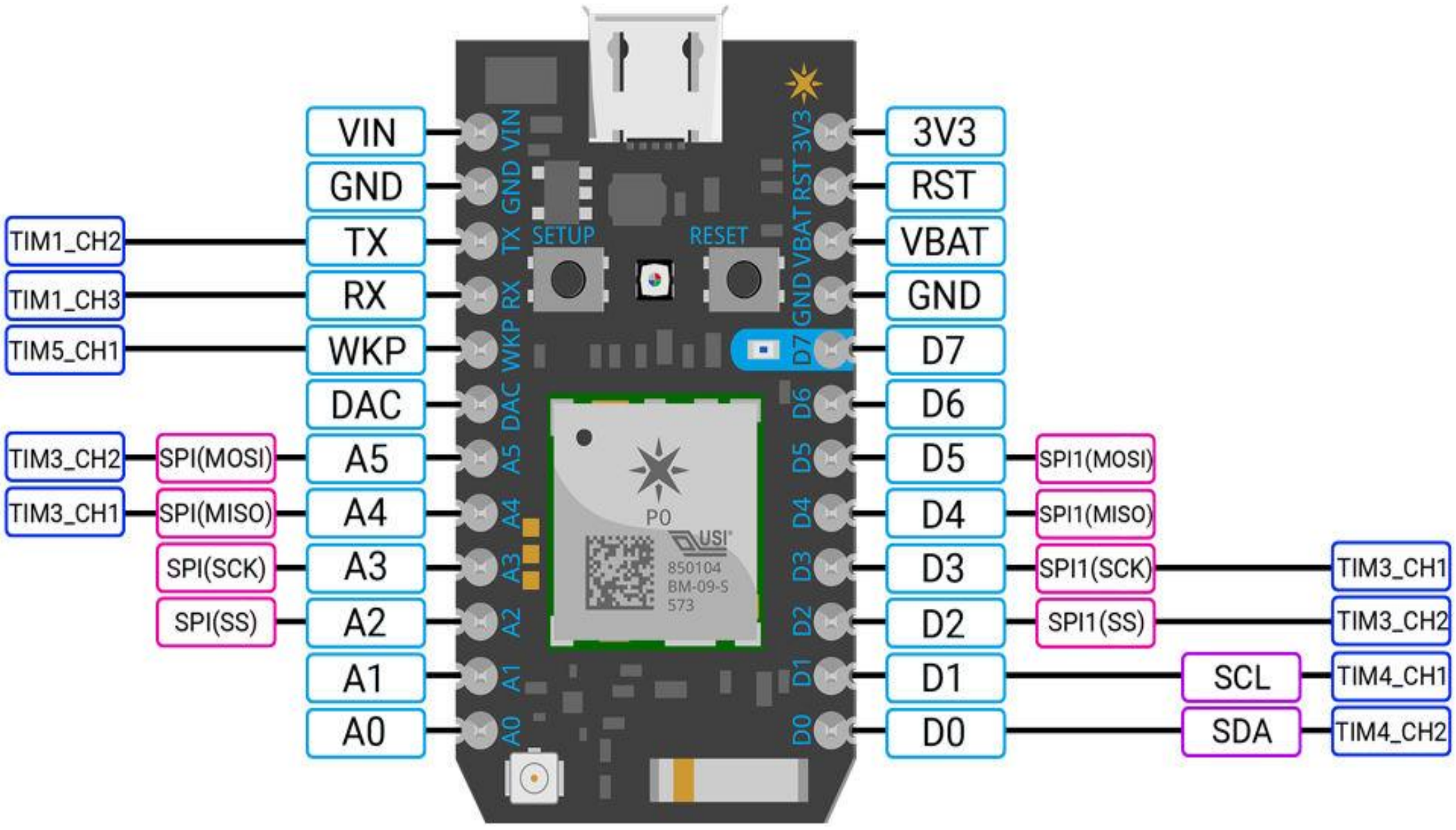

Figure 14 : Particle Photon Pin Output.[68]

Table 3: Particle Photon Pin Description.[68]

| Pin | Description |
|---|---|
| VIN | This pin can be used as an input or output. As an input, supply 3.6 to 5.5VDC to power the Photon. When the Photon is powered via the USB port, this pin will output a voltage of approximately 4.8VDC due to a reverse polarity protection series Schottky diode between VUSB and VIN. When used as an output, the max load on VIN is 1A. |
| RST | Active-low reset input. On-board circuitry contains a 1k ohm pull-up resistor between RST and 3V3, and 0.1uF capacitor between RST and GND. |
| VBAT | Supply to the internal RTC, backup registers and SRAM when 3V3 is not present (1.65 to 3.6VDC). |
| 3V3 | This pin is the output of the on-board regulator and is internally connected to the VDD of the Wi-Fi module. When powering the Photon via VIN or the USB port, this pin will output a voltage of 3.3VDC. This pin can also be used to power the Photon directly (max input 3.3VDC). When used as an output, the max load on 3V3 is 100mA. NOTE: When powering the Photon via this pin, ensure power is disconnected from VIN and USB. |
| RX | Primarily used as UART RX, but can also be used as a digital GPIO or PWM. |
| TX | Primarily used as UART TX, but can also be used as a digital GPIO or PWM. |
| WKP | Active-high wakeup pin, wakes the module from sleep/standby modes. When not used as a WAKEUP, this pin can also be used as a digital GPIO, ADC input or PWM. Can be referred to as A7 when used as an ADC. |
| DAC | 12-bit Digital-to-Analog (D/A) output (0-4095), referred to as DAC or DAC1 in software. Can also be used as a digital GPIO or ADC. Can be referred to as A6 when used as an ADC. A3 is a second DAC output used as DAC2 in software. |
| A0~A7 | 12-bit Analog-to-Digital (A/D) inputs (0-4095), and also digital GPIOs. A6 and A7 are code convenience mappings, which means pins are not actually labeled as such but you may use code like analogRead(A7). A6 maps to the DAC pin and A7 maps to the WKP pin. A4,A5,A7 may also be used as a PWM output. |
| D0~D7 | Digital only GPIO pins. D0~D3 may also be used as a PWM output. |

ESP32 is a low-cost, low-power system on a chip (SoC) series with Wi-Fi & dual-mode Bluetooth capabilities. Both ESP32 and Particle Photon have their pros and cons, but the main feature of Particle Photon that will automatically generate it and an ideal IoT device is that it has its own separate cloud storage. Each Particle microcontroller comes with its own cloud storage, so there is no need to use limited cloud storage or paid cloud storage.

Table 4 : Comparison of ESP32 and Particle Photon.[68-69]

| Specifications | ESP32 | Particle Photon |
|---|---|---|
| CPU Frequency | 160MHz | 120Mhz |
| WiFi | Yes | Yes |
| RAM | 512KB | 128KB |
| Flash | 16MB | 1MB |
| GPIO PINS | 36 | 25 |
| ADC PINS | 18 | 13 |
| DAC PINS | 2 | 2 |
| Separate Cloud Server | No | Yes |
| Size | Bigger | Smaller |
| Cost BDT (Date) | 770 (14/12/2020) | 1611 (14/12/2020) |

### 3.4.2. Electret Microphone & its Working Principle

A microphone is a transducer that converts sound into an electrical signal. Microphones are used in many applications such as Voice over Internet Protocol (VoIP), telephony, hearing aids, film production, live sound engineering and recording, voice recording, radio two-way, radio, radio and television through loudspeakers. Several types of microphones are in use today, using different methods to convert changes in atmospheric pressure from sound waves into electrical signals. [70]

The type of microphone used to build and test our VoIP system was the Electret microphone. An electret microphone, also known as a condenser or condenser microphone, is the most common type of transducer used to sense or measure audio signals. It works exactly the opposite of conventional speakers, i.e. when an audio signal is detected, it generates an electrical signal. Electronic microphones are a type of condenser microphone that eliminates the need for a polarized power source by using a permanently charged material.

Condenser (or condenser) microphones use a lightweight diaphragm and a fixed plate that act as opposite sides of the capacitor. The sound pressure against this thin polymer film makes it move. This movement changes the capacitance of the circuit, which is received by the FET (Field Effect Transistor) impedance converter and thus causes a change in the electrical output. The low mass of the diaphragm allows for a prolonged high-frequency response, while the very nature of the design also ensures exceptional low-frequency reception. The resulting sound is natural, clean and clear, with excellent clarity and detail. [71]

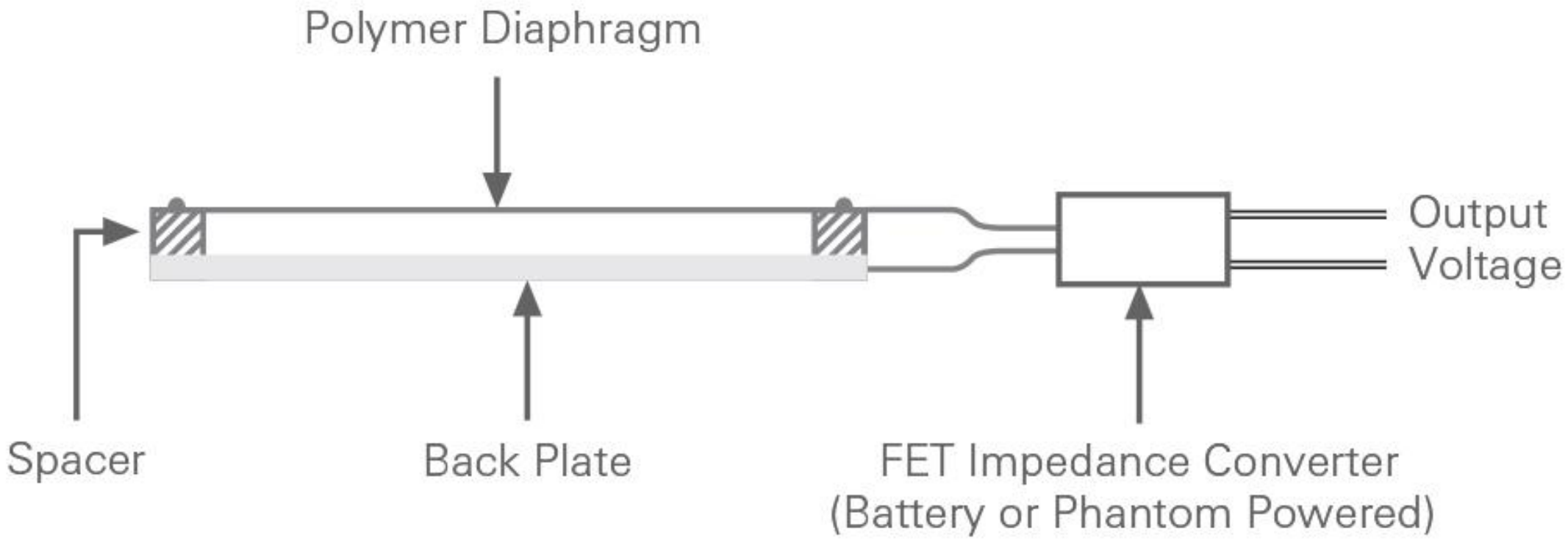

Figure 15 : Electret Condenser Microphone Element.[71]

While the electret condenser microphone doesn't need a power supply to provide polarizing voltage, an FET impedance matching circuit inside the microphone does require some power.

Electret Microphone Specifications [72]:

- Operating Voltage: 2V to 10V
- Current consumption: 0.5mA (max)
- Operating Frequency: 20Hz to 16,000Hz
- Impedance: <2.2kΩ
-

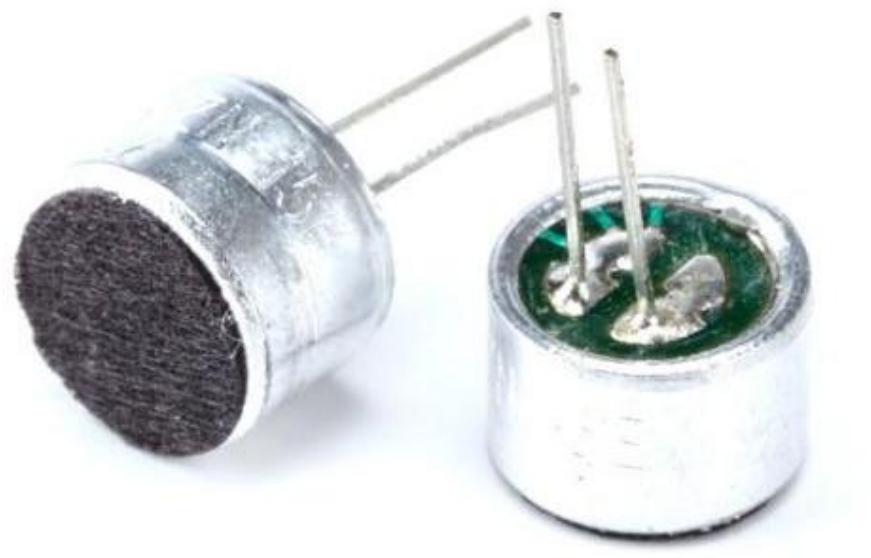

Figure 16 : Electret Microphone.

The Electret microphone also needs a filter to make the sound clearer, reduce noise and eliminate aliasing. For this reason in the system we have used a Voice Module which consist of a Electret condenser microphone to receive sound waves.

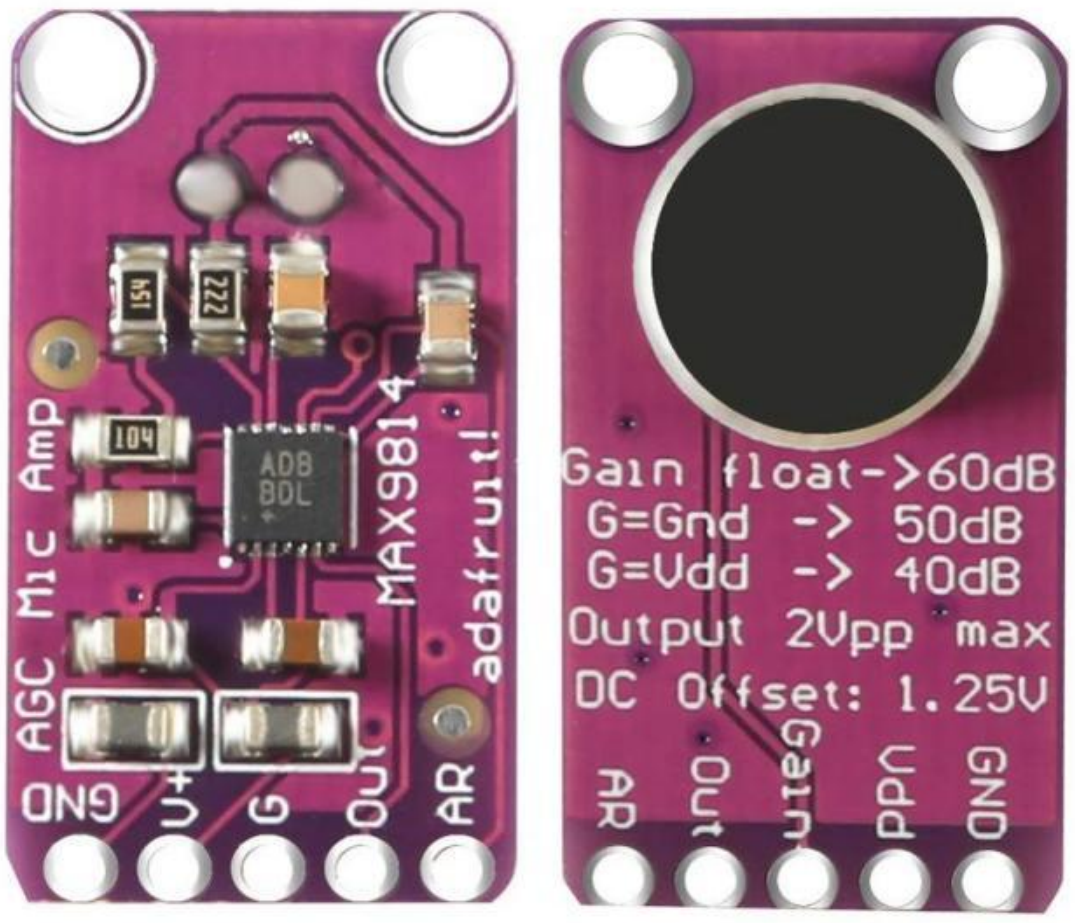

Figure 17 : MAX9814 Electret Microphone Module.

The MAX9814 is a low-cost, high-quality microphone amplifier with automatic gain control (AGC) and low-noise microphone bias. The device features a low-noise preamplifier, variable gain amplifier (VGA), output amplifier, microphone-bias-voltage generator, and AGC control circuitry.[73]

The features of MAX9814 Electret Microphone Module are given below[73]:

1. Automatic Gain Control (AGC)
2. Three Gain Settings (40dB, 50dB, 60dB)
3. Programmable Attack Time
4. Programmable Attack and Release Ratio
5. 2.7V to 5.5V Supply Voltage Range
6. Low Input-Referred Noise Density of 30nV/√Hz
7. Low THD: 0.04% (typ)
8. Low-Power Shutdown Mode
9. Internal Low-Noise Microphone Bias, 2V
10. Available in the Space-Saving, 14-Pin TDFN (3mm x 3mm) Package
11. -40°C to +85°C Extended Temperature Range.

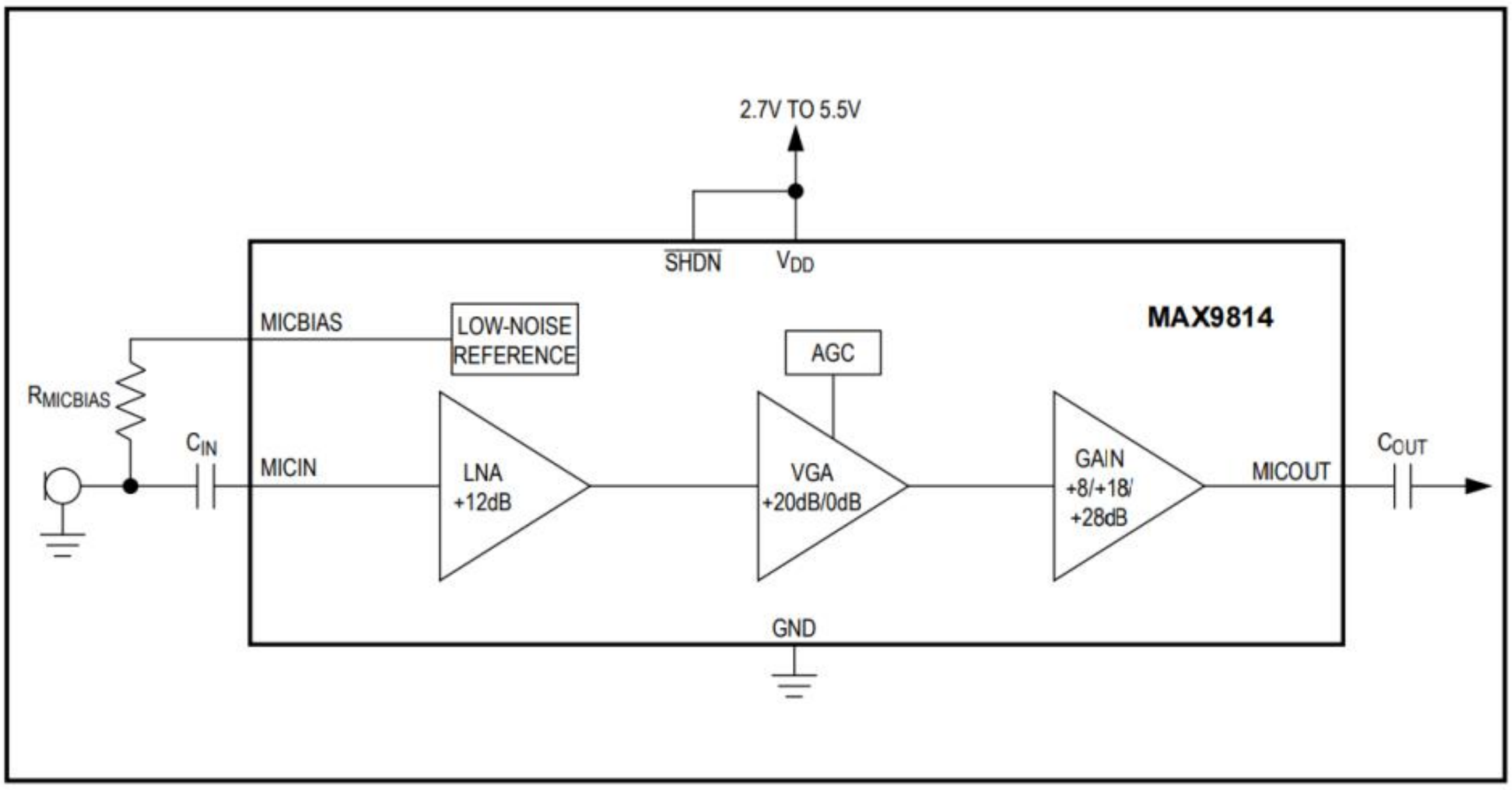

Figure 18 : MAX9814 simplifed block diagram.[73]

The low-noise preamplifier has a fixed 12dB gain, while the VGA gain automatically adjusts from 20dB to 0dB, depending on the output voltage and the AGC threshold. The output amplifier offers selectable gains of 8dB, 18dB, and 28dB. With no compression, the cascade of the amplifiers results in an overall gain of 40dB, 50dB, or 60dB. A trilevel digital input programs the output amplifier gain. An external resistive divider controls the AGC threshold and a single capacitor programs the attack/release times. A trilevel digital input programs the ratio of attack-to-release time. The hold time of the AGC is fixed at 30ms. The lownoise microphone-bias-voltage generator can bias most electret microphones.[73]

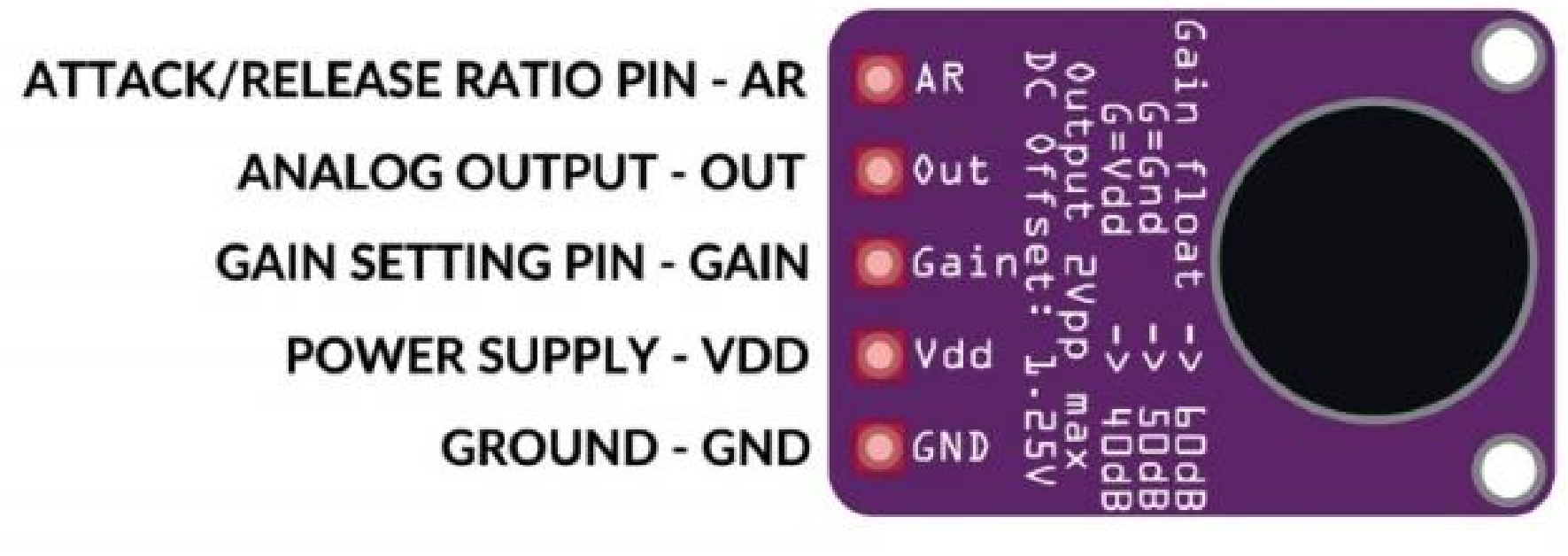

Figure 19 : MAX9814 Pinout.[73]

Typical Operating Characteristics curves are given below for the following working conditions VDD = 5V, CCT = 470nF, CCG = 2.2µF, VTH = VMICBIAS x 0.4, GAIN = VDD (40dB), AGC disabled, no load, RL = 10kΩ, COUT = 1µF, TA = +25°C, unless otherwise noted[70].

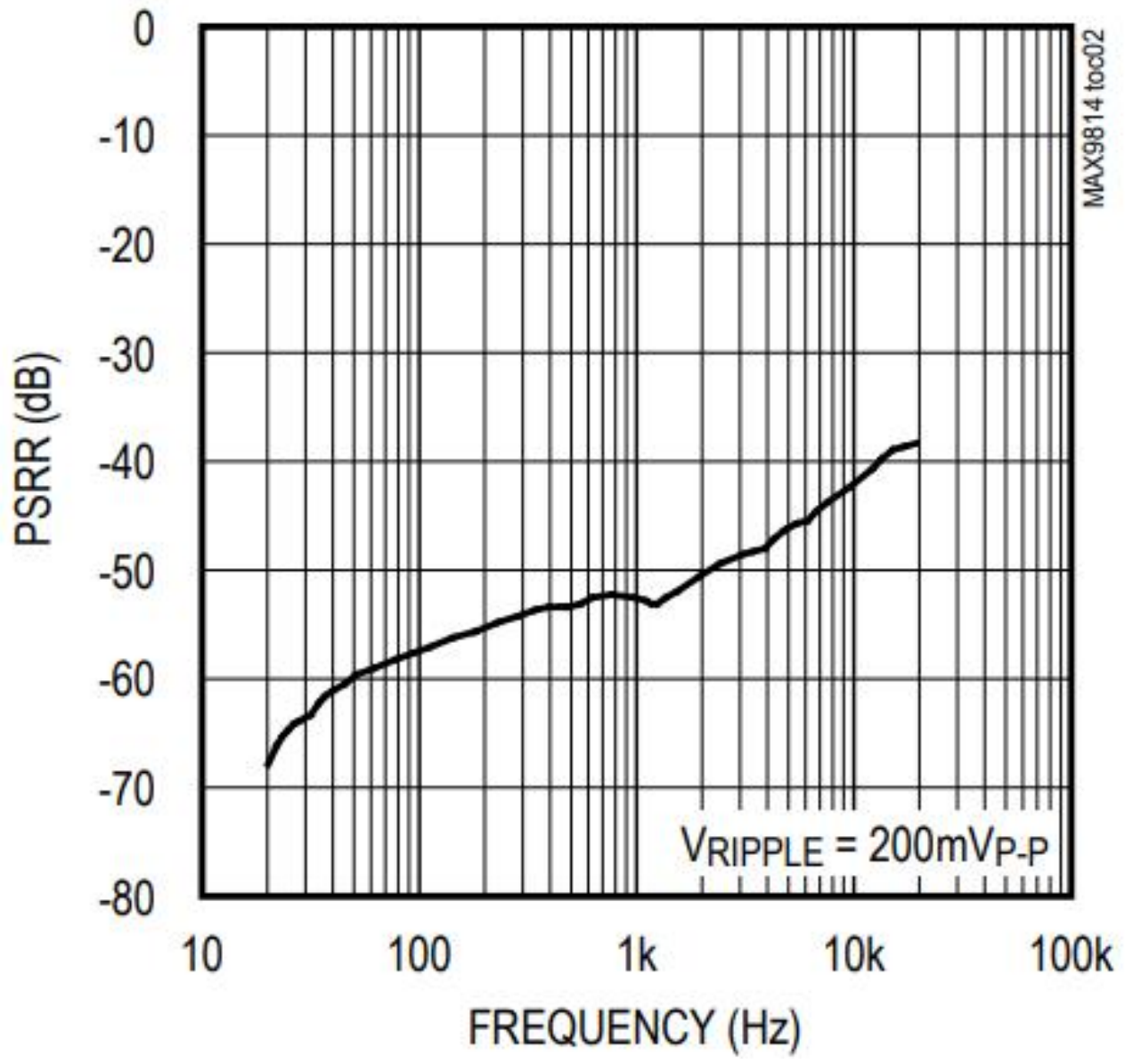

Figure 20 : Power-Supply Rejection Ratio vs Frequency Curve.[73]

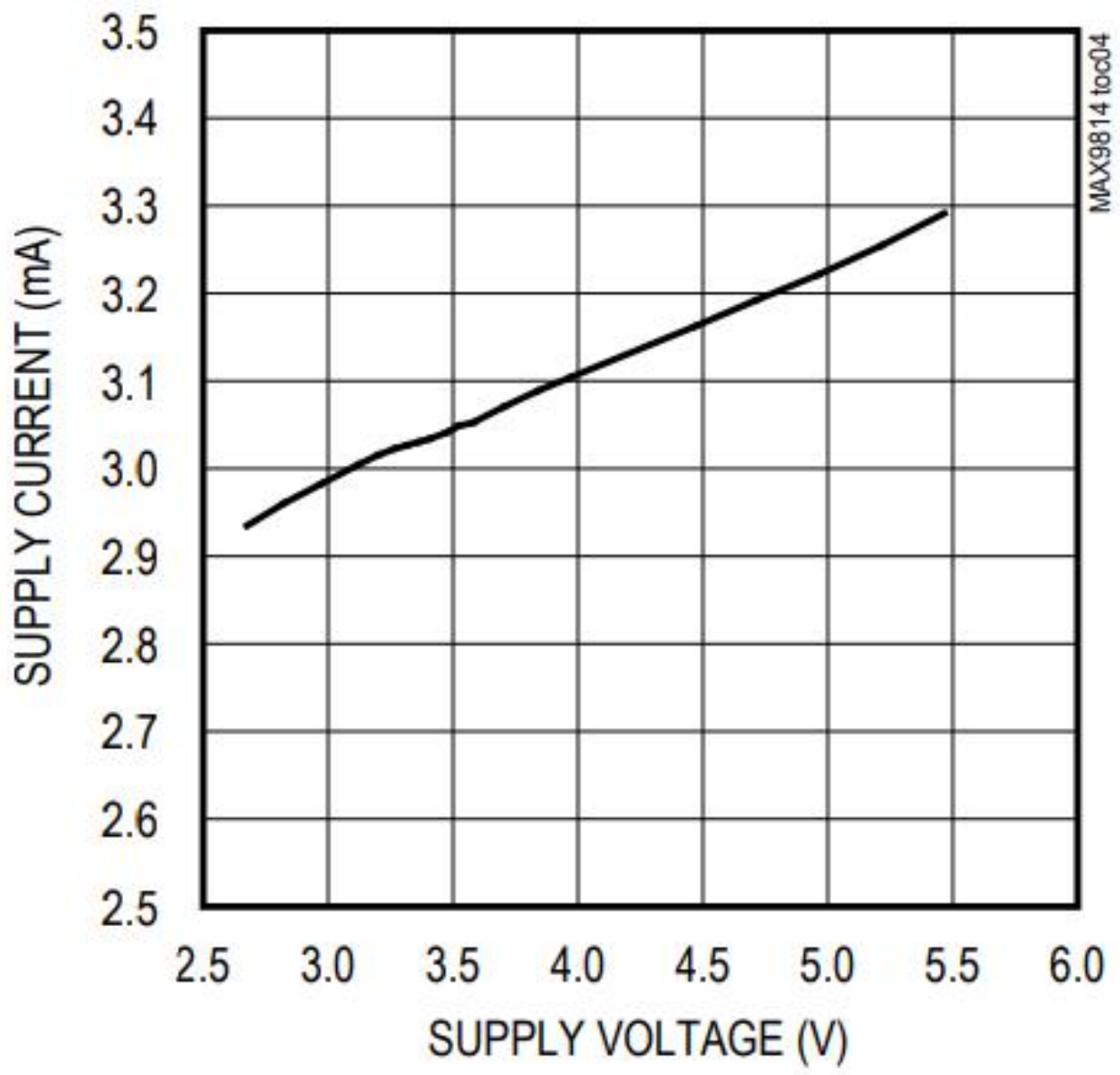

Figure 21 : Supply Current vs Supply Voltage Curve.[73]

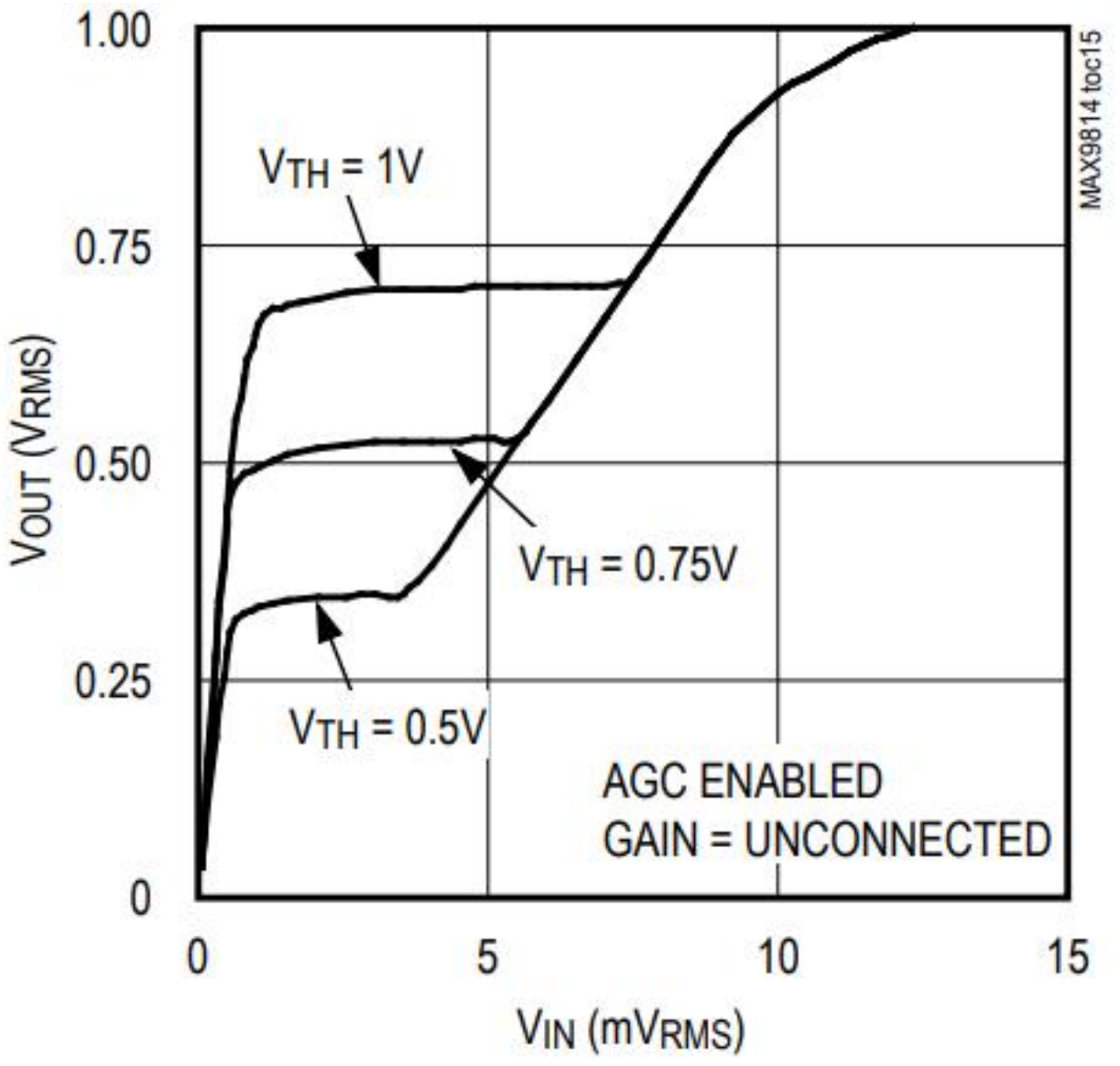

Figure 22 : Voltage IN vs Voltage OUT Curve.[73]

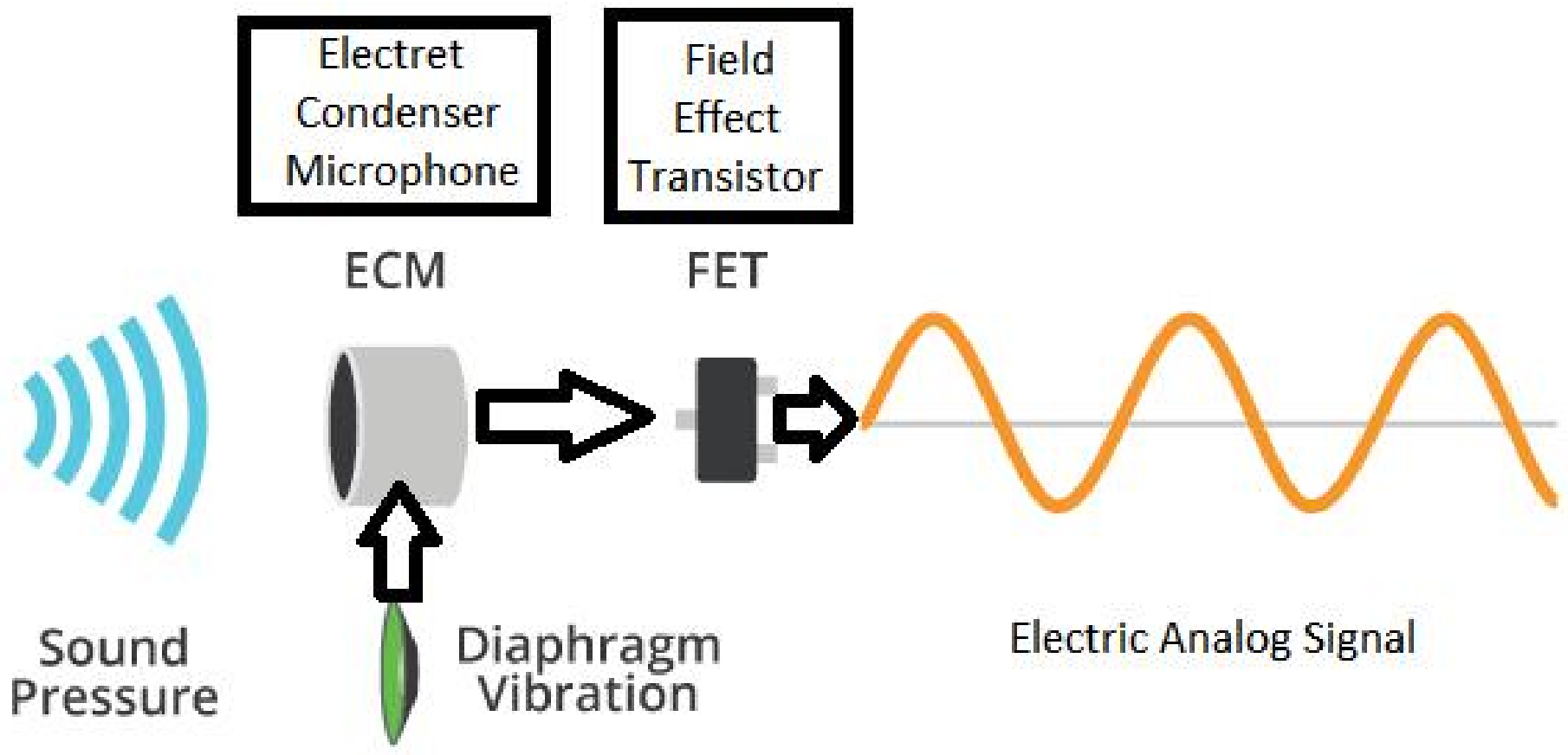

Figure 23 : Electret Microphone converting sound intensity to electric signal.

The principle of operation of condenser microphones is that the diaphragm acts like a plate of a condenser. The vibrations create changes in the distance between the diaphragm and the base plate. The voltage maintained across the diaphragm and the substrate varies with oscillations in the air, depending on the capacitance equation,

$C=Q/V$……………………………………………………………………………….(i)

where ,

Q = charge in coulombs,

C = capacitance in farads and

V = potential difference in volts.

This voltage change is amplified by the FET, and the audio signal as an electrical analog appears at the output, after a DC blocking capacitor.[74]

A capacitor is used as here a low-pass filter to reduce the noise of the audio signal.A low pass filter is a filter which passes low-frequency signals and blocks, or impedes, high-frequency signals.A low-pass filter (LPF) is an audio signal processor that removes unwanted frequencies from a signal above a determined cutoff frequency. It progressively filters out (attenuates) the high-end above its cutoff frequency while allowing the low-end to pass through, ideally without any changes.[75]

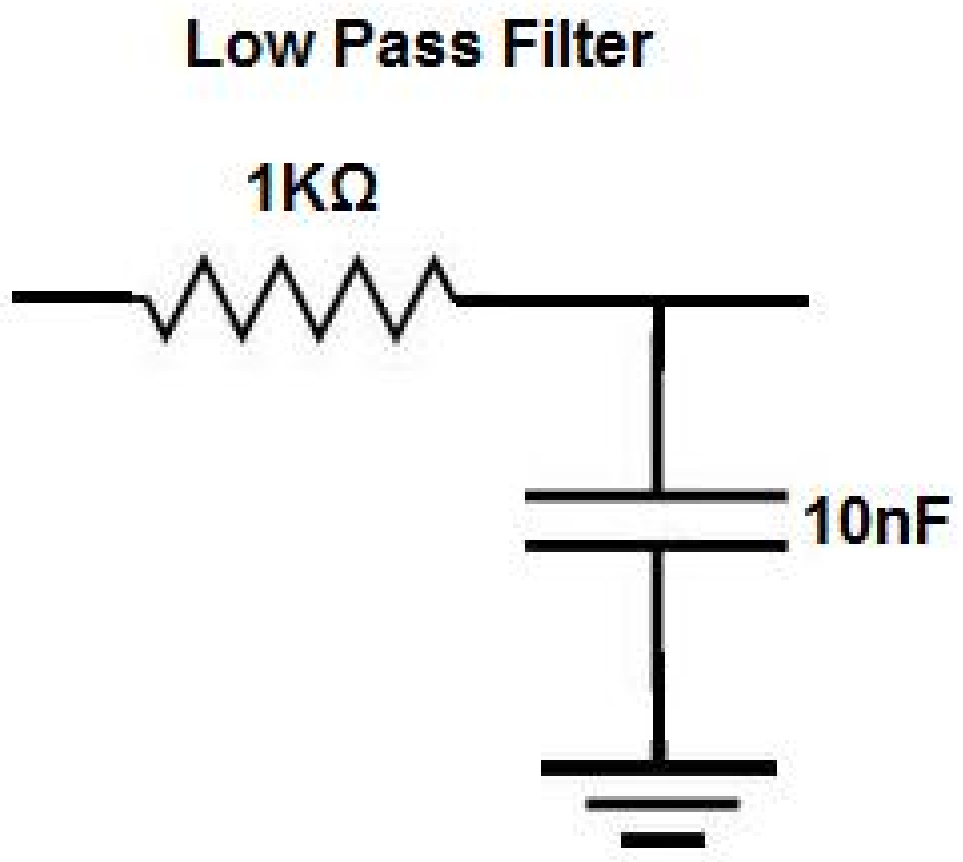

Figure 24 : Capacitor low-pass filter.

### 3.4.3. Analog to Digital Converter(ADC) & Digital to Analog Converter(DAC)

An analog signal is continuous over time and it is necessary to convert this signal into a stream of digital values. It is therefore necessary to determine the frequency at which new digital values are sampled from the analog signal. The ratio of the new values is called the sample rate or the converter sample rate. A signal with a continuously variable band boundary can be sampled, and then the original signal can be reconstructed from discrete time values using a reconstruction filter. The Nyquist-Shannon sampling theorem implies that the original signal can only be faithfully reproduced if the sampling rate is greater than twice the highest frequency of the signal.

In electronics, an analog-to-digital converter (ADC) is a system that converts an analog signal, such as the sound received from a microphone, into a digital signal. An ADC can also provide an isolation measurement such as an electronic device that converts the input analog voltage or current into a digital number representing the magnitude of the voltage or current. Usually, the digital output is a binary number of two's complement proportional to the input, but there are other possibilities. A digital-to-analog converter (DAC) performs the opposite function; it converts digital signal to analog signal.

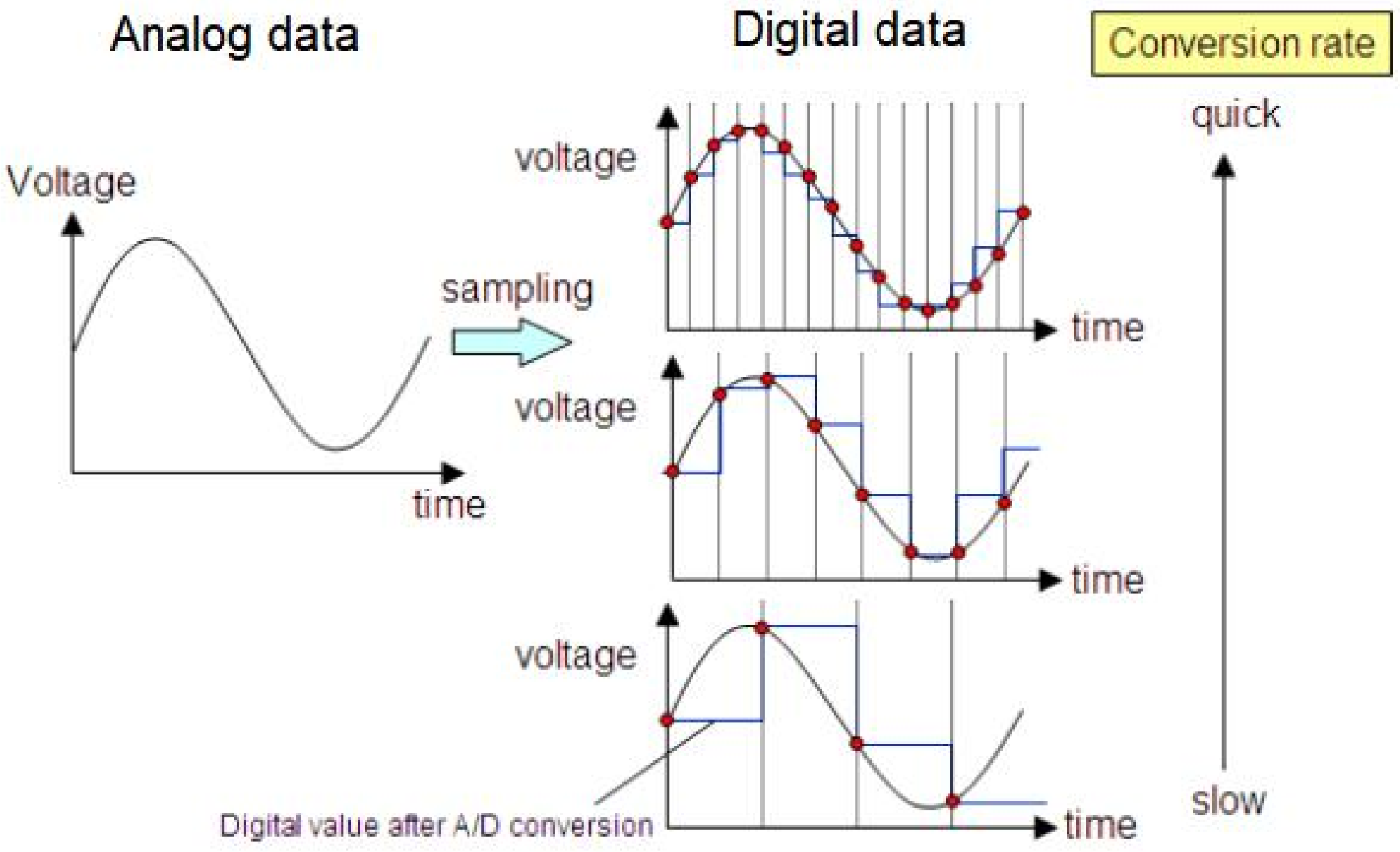

Figure 25 : Sampling Through ADC.[76]

The analog electrical signal is sent from the microphone to an analog-to-digital converter (ADC). ADC converts analog electrical signals into digital values that can be stored on a computer. This digital value can easily be stored or broadcast as needed. Once archived, the file can be edited with software like audacity.

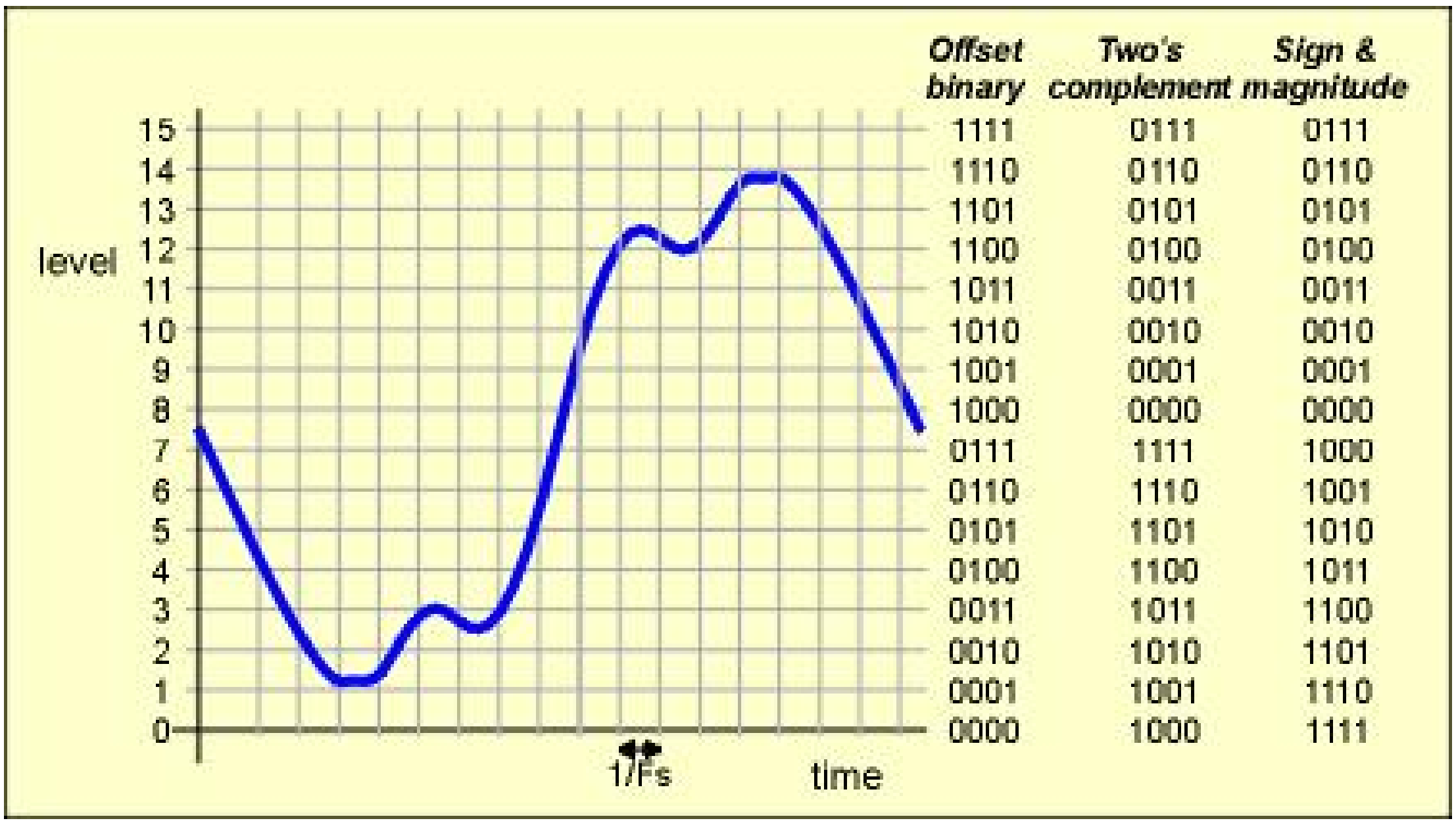

Figure 26 : Conversion of binary analog data to digital data.[77]

Analog data, in most cases, in binary is converted to numeric data by the complement of those two binary numbers. For example, if binary analog data is 1111, two's complement of 0111 will be the value of the digital data. This digital data can then be stored or transmitted as desired.

## 3.5. Software Configuration

Software that are used in the project are

1) Particle Web IDE(Integrated development environment)
2) Node.js JavaScript

### 3.5.1. Particle Web IDE

The Particle Web IDE is an Integrated Development Environment, or IDE; thus one can do software development in an easy-to-use application, which just so happens to run in a web browser [78].

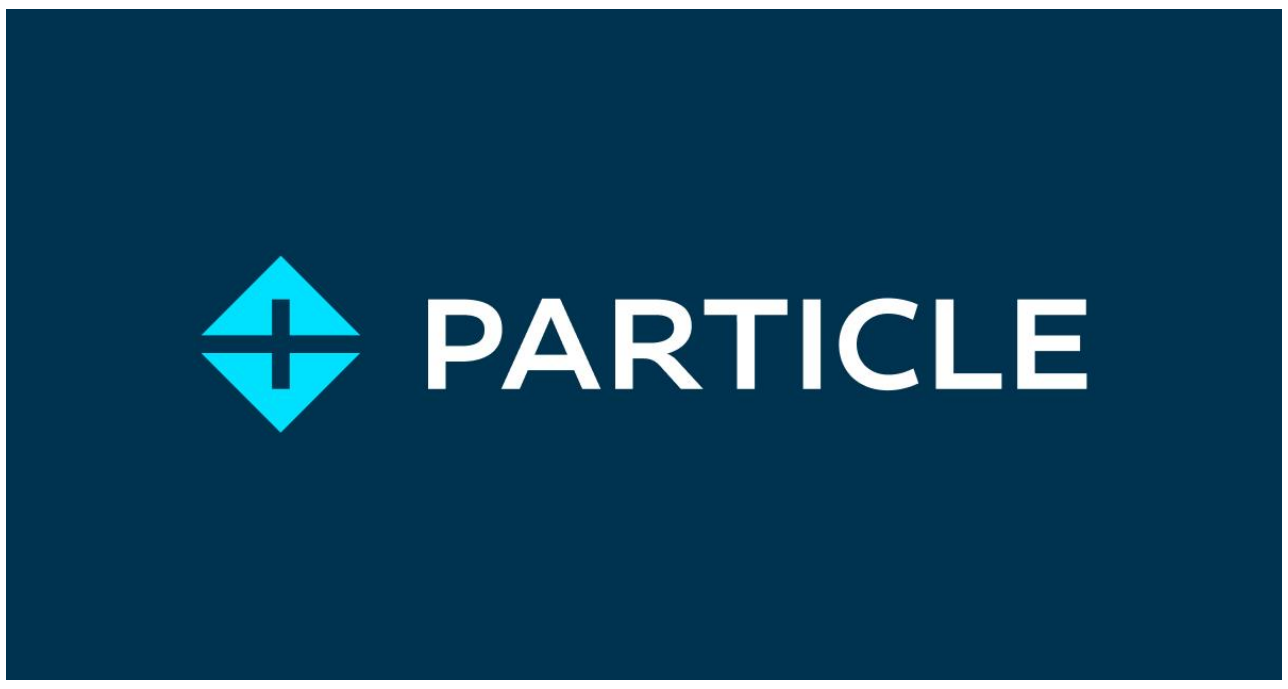

Figure 27 : Particle Web IDE .

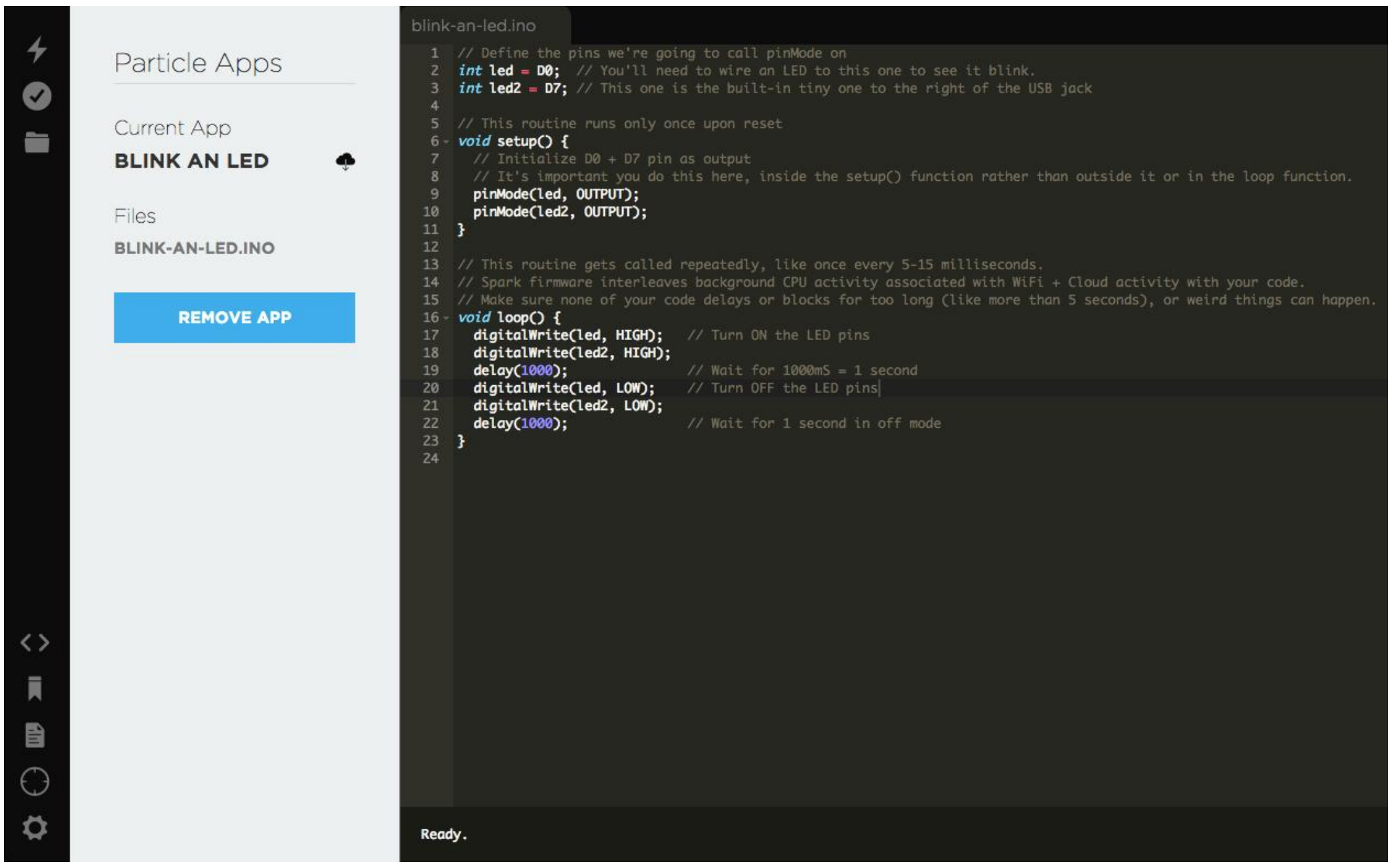

Figure 28 : Particle Web IDE outlook.

The programs are written in C or C++. A typical first program for a microcontroller simply blinks an LED on and off. In the IDE, the user might write a program like this:

```
int led1 = D0;
void setup() {
  pinMode(led1, OUTPUT);
}
void loop() {
  digitalWrite(led1, HIGH);
  delay(1000);
  digitalWrite(led1, LOW);
  delay(1000);
}
```

If the code if uploaded to the particle photon the LED integrated on the board which is connected to pin0 will turn on for 1sec and turn off for 1 sec in a continuous loop.

### 3.5.2. Node.js JavaScript

Node.js is an open source, cross-platform runtime environment and library that is used for running web applications outside the client's browser.

It is used for server-side programming, and primarily deployed for non-blocking, event-driven servers, such as traditional web sites and back-end API services, but was originally designed with real-time, push-based architectures in mind. Every browser has its own version of a JS engine, and node.js is built on Google Chrome's V8 JavaScript engine.

Figure 29 : Node.JS Logo.

In simple terms, this means that entire sites can be run using a unified 'stack', which makes development and maintenance quick and easy, allowing you to focus on other goals of a project.The fact that Node.js is open source means that it is free to use and constantly being tweaked and improved by a global community of developers.

### 3.6. VoIP Systems Security

The VoIP system includes the following three security methods WPA, Transmission Control Protocol (TCP) and Symmetric Encryption. As the both VoIP system and device that wants to download the data needs to be connected to the same WiFi signal of a router which has WPA system in it,thus requiring Service Set IDentifier(SSID) and password to connect to it.

Internet Protocol(IP) addresses are the identifier that allows information to be sent between devices on a network. TCP are the main protocol or set of rules implemented upon an IP address for organizing data in a way that ensures secure transmission between the server and client. In the VoIP system to access the audio data from particle photon other devices need to have the correct IP address and correct port. A port is a number used to uniquely identify a transaction over a network by specifying both the host and the service.

Also when the particle photon transmit the binary digital data it is converted into hexadecimal data thus data is encrypted. Before generating the audio data the hexadecimal data needs to be decrypted to again binary digital data. For this reason if anyone is able to access the hexadecimal data they would not be able to make it into a understandable file without knowing to which format and file this needs to converted to.

# CHAPTER 4

# RESULTS & DISCUSSION

## 4.1. Testing of  VoIP system

The Bangladesh national anthem was played and 15 second part of it was recorded through the VoIP system. The resultant audio files signal is given below :

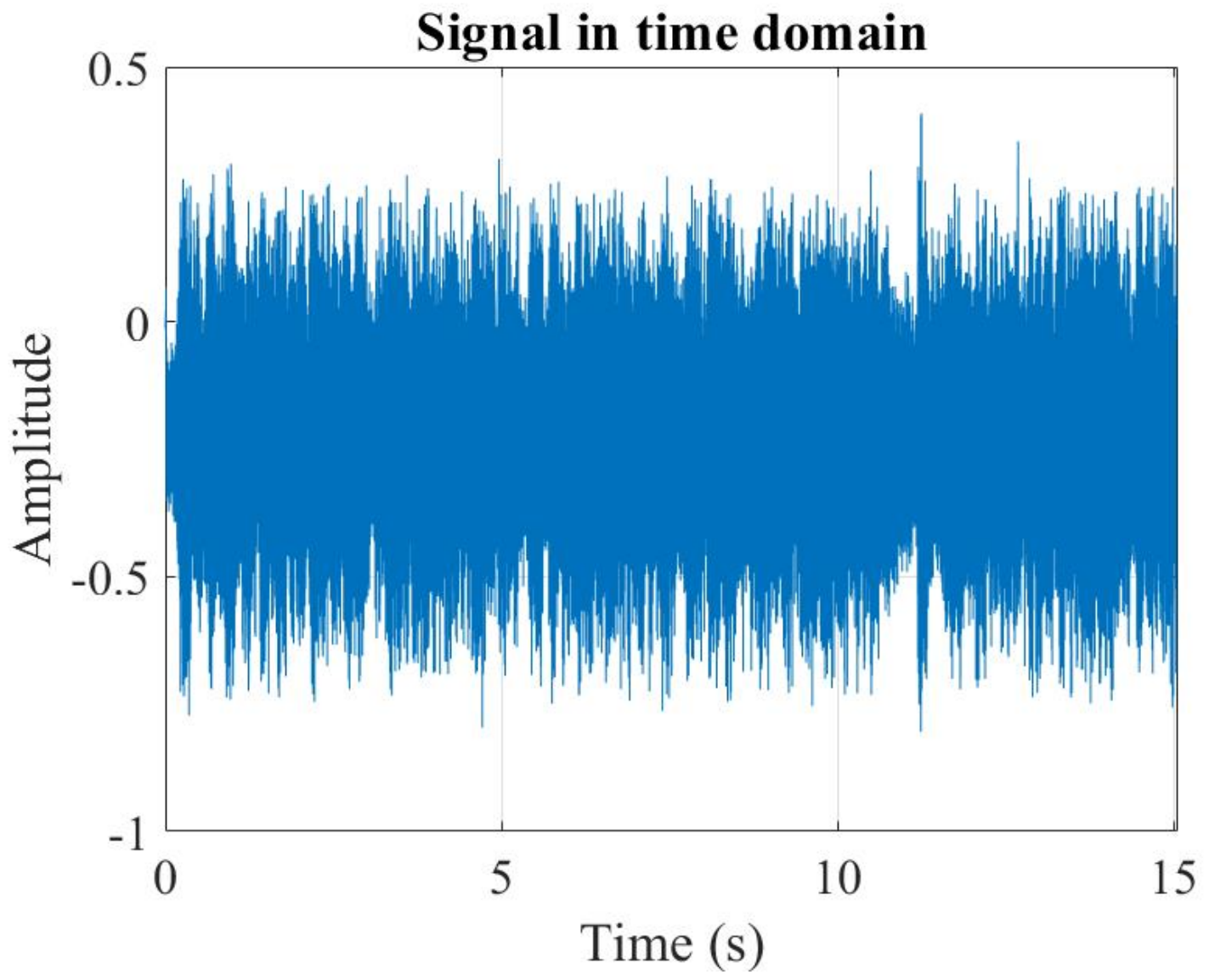

Figure 30 : VoIP system output sound signal in Amplitude vs Time graph (test-1 15sec).

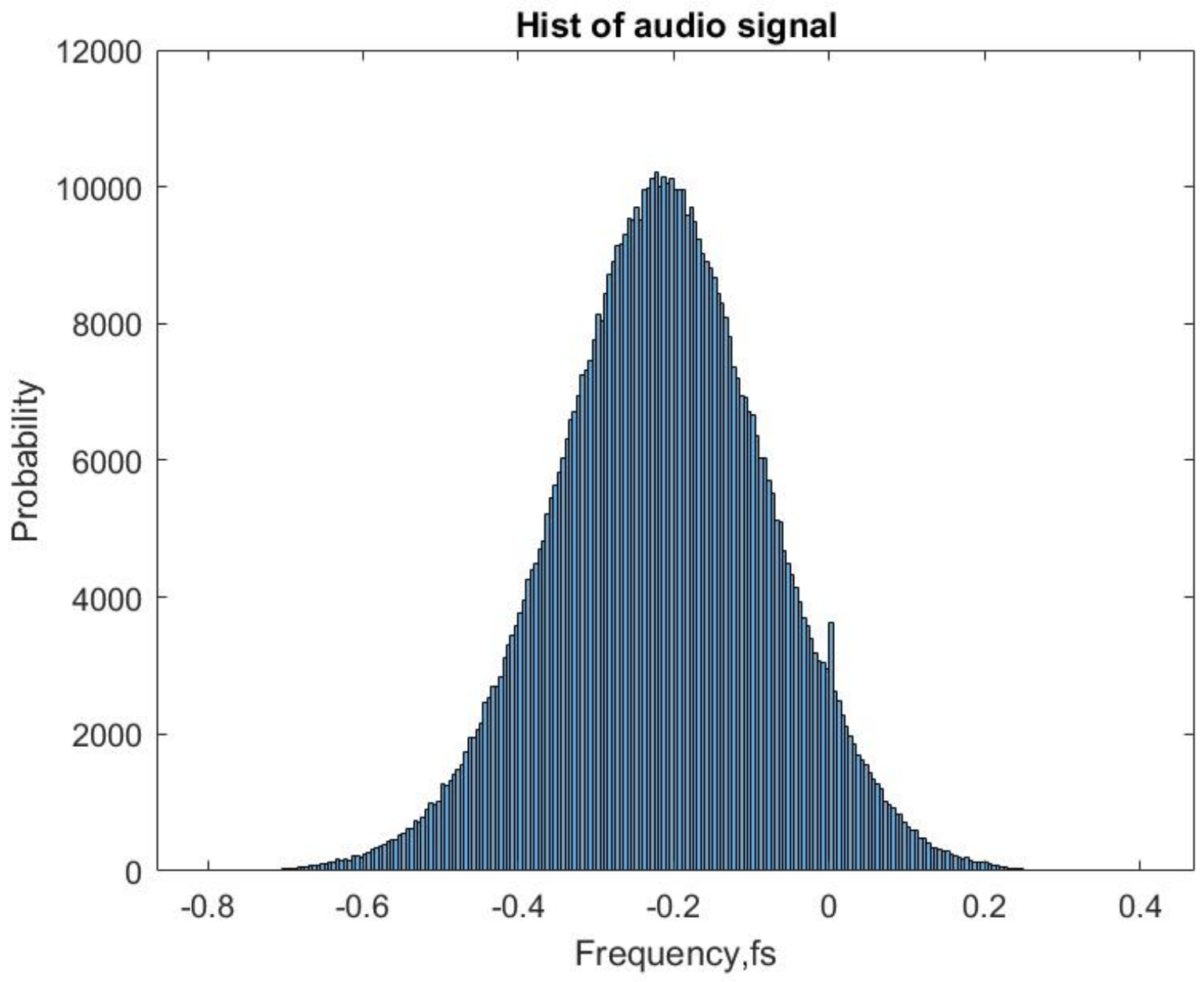

Figure 31 : Histogram of VoIP system output sound signal (test-1 15sec).

Figure 31 represents the number of times a frequency repeated itself in the audio signal. It is seen from both graphs that there is a DC offset which caused the audio waveform to be shifted from the zero position to near 0.2 position. This issue can be resolved by processing audio file in a software called "Audacity".

## 4.2. Sending of Audio Data

The code to run the system is uploaded to the Particle Photon via the build.particle.io website. This code is uploaded to the particle photon via the Internet. Therefore particle photon must be connected to the Internet.

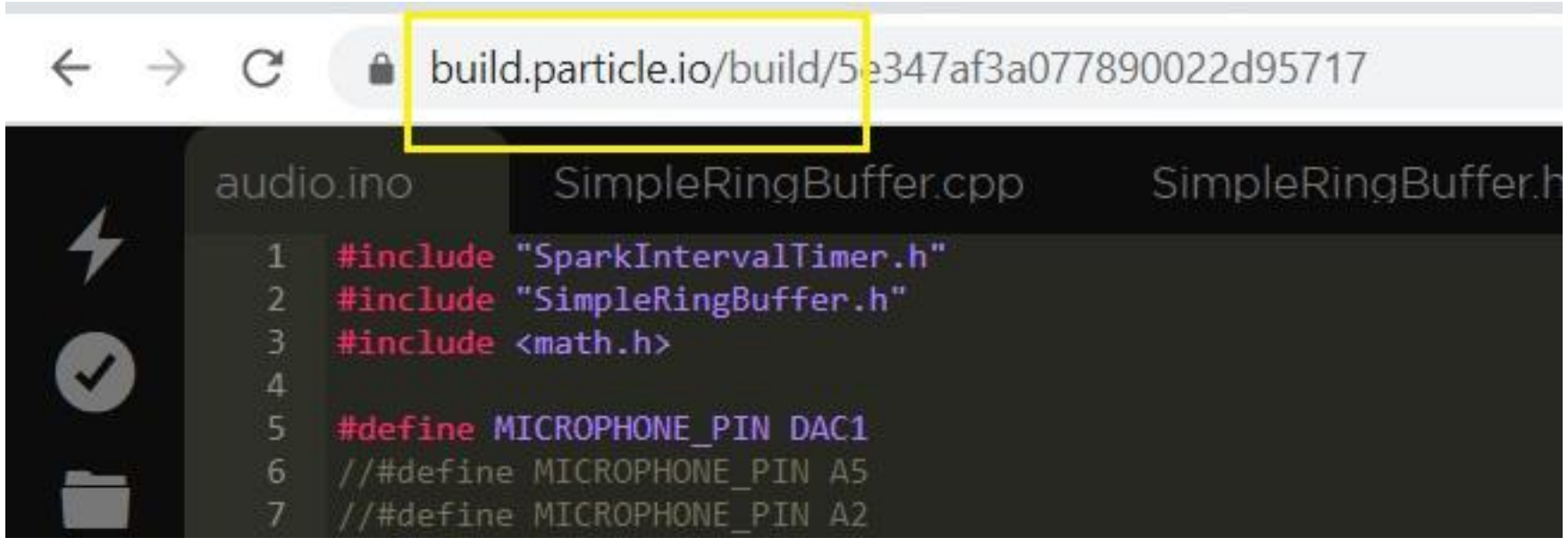

Figure 32 : Build.particle.io website.

After finishing upload the particle photon receives analog data from the microphone which is converted into digital data in the ADC. This data is encrypted by converting the digital binary data to hexadecimal then buffered so that it can send to TCP server.

## 4.3. Receiving Data from Particle Photon

Now to listen or download the audio the particle photon is uploading to its TCP server. A laptop needs to be connected to the same WiFi as particle photon and also must have the JavaScript code, in that code the IP of the particle photon needs to be set to access the TCP server.

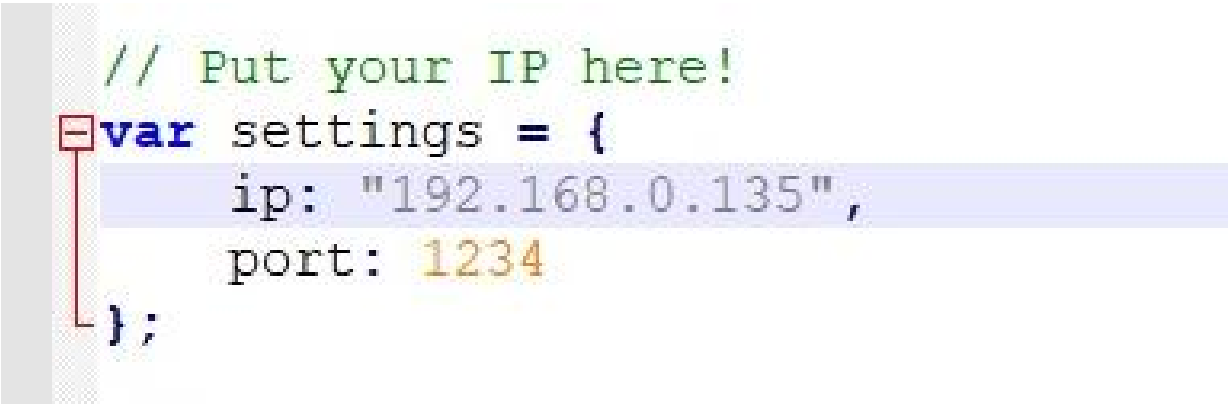

Figure 33 : IP replacement in the java script.

If the IP address specified in the script is not the same as that of the particle photon, it will not return any data. After setting the IP address, Windows PowerShell is opened in the directory where the script is located.After opening Windows Powershell, running the below command line. Therefore, the server.js file is located in the startup directory.

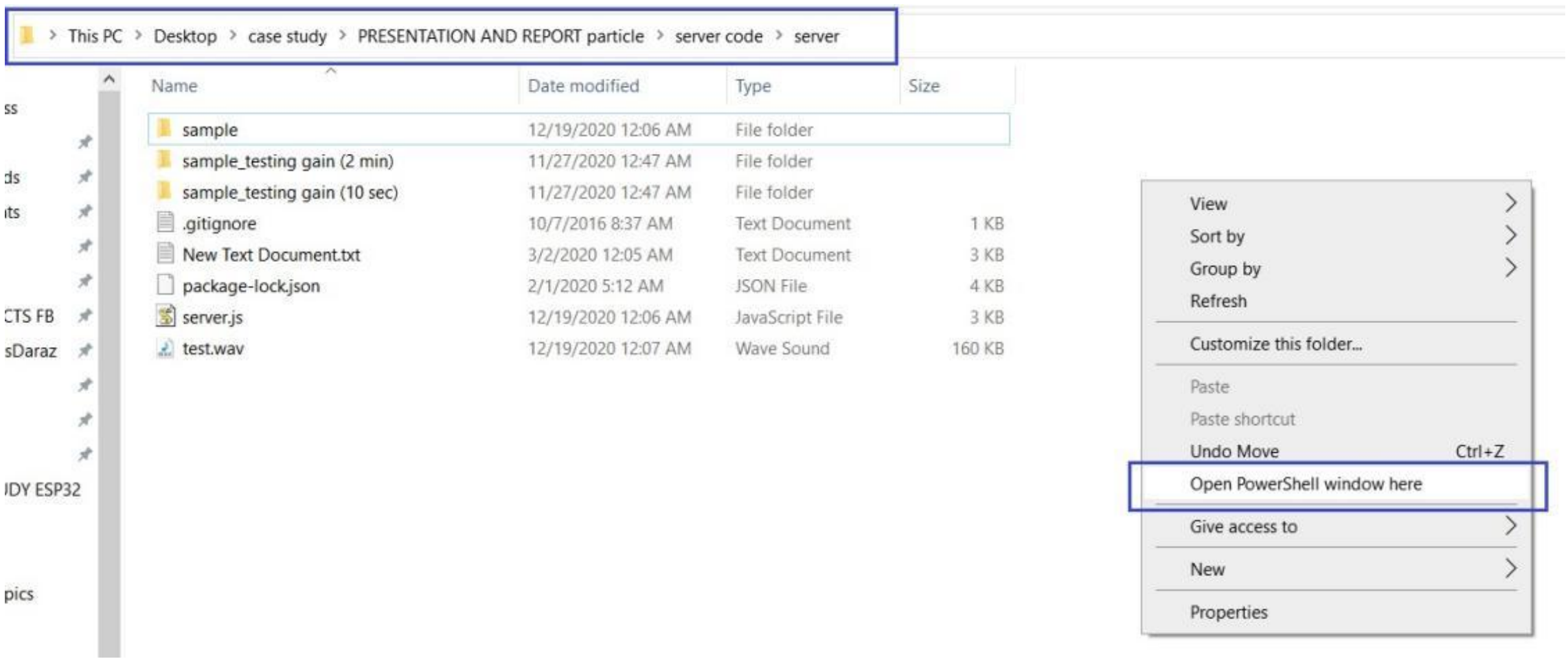

Figure 34 : JavaScript located folder.

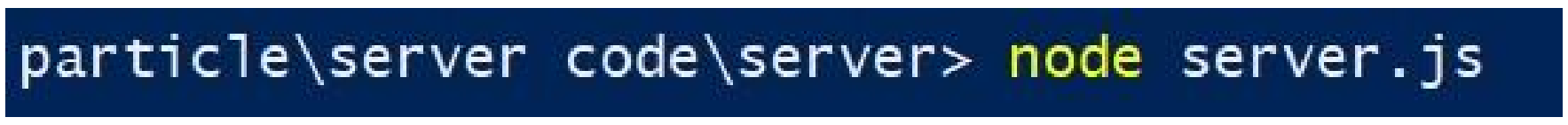

Figure 35 : Command Line of windows PowerShell.

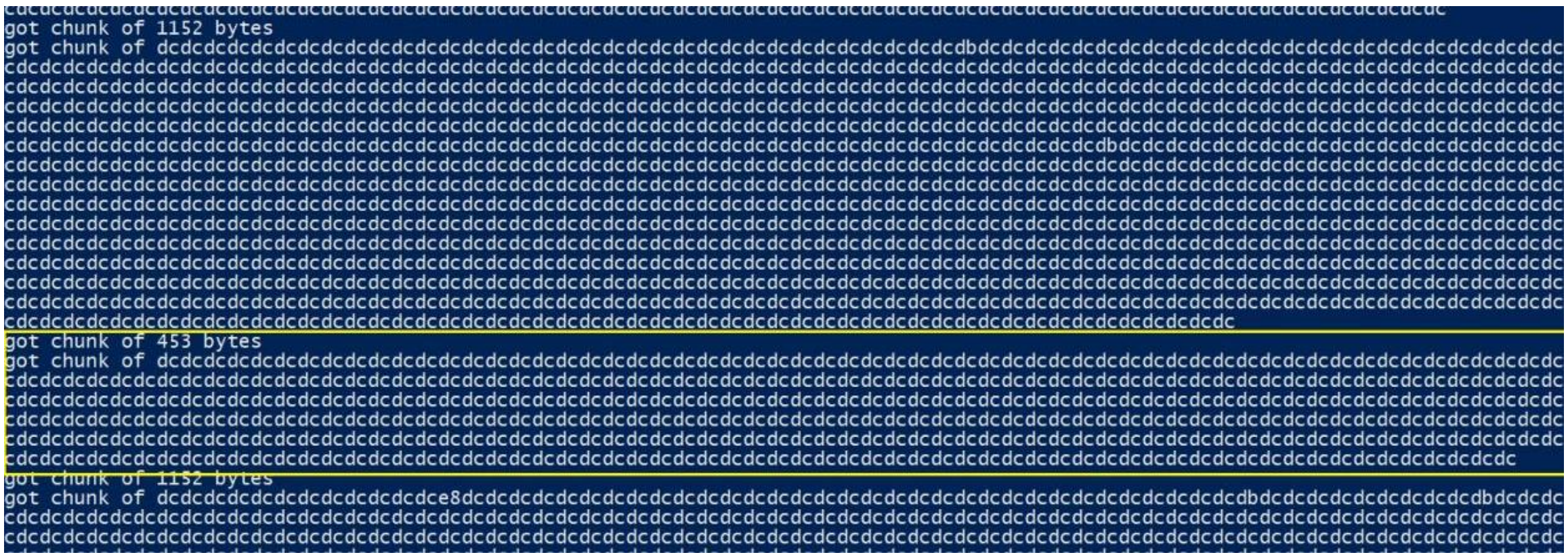

Figure 36 : Receiving data from Particle Photon.

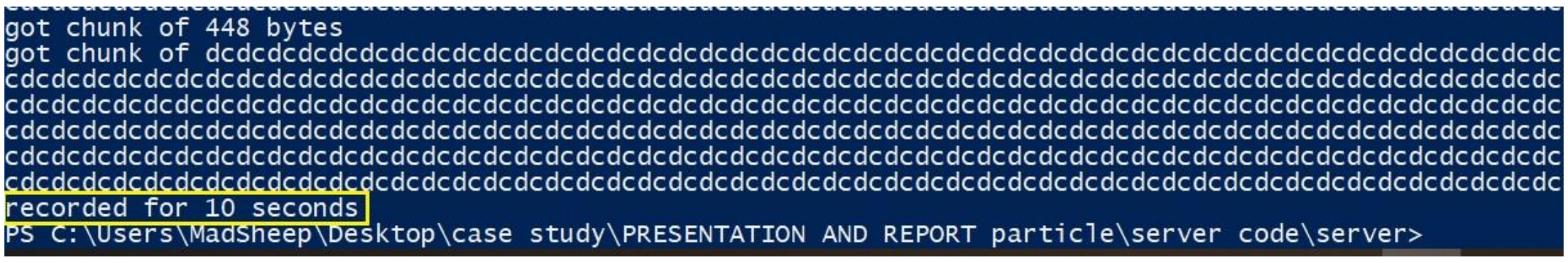

Figure 37 : Data receiving finished and recorded.

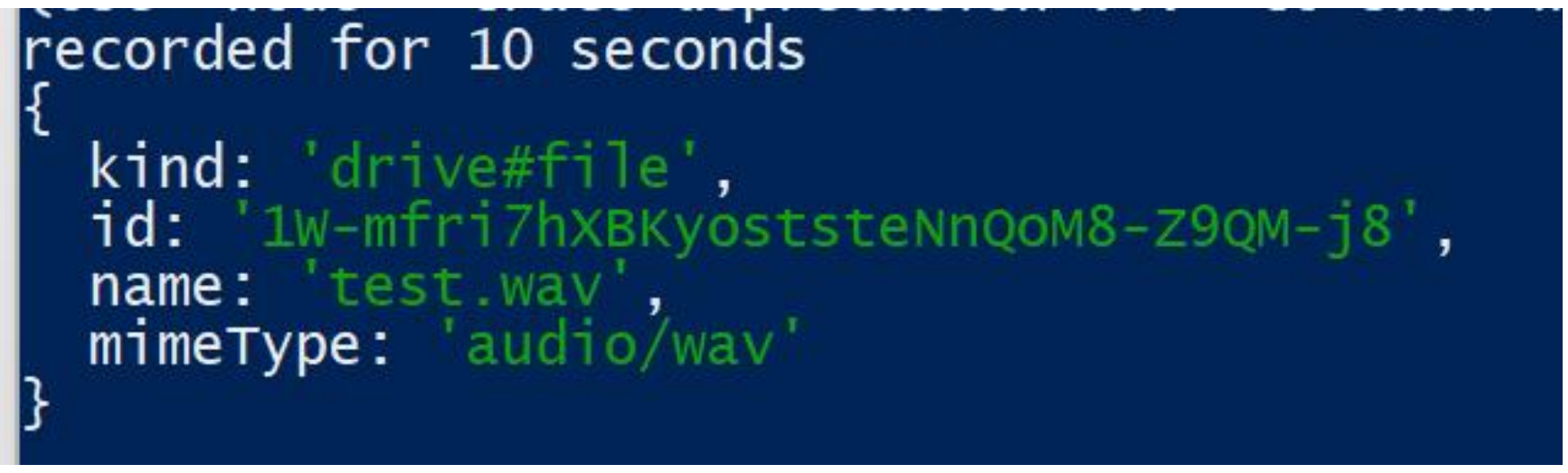

Figure 38 : File uploaded to Google Drive.

Once the data transfer is complete, the scripts combine all the data in the packet and produce an audio file which is stored in the directory where the script is located and also it is uploaded to Google Drive.

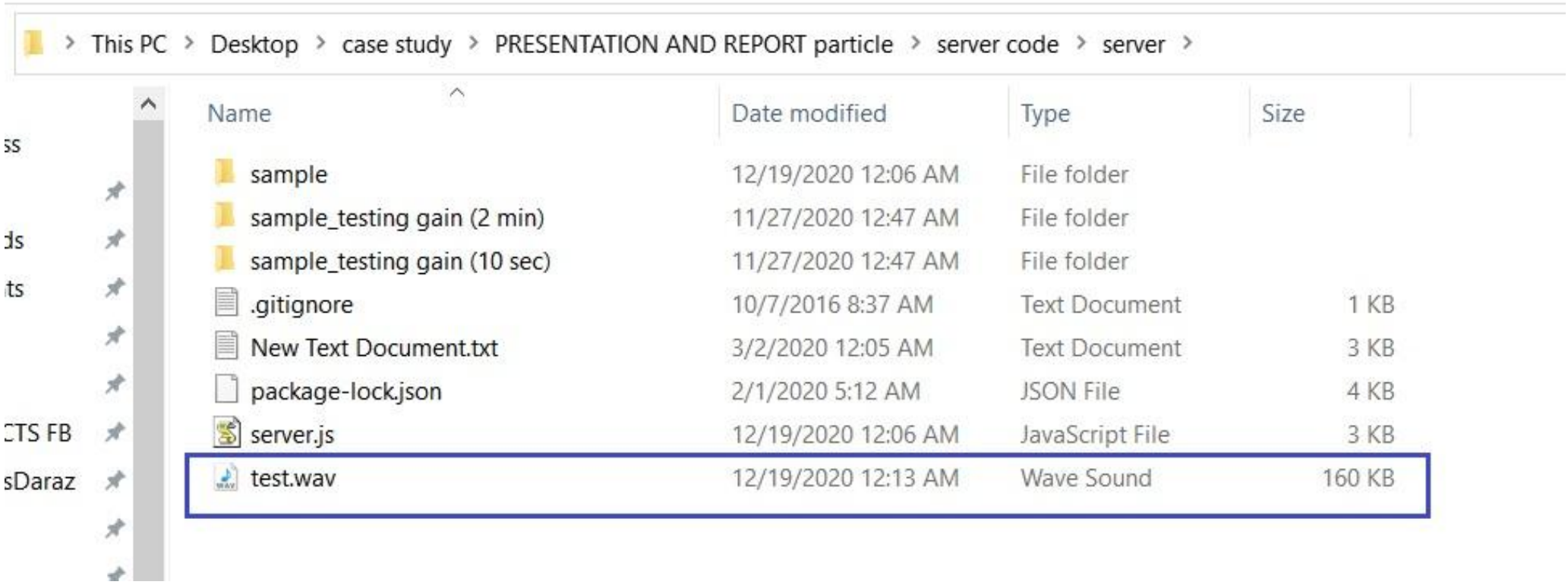

Figure 39 : Recorded audio file.

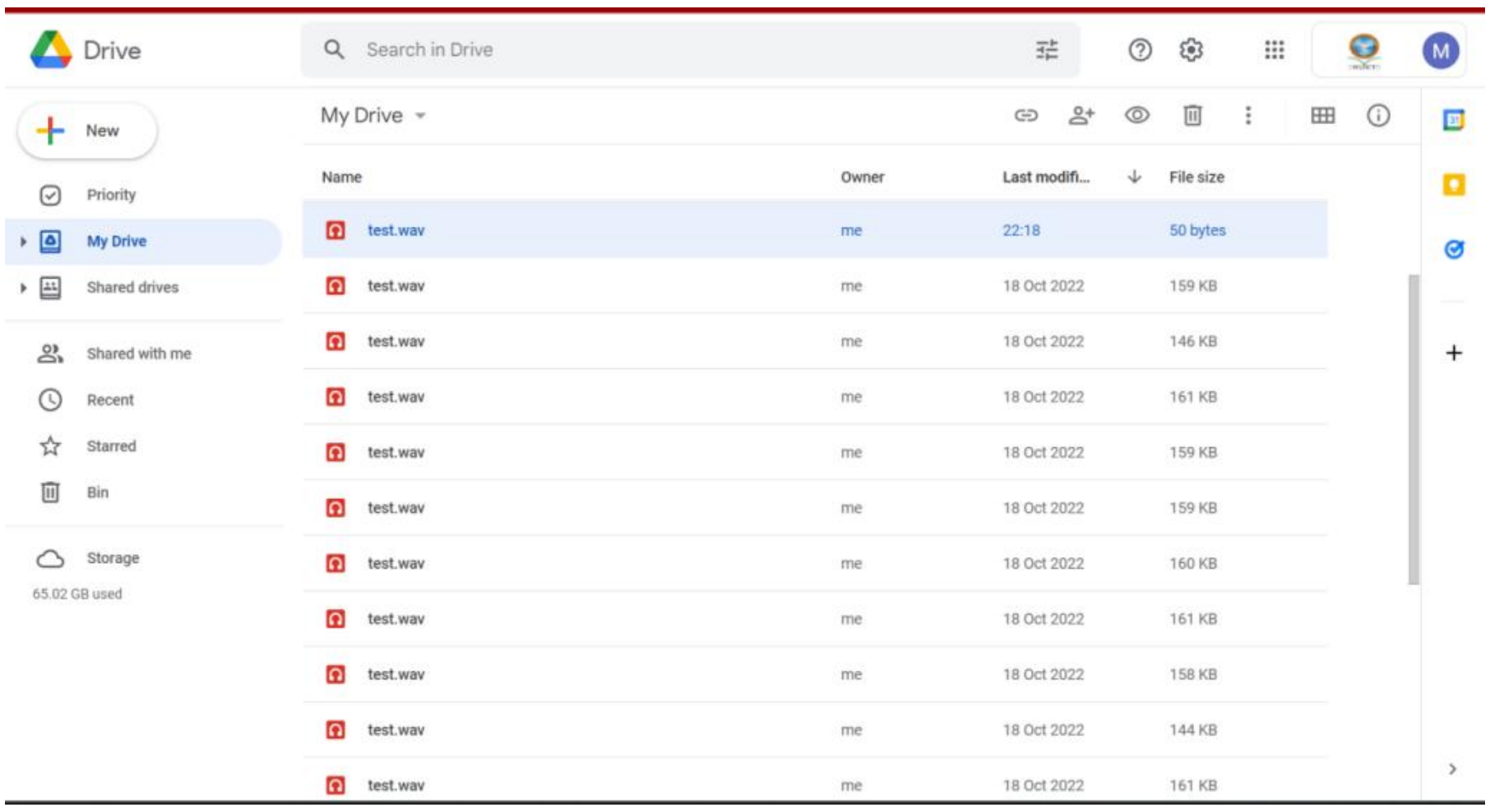

Figure 40 : Recorded audio file in Google Drive.

The audio files stored in Google Drive are uploaded with same name but with different time of upload and file ID thus it does not get replaced.

### 4.4. Comparative Signal Analysis

To see how well the VoIP systems electret microphone received the audio it is compared with the original audio as well as an audio recorded by conventional means. The audio played is a part of Bangladesh national anthem, it is about 10 sec. The audio files were added to “Matlab for acoustic analysis.

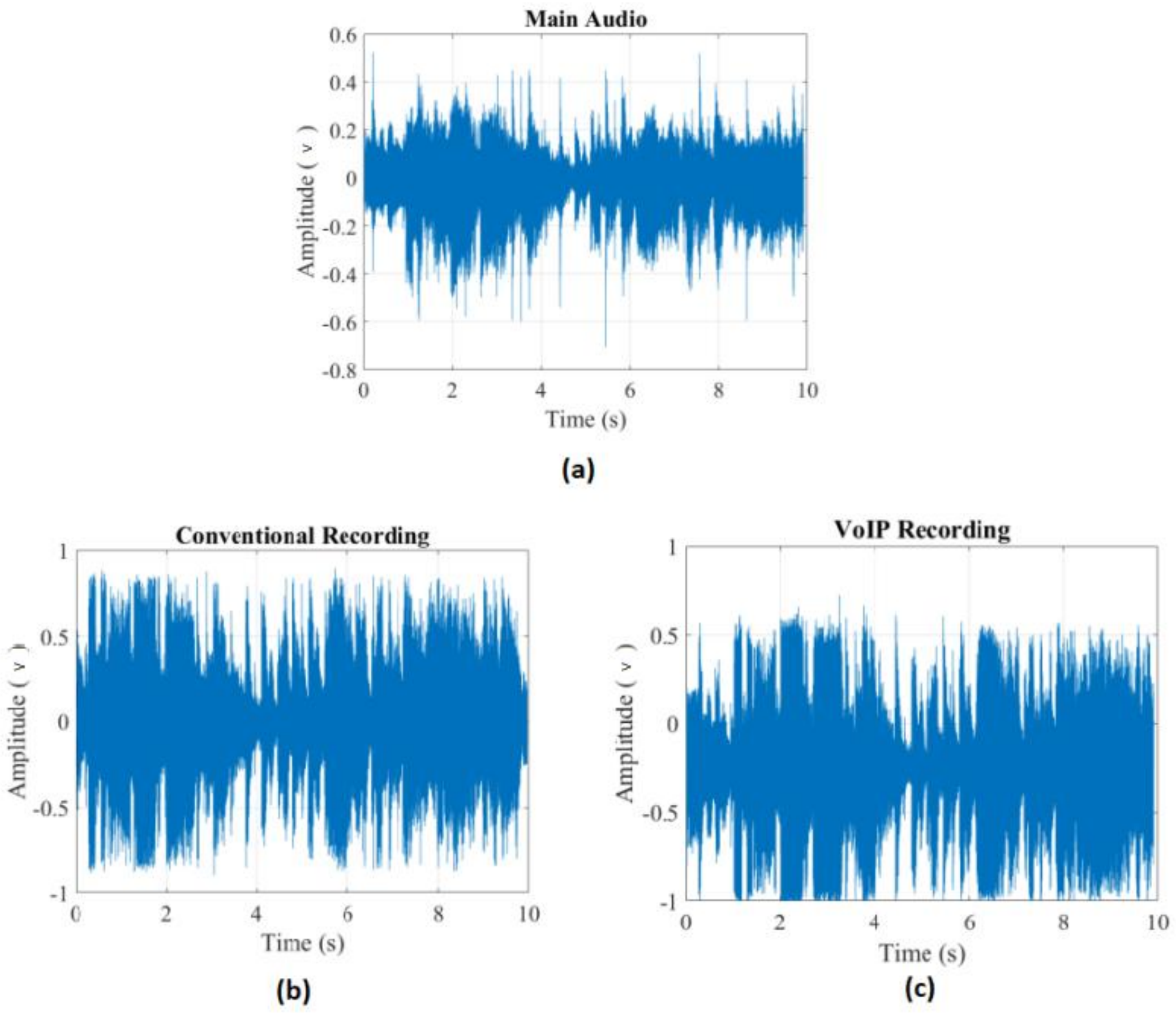

Figure 41 : Plot of signal in time domain (a) main audio, (b) conventional recording, (c) VoIP signal.

Figure 41 (a) shows 10sec of Bangladesh national anthem audio signal that was played in speaker for the conversational recorder and VoIP system to record. Figure 41 (b) and figure 41 (C) shows the recorded audio signal by conventional recorder and VoIP system Respectively.The reasons for VoIP system signal to show DC offset or an offsetting of a signal from zero can come from any number of issues from a bad analog-to-digital converter(ADC), broken hardware, or powering issues.

Now the audio signals recorded in conventional recording and VoIP recording is comparatively analyzed below[79]:

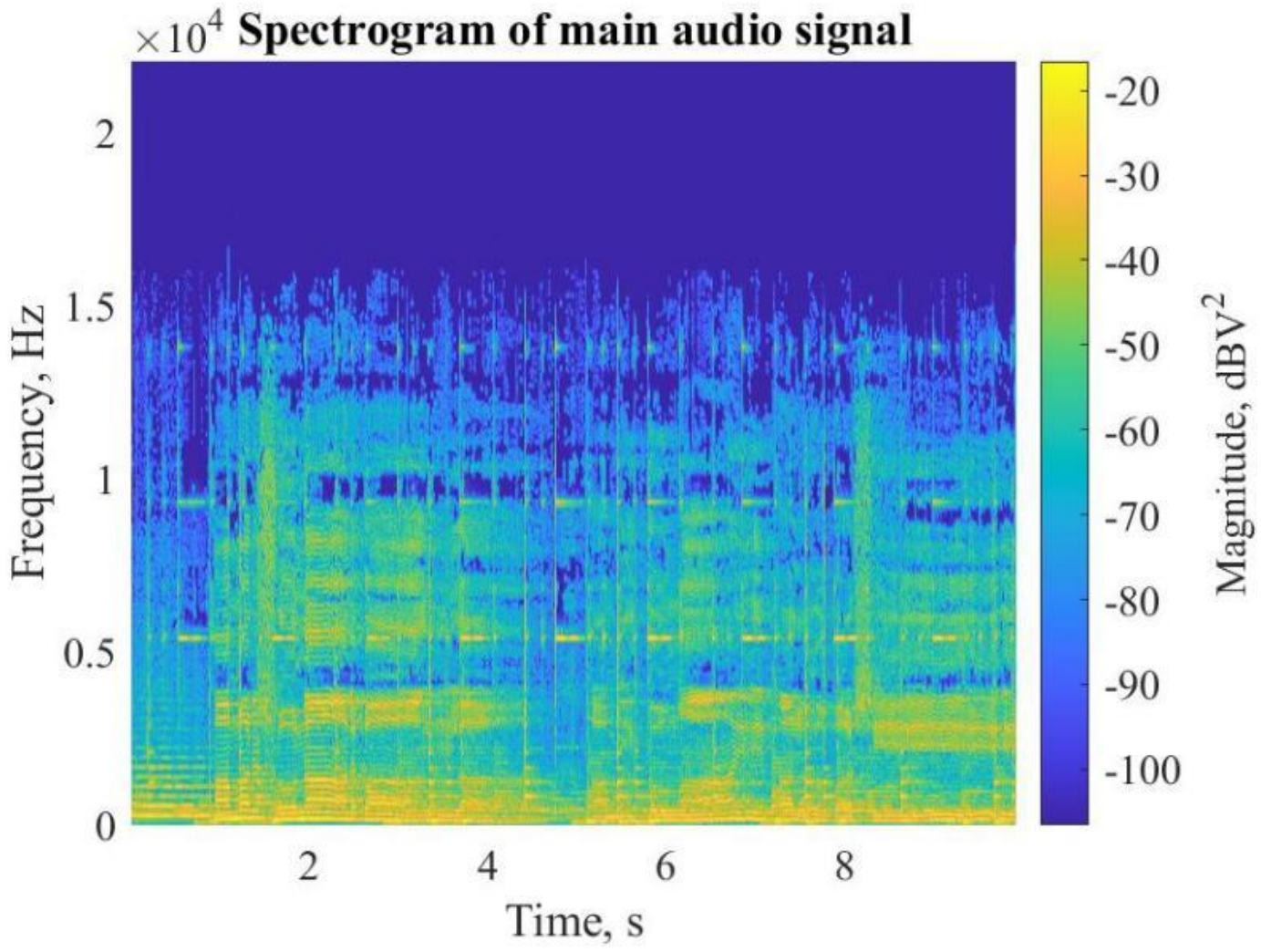

(a)

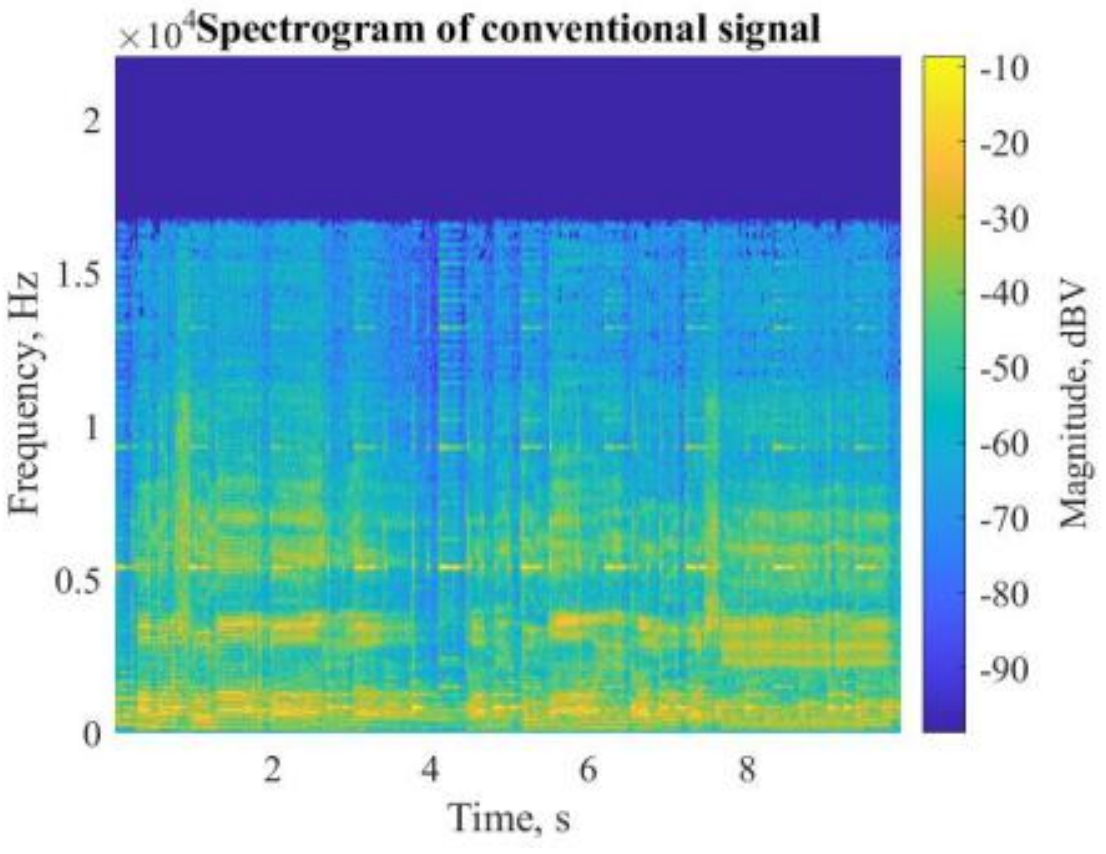

(b)

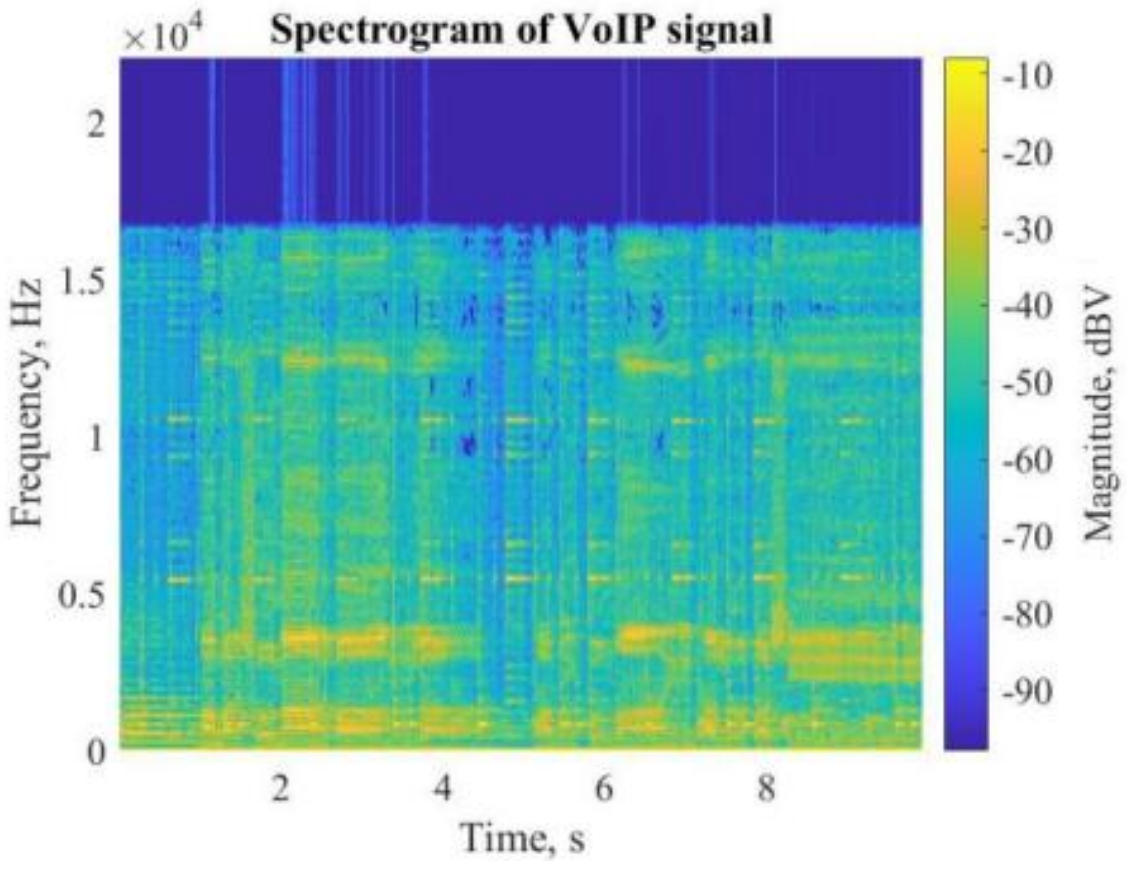

(c)

Figure 42 : Spectrogram of (a) main audio, (b) conventional recording & (c) VoIP signal.

Figure 42 (a) and figure 42 (b) shows the spectrogram or sonographs of the conventional recording signal and VoIP signal respectively. The spectrogram is the representation of the spectrum of frequencies of a signal as it varies with time. It is seen from both spectrogram that high intensity power/frequency ratio are seen up to 16 kHz, but above 16kHz frequency intensity of power/frequency ratio suddenly decreases because in both cases the audio was recording in a sampling rate of 16kHz minimum. However, comparing the power/frequency ratio intensity of the conventional recording and VoIP signal from the above figures it is seen that the audio recorded by the VoIP system consisted of more intensity of power/frequency ratio in the range 0 Hz to 16 kHz during the period of 10 sec. This occurred due to the VoIP system microphone to have a much greater sensitivity then the conventional recorder.

If signal has a power of $P_1$ watts, and a second signal has a power of $P_2$ watts. Then the power amplitude difference in decibels is:

Decibel-Watt $(dBW) = 10 \log_{10} (P_2 / P_1)$…………………………………………………….(ii)

But, $P_2/P_1 = (V_2 / V_1)^2$

Thus, Decibel-Volt $(dBV)=20 \log_{10} (V_2 / V_1)$………………………………………………(iii)

Equation (iii) is used to plot the logarithmic spectrum of power for both conventional recording and VoIP signal which describes the signal's magnitude and phase characteristics as a function of frequency. The below figure 43 (a) and figure 43 (b) represents the spectram of conventional recording & VoIP signal respectively. Both signals magnitude are in negative values due to the fact that the developed voltage is very small but positive which are presented in logarithmic scale. Both systems developed similar magnitude but the VoIP system developed slightly more magnitude within the range of frequency.

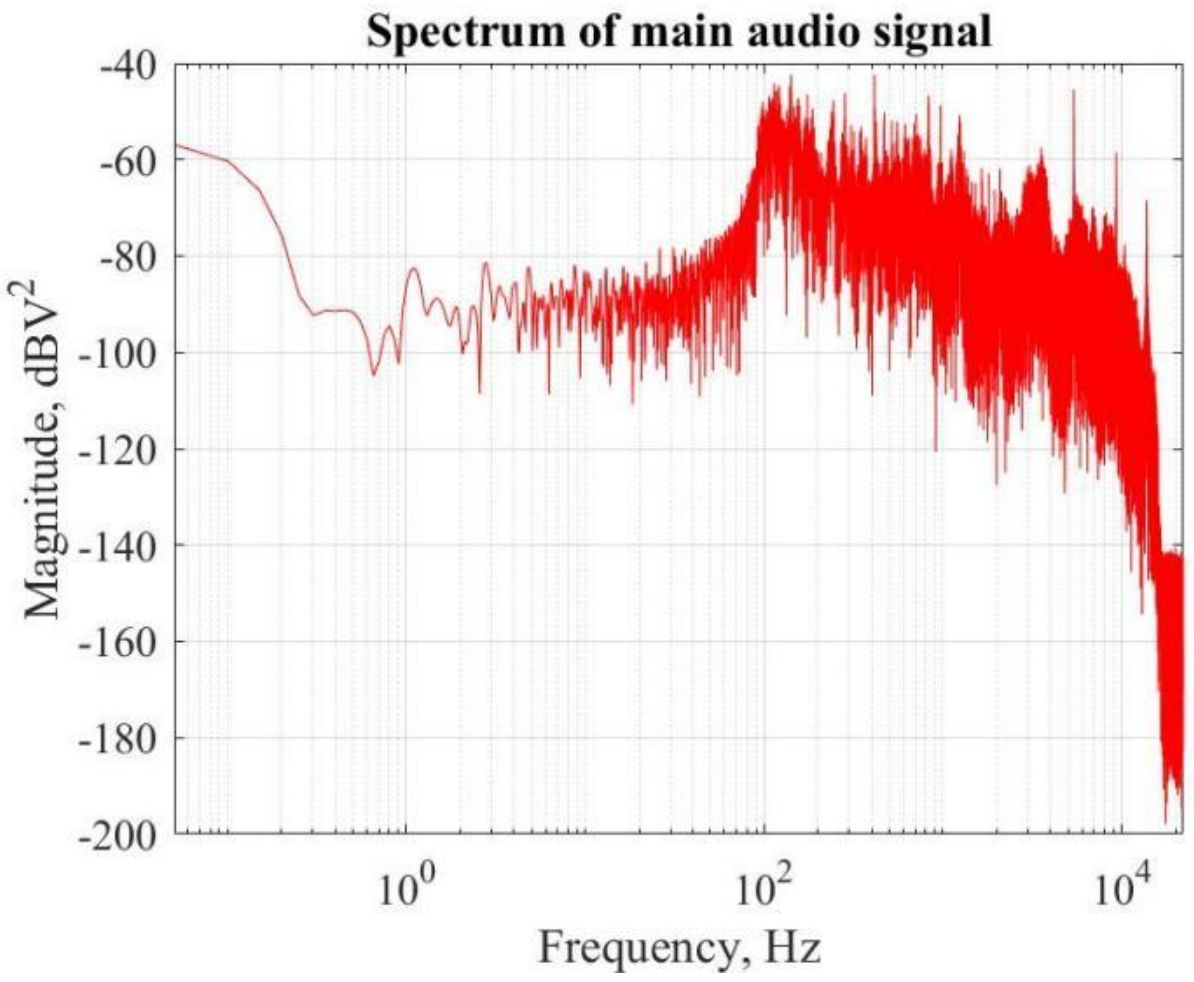

(a)

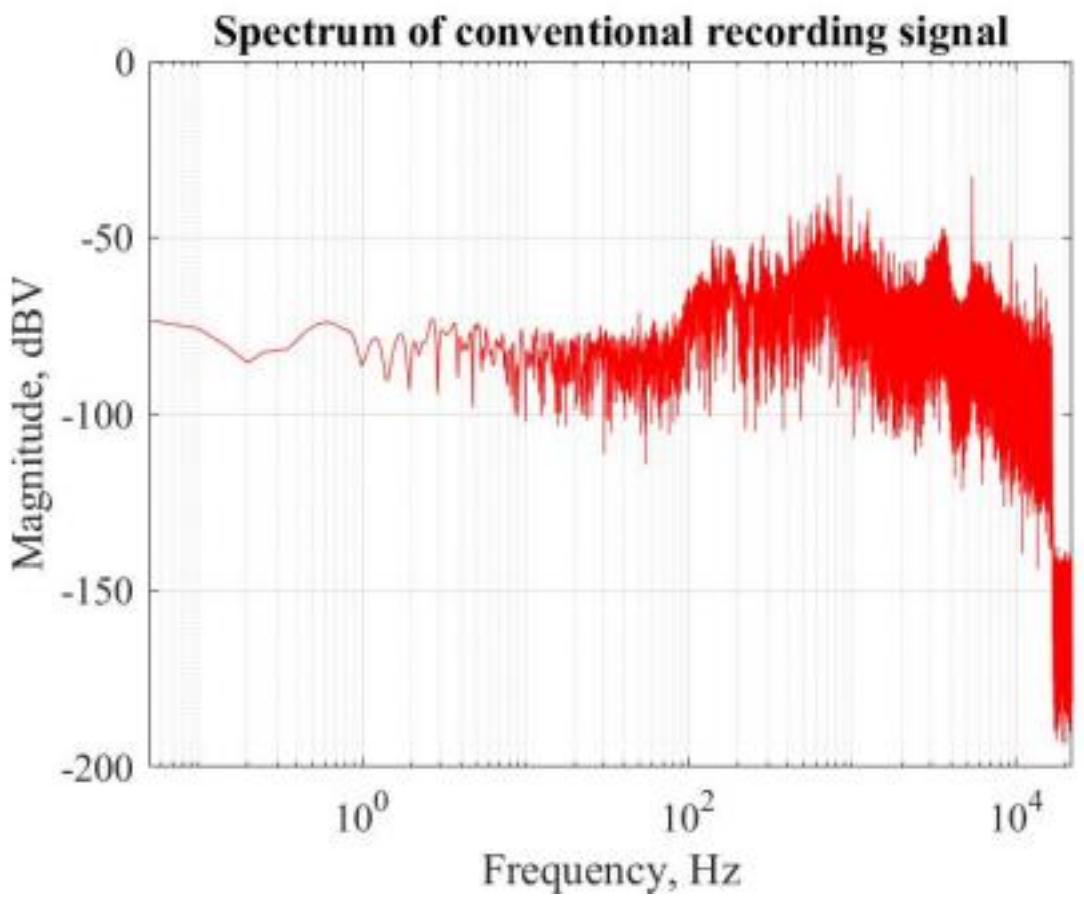

(b)

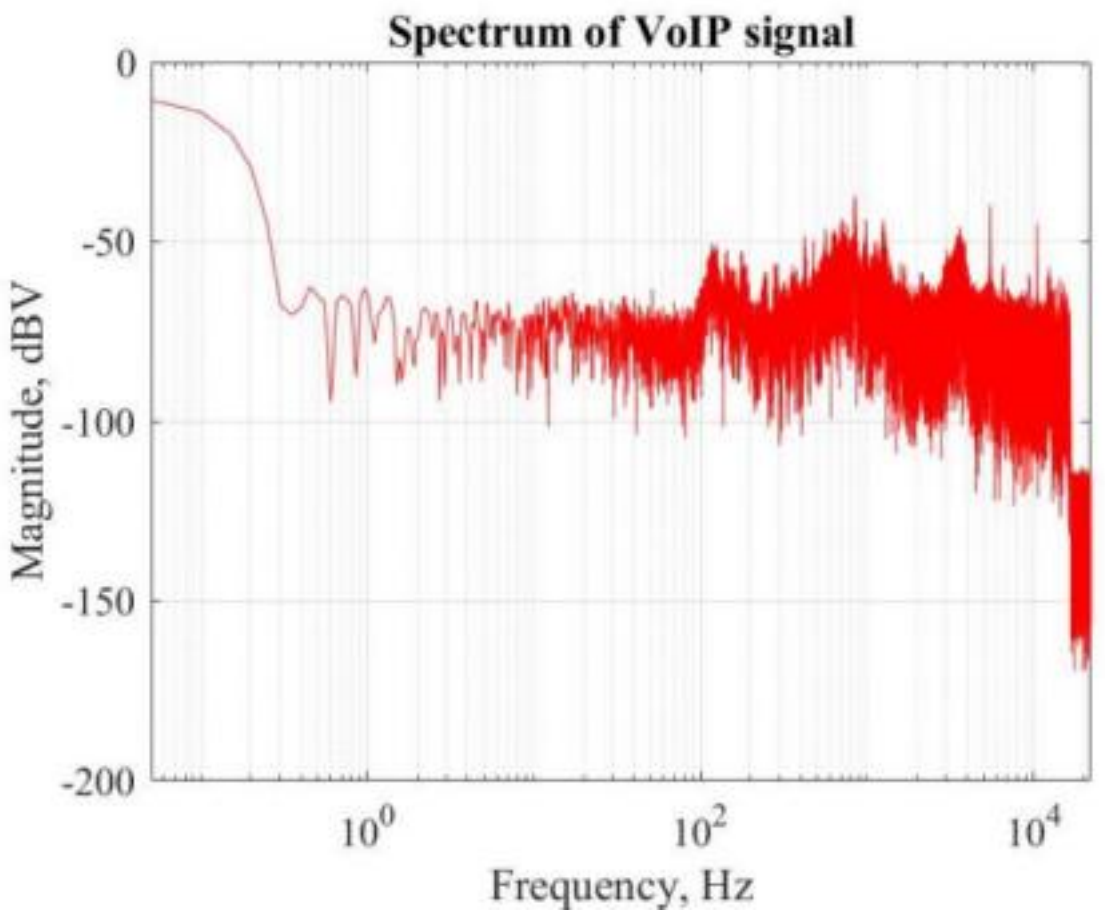

(c)

Figure 43 : Spectrum of (a) main audio, (b) conventional recording & (c) VoIP signal.

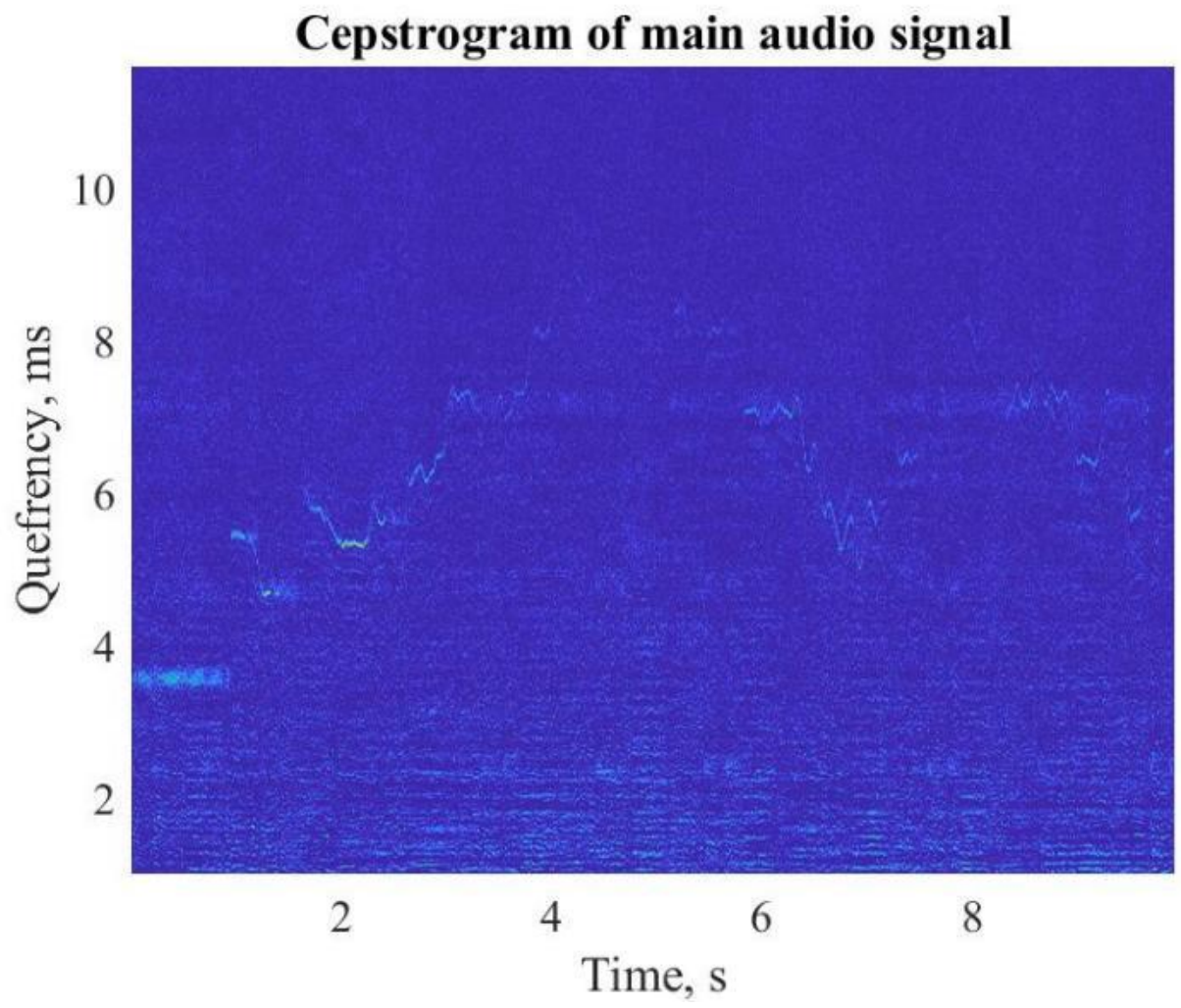

(a)

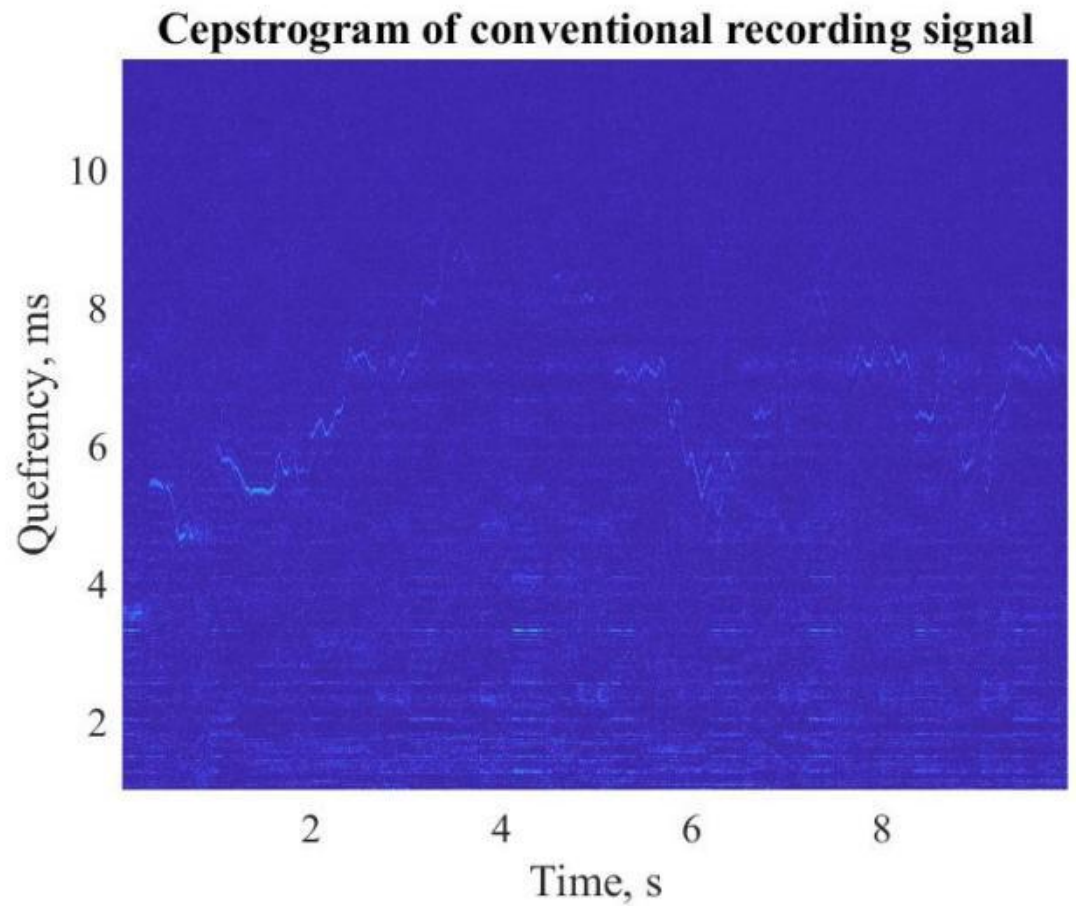

(b)

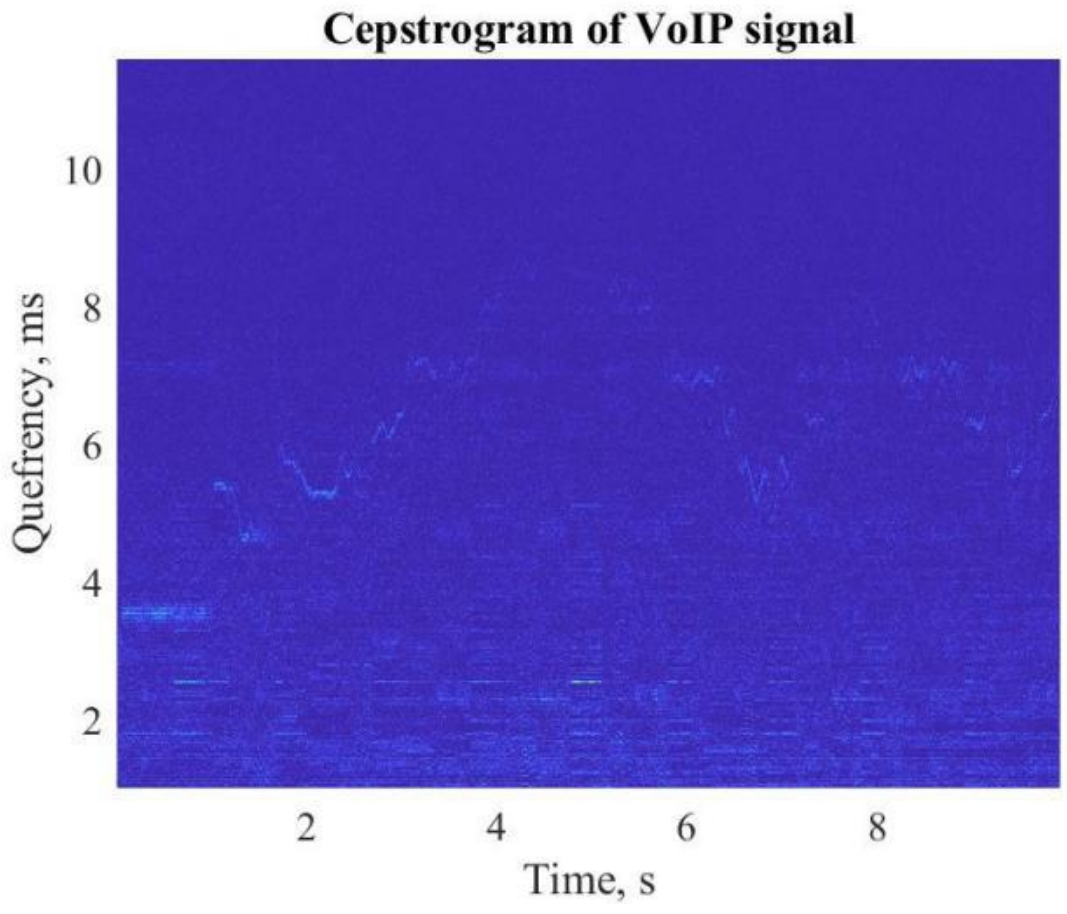

(c)

Figure 44 : Cepstrogram of (a) main audio, (b) conventional recording & (c) VoIP signal.

Cepstrogram is the graphical representation of cepstrum which is the result of computing the inverse Fourier transform of the logarithm of the estimated signal spectrum. Figure 44 (a) and figure 44 (b) represents the Cepstrum of conventional recording & VoIP signal respectively.

The histogram of the signals represent the number of times a typical amplitude occurs during the period of the audio signal. The figure 45 (b) shows that VoIP signal has a DC offset thus the mean value has shifted to -0.23831from the zero position but in case of the conventional recording from figure 45 (a) the mean is at zero.

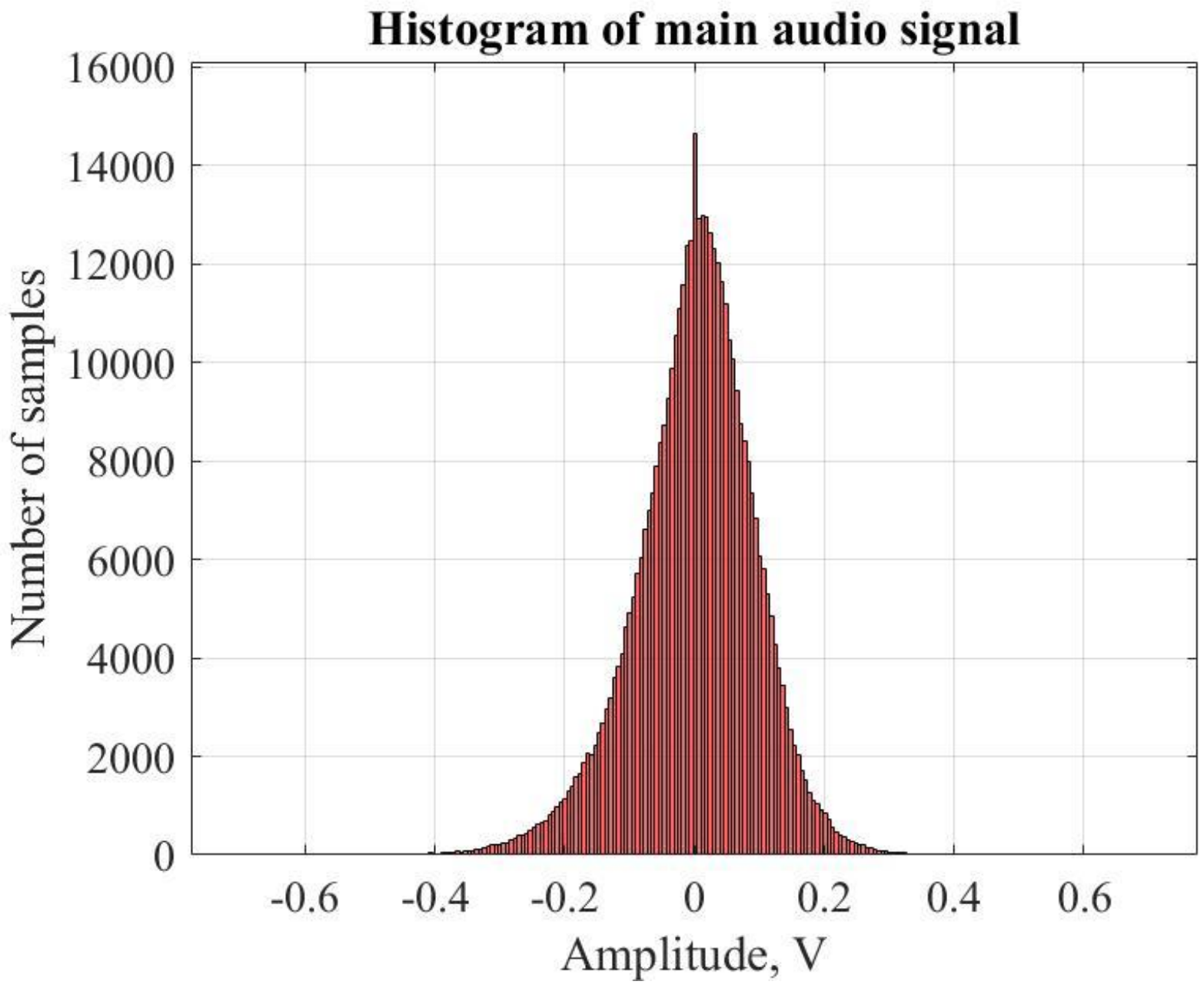

(a)

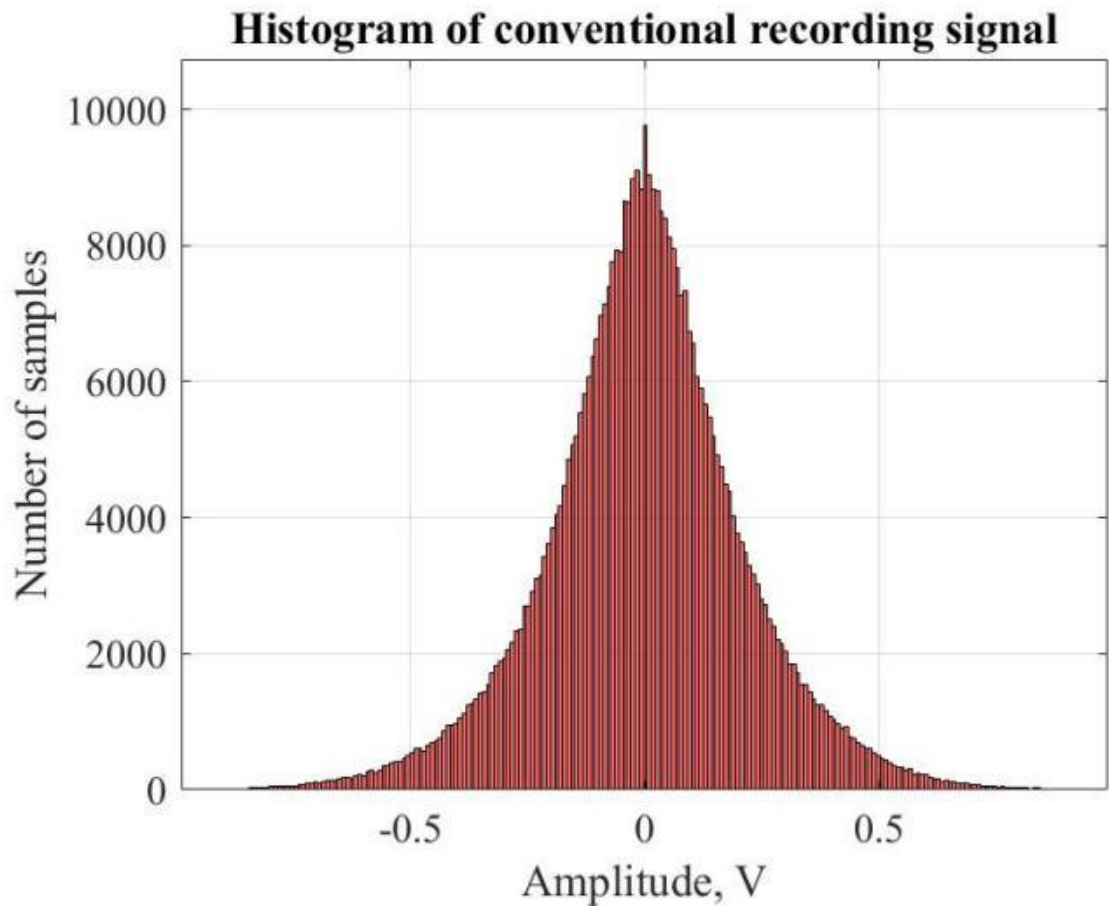

(]b)

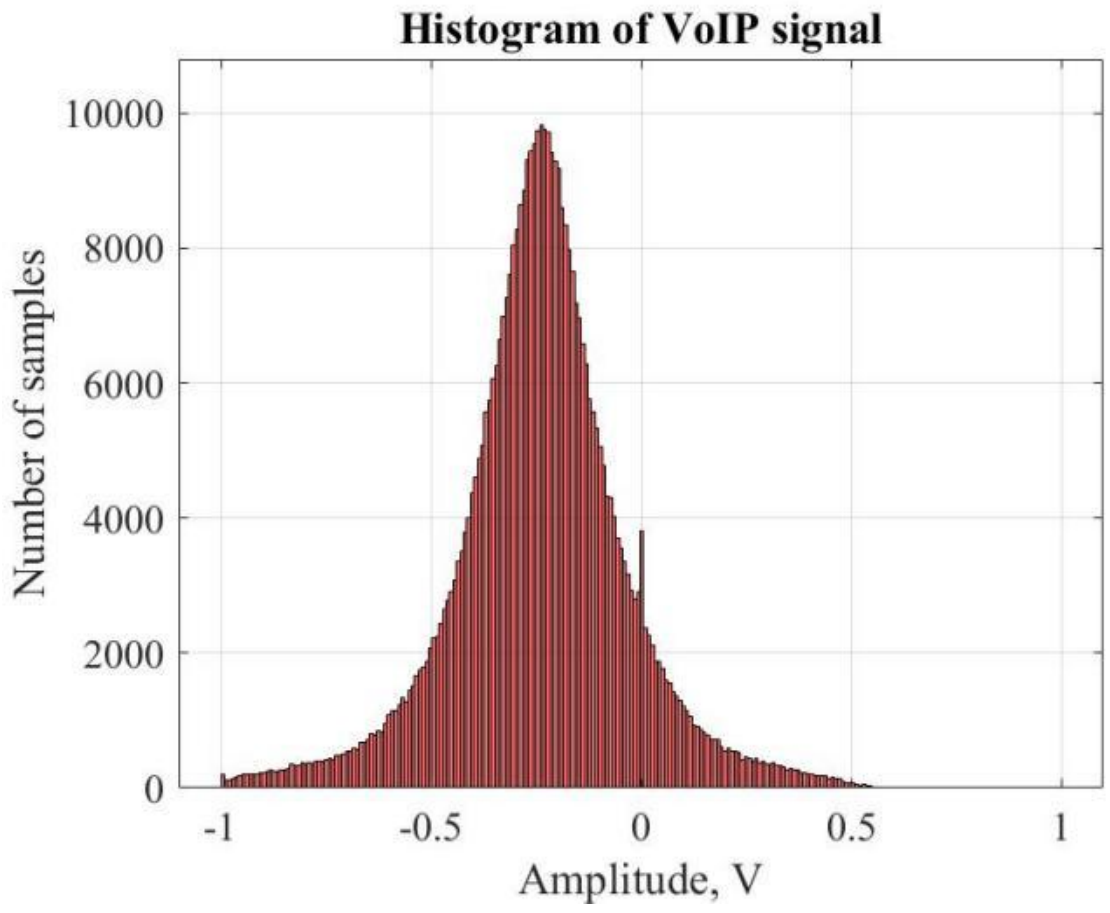

(c)

Figure 45 : Histogram of (a) main audio, (b) conventional recording & (c) VoIP signal.

A correlogram, also known as Auto Correlation Function (ACF) plot, is a graphic way to demonstrate serial correlation in data that doesn't remain constant with time. It's used to check randomness in a data set which is done by computing auto-correlations for data values at different time lags.

From the figure 46 (a) it is seen that the autocorrelation coefficient of the conventional recording signal is near zero, while figure 46 (b) shows that the VoIP signal autocorrelation coefficient varies from 0 to 0.6 within the delay period.

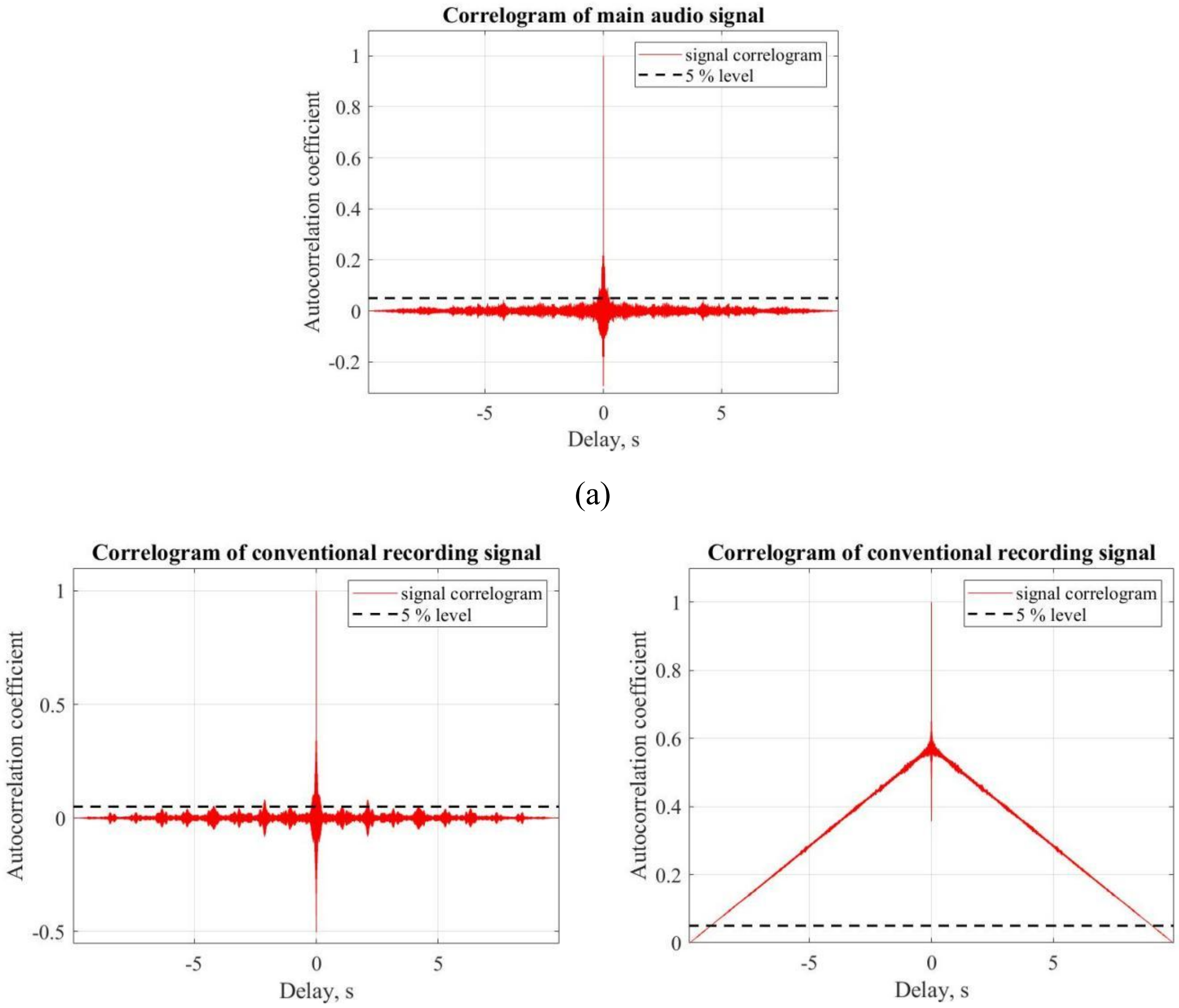

(a)

(b) (c)

Figure 46 : Correlogram of (a) main audio, (b) conventional recording & (c) VoIP signal.

Different properties derived from the graphs of both signal are compared and given below :

Table 5 : Comparison of audio signal properties.

| Properties | Main audio | Conventional recording | VoIP recording |
|---|---|---|---|
| Max value (V) | 0.52217 | 0.89709 | 0.72378 |
| Min value (V) | 0.70547 | -0.8995 | -1 |
| Mean value (V) | 0.0011556 | -4.748e-05 | -0.23831 |
| RMS value (V) | 0.092756 | 0.21002 | 0.20649 |
| Dynamic range, DR (dB) | 87.2783 dB | 89.3888 dB | 90.3087 dB |
| Crest factor, CF (dB) | 17.6227 dB | 12.6346 dB | 13.7022 dB |
| Signal duration | 9.9407 s | 9.9668 s | 9.9407 s |
| Autocorrelation time (s) | 0.2585 s | 4.2252 s | 9.0543 s |

Table 6: Ratio of properties to main audio.

| Properties | Similarity of Converntional Recording to Main Audio | Similarity of VoIP recording to Main Audio |
|---|---|---|
| Max value (V) | 1.71800371526514 | 1.38610031215888 |
| Min value (V) | -1.27503650048904 | -1.41749471983217 |
| Mean value (V) | -0.0410868812737972 | -206.221876081689 |
| RMS value (V) | 2.26422010435983 | 2.22616326706628 |
| Dynamic range, DR (dB) | 1.02418126842526 | 1.03472111624539 |
| Crest factor, CF (dB) | 0.716950297060042 | 0.777531252305265 |
| Signal duration | 1.00262556962789 | 1 |
| Autocorrelation time (s) | 16.3450676982592 | 35.0263056092843 |

Table 6 show the raito of conventional recording to main audo and VoIP recording to main audio, the values that are near "1" determines that it is more close to the main audio signal.

## 4.5. Signal analysis for various microphone distance

The microphone is not always directly attached to the particle photon microcontroller, but also it can be attached through a long wire which would partially isolate the control unit from the sensor that is the microphone.

The microphone is connected to the source microcontroller at various distances and audio data is to be compared with each other. From the graphs below it can be seen that the amplitude deteriorated as the wire length is increased resulting in a signal where sound can not be easily heard. Thus wire length should be kept such that it gives the most effective signal output.

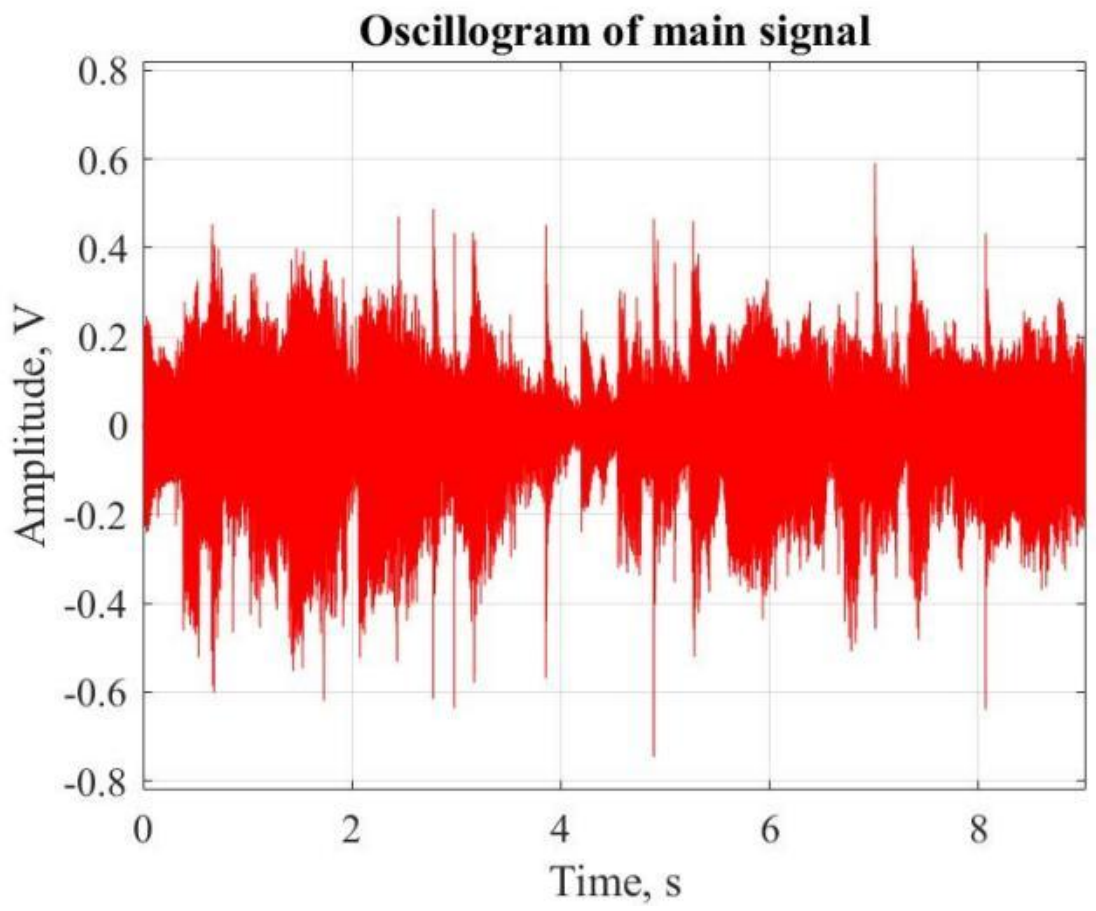

(a)

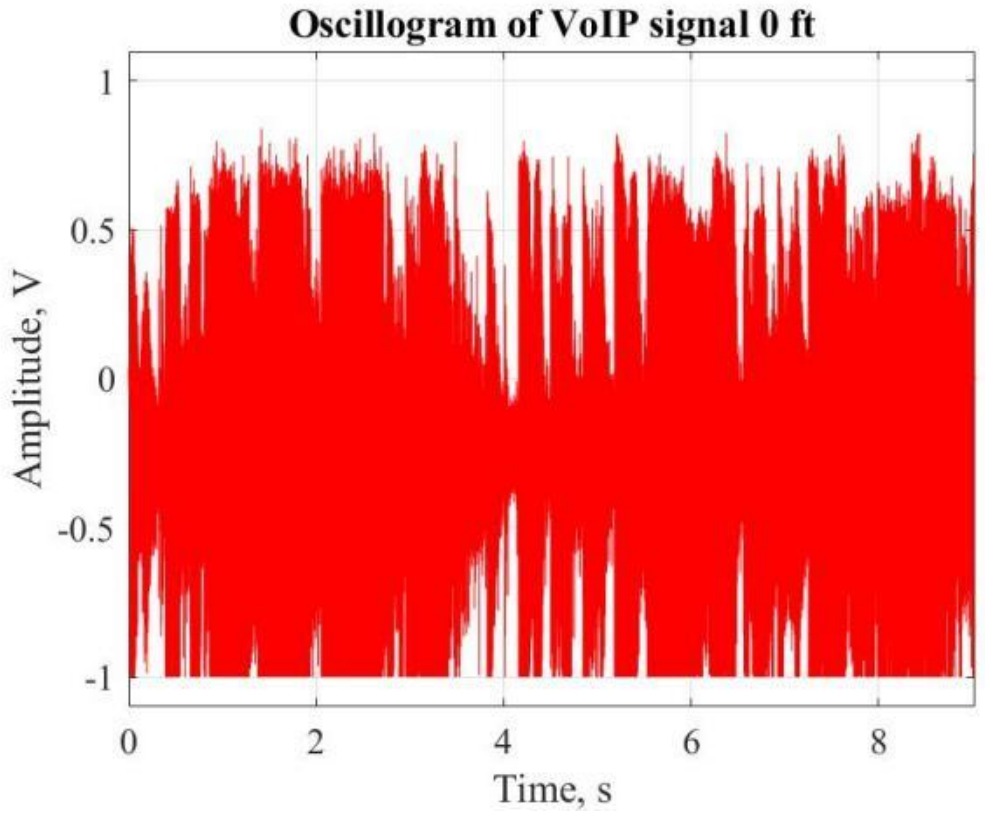

(b)

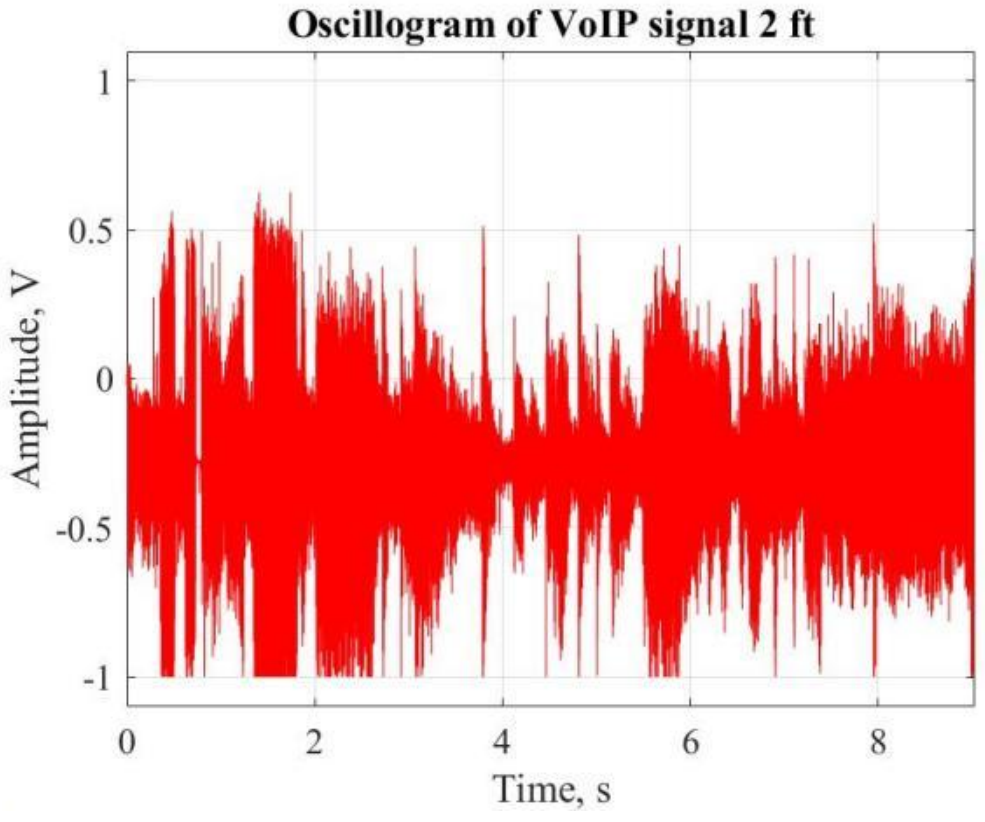

(c)

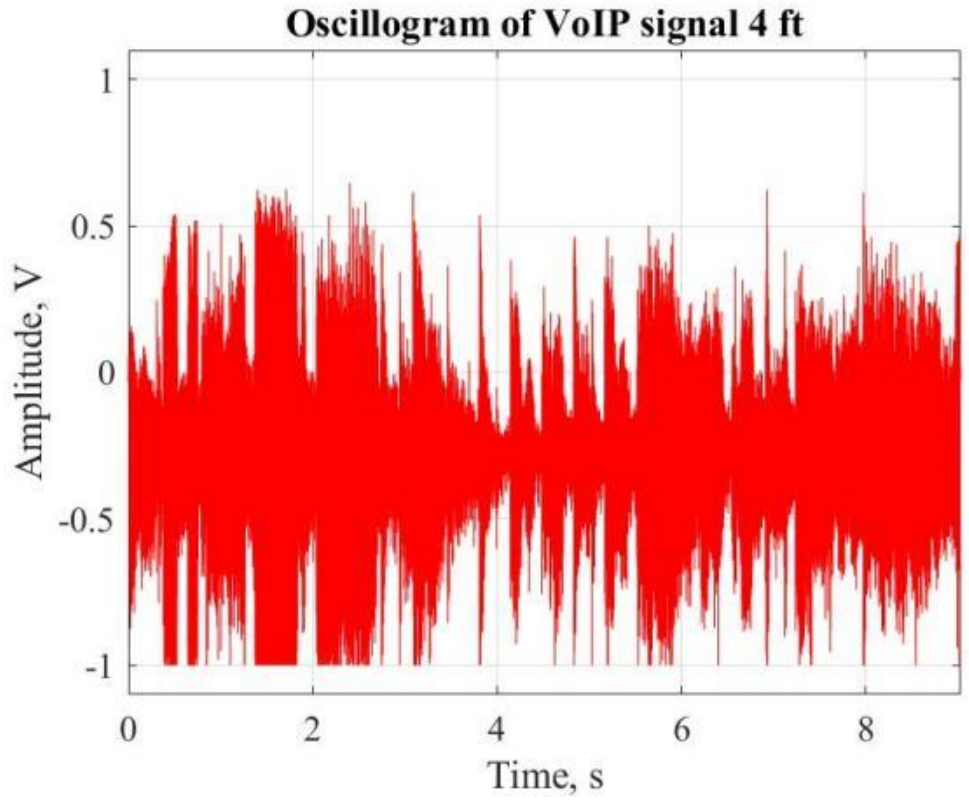

(d)

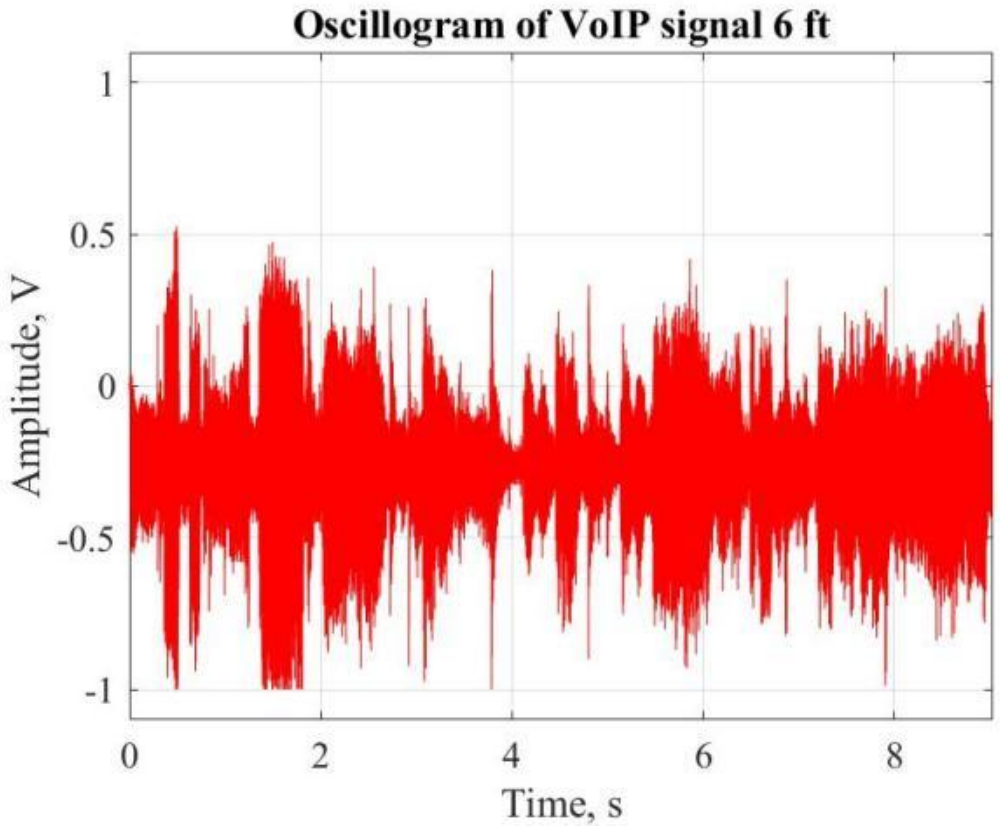

(e)

Figure 47 : Time domain graph (a) main signal, (b) VoIP signal at 0 ft, (c) VoIP signal at 2 ft, (d) VoIP signal at 4 ft, (e) VoIP signal at 6 ft.

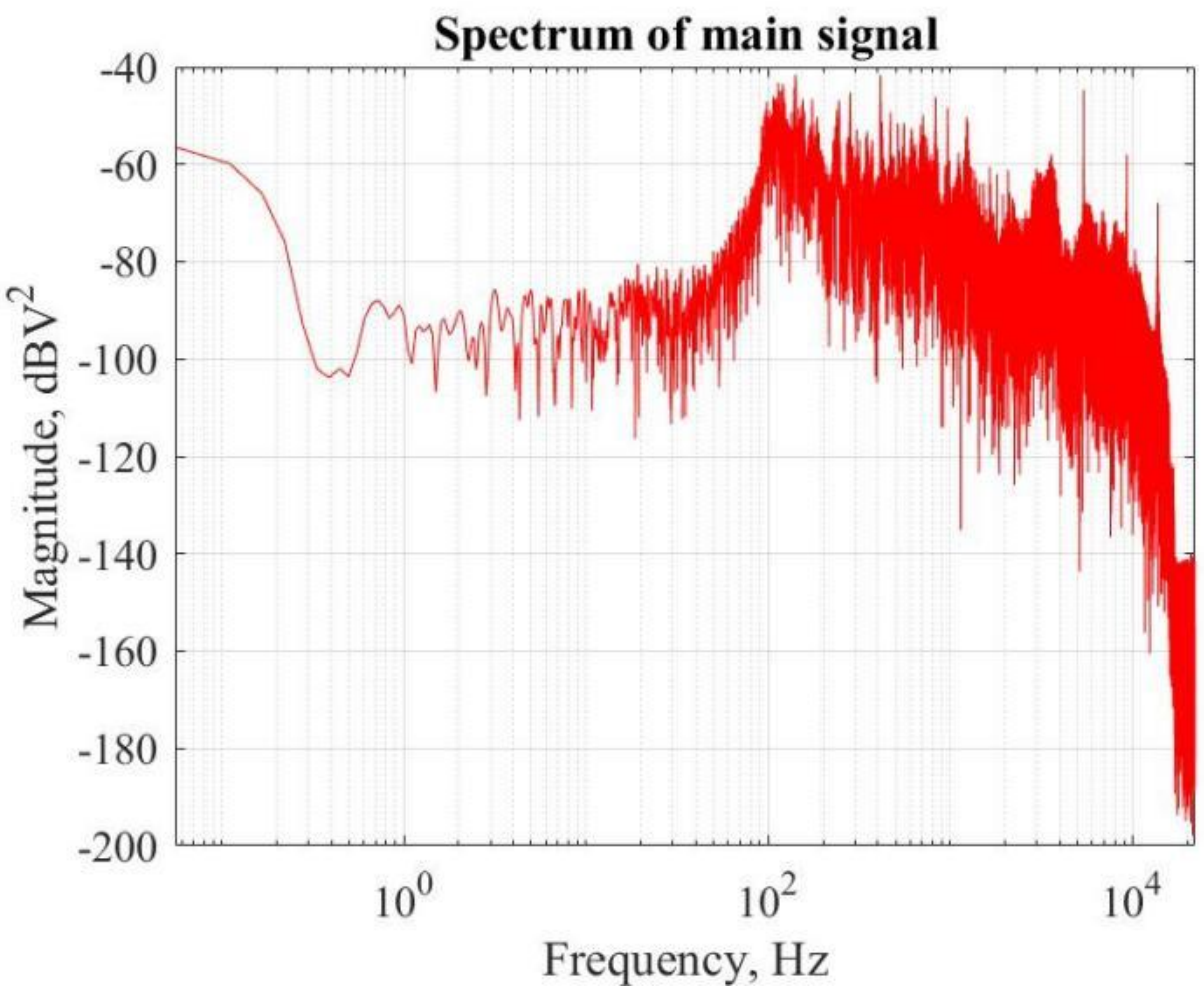

(a)

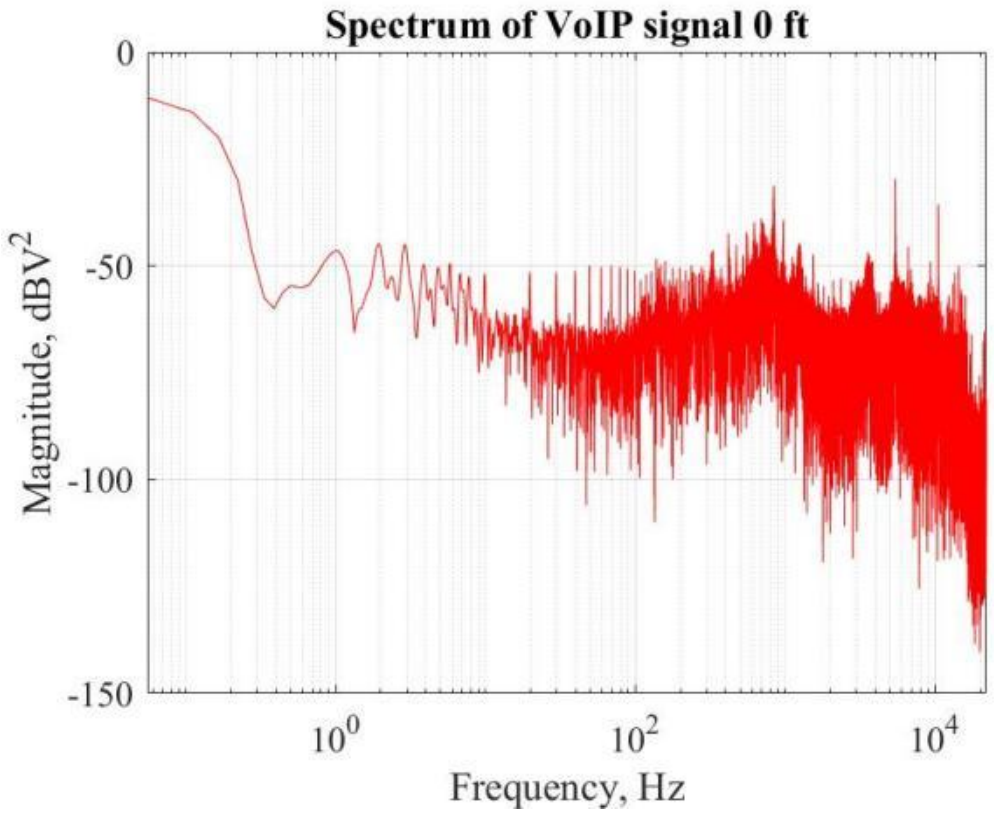

(b)

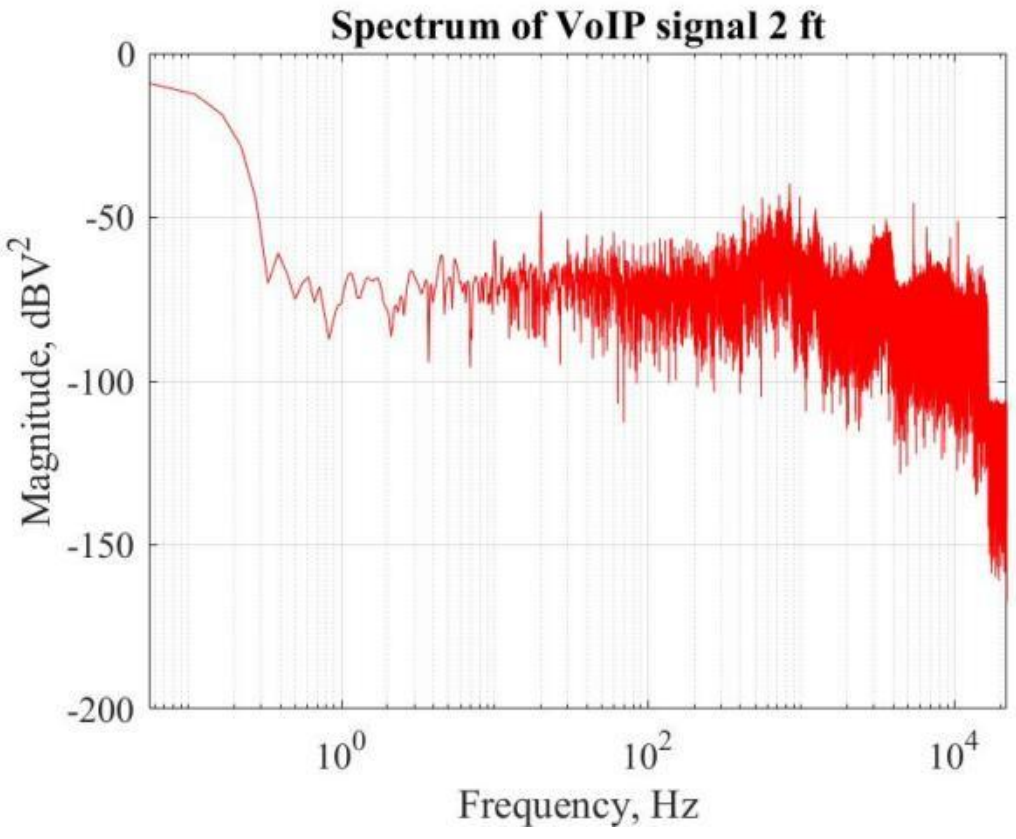

(c)

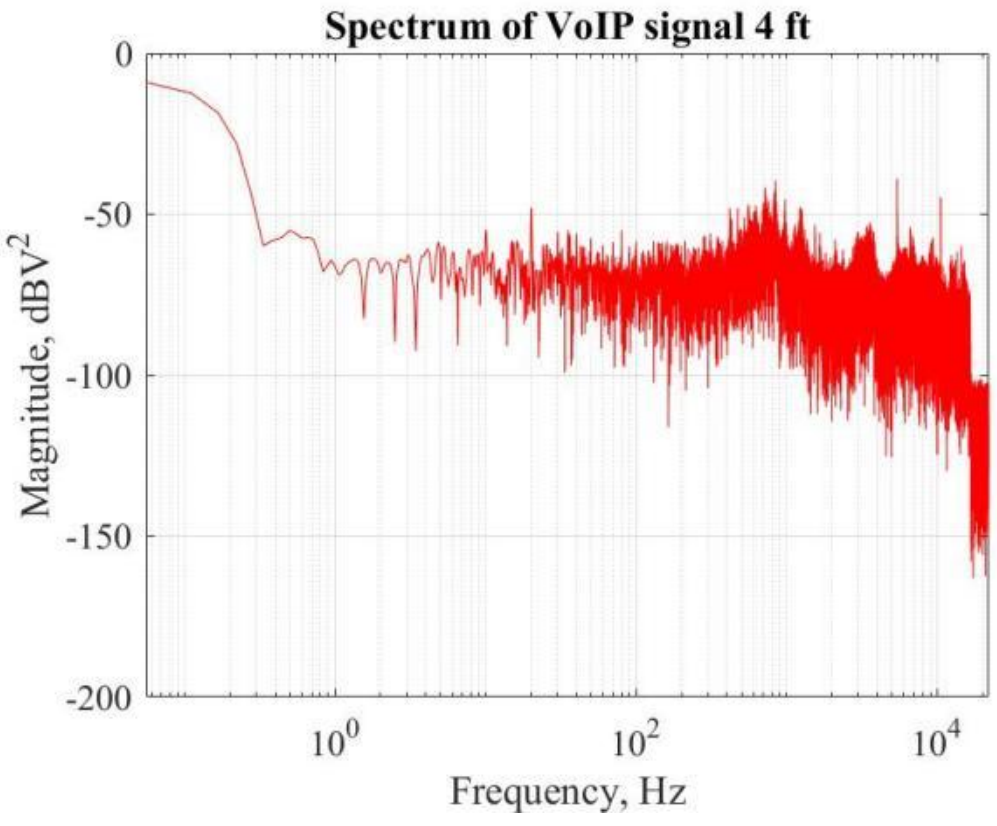

(d)

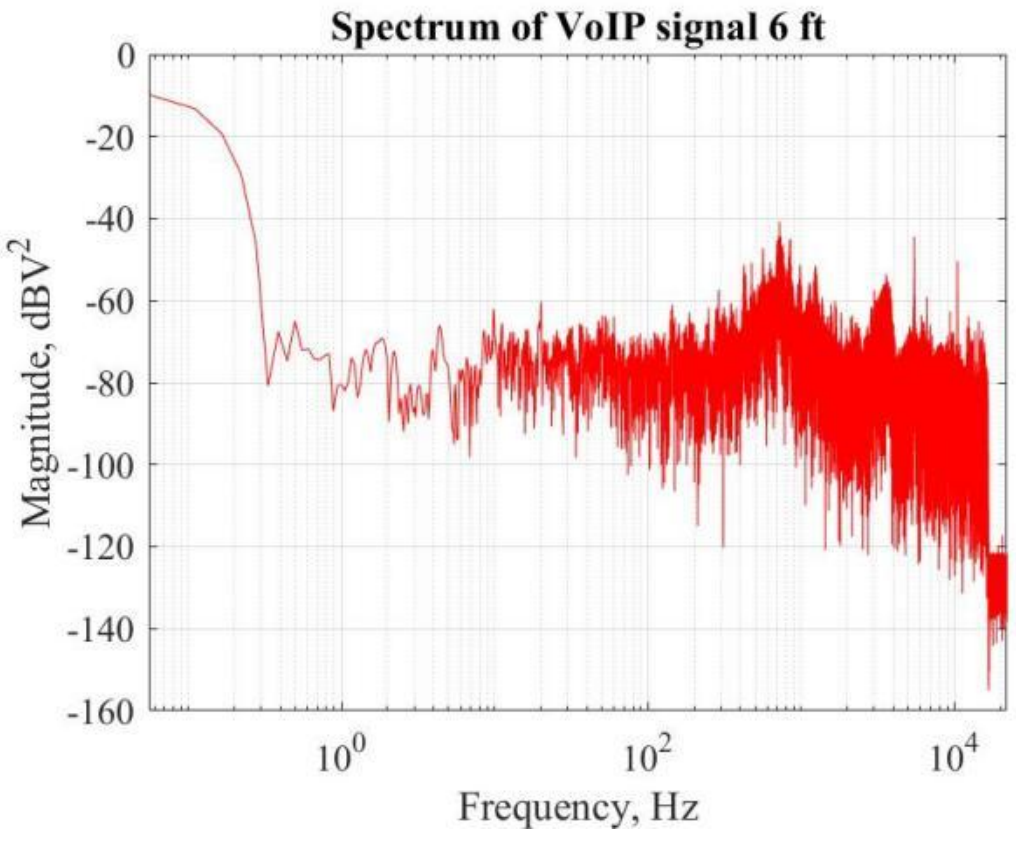

(e)

Figure 48 : Spectrum of magnitude graph of (a) main signal, (b) VoIP signal at 0 ft, (c) VoIP signal at 2 ft, (d) VoIP signal at 4 ft, (e) VoIP signal at 6 ft.

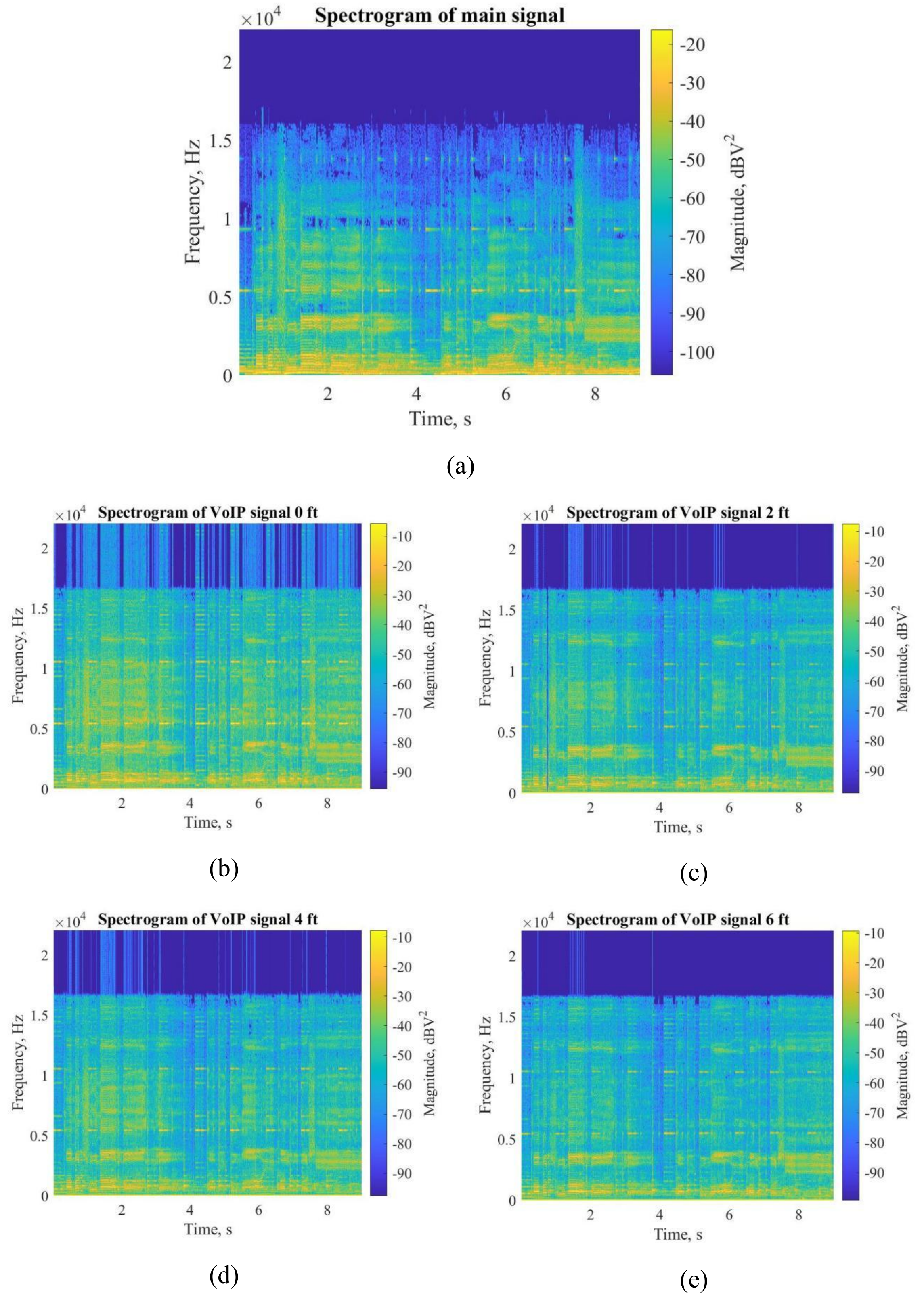

Figure 49 : Spectrogram of (a) main signal, (b) VoIP signal at 0 ft, (c) VoIP signal at 2 ft, (d) VoIP signal at 4 ft, (e) VoIP signal at 6 ft.

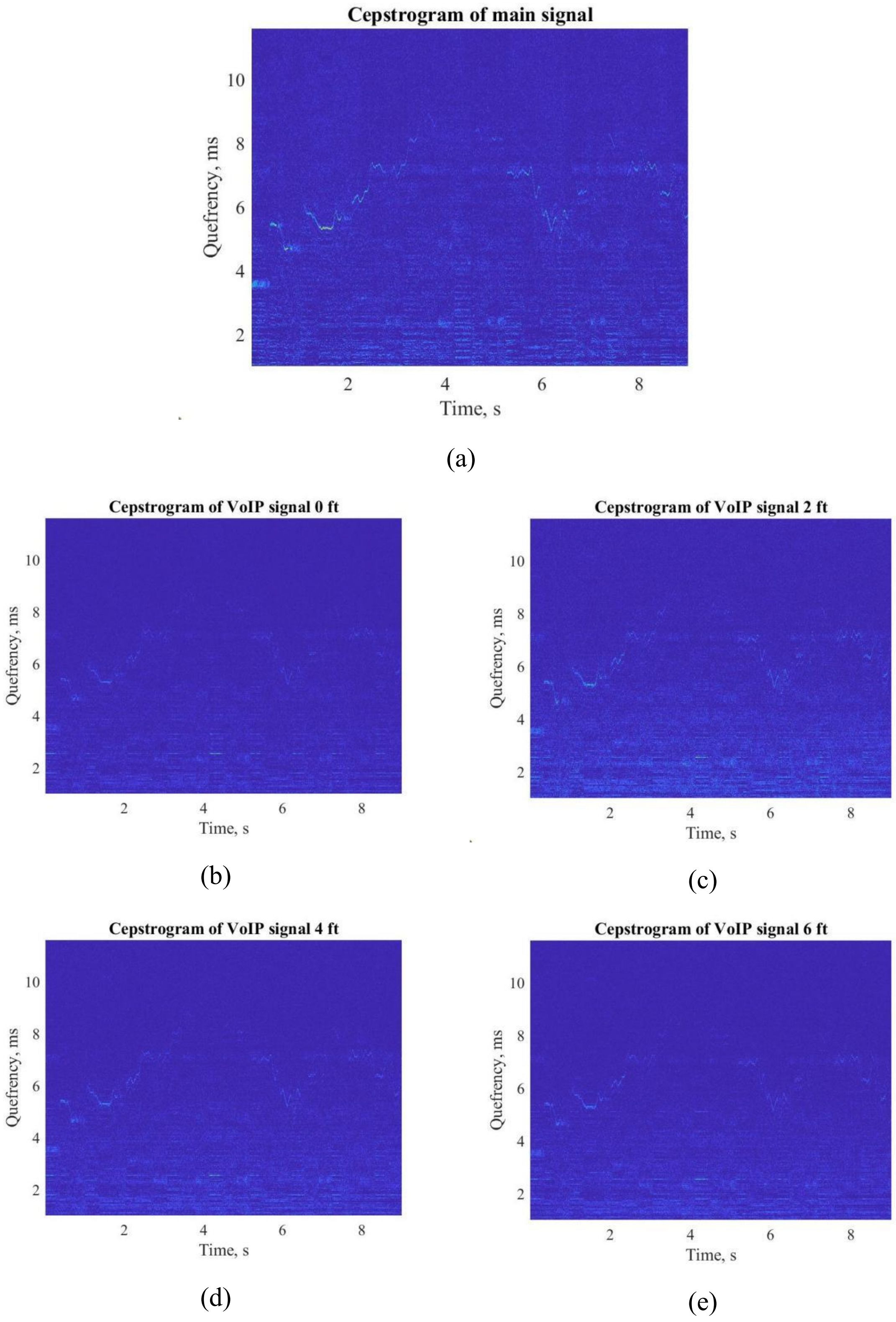

Figure 50 : Cepstrogram of (a) main signal, (b) VoIP signal at 0 ft, (c) VoIP signal at 2 ft, (d) VoIP signal at 4 ft, (e) VoIP signal at 6 ft.

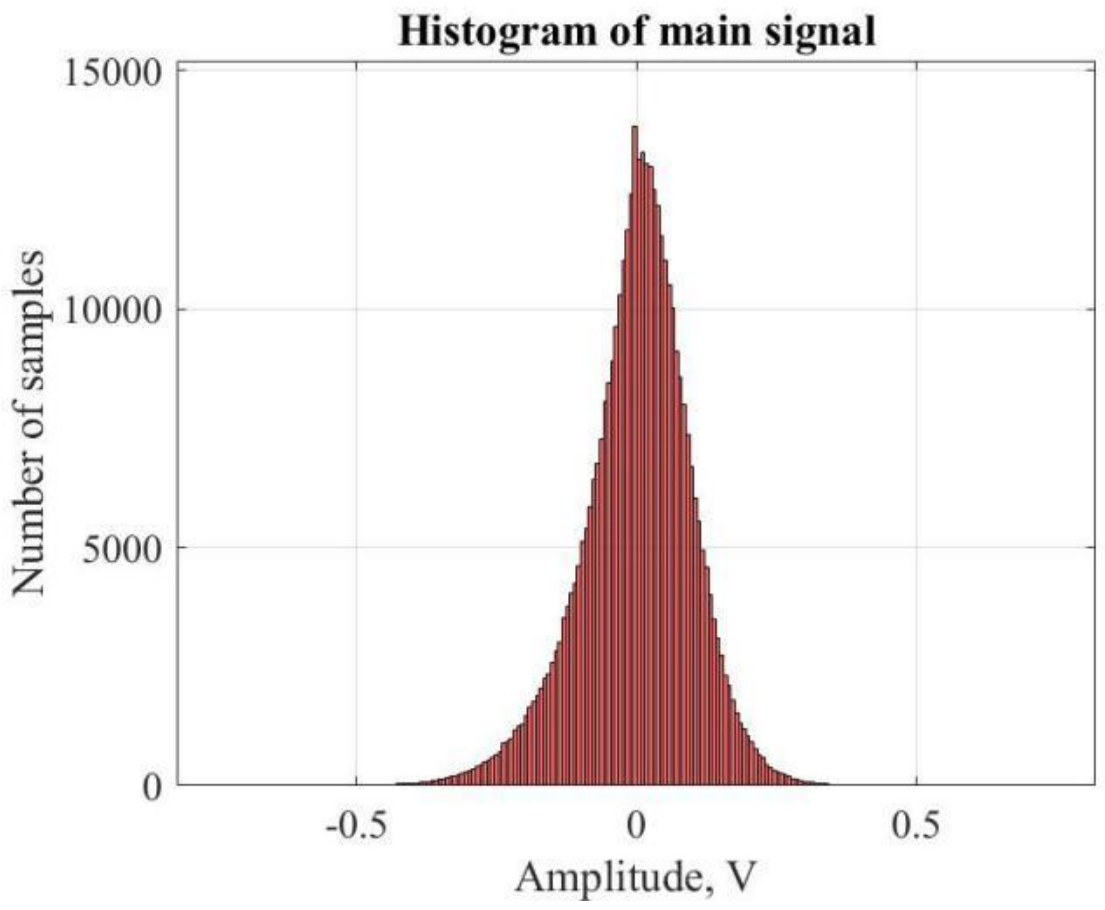

(a)

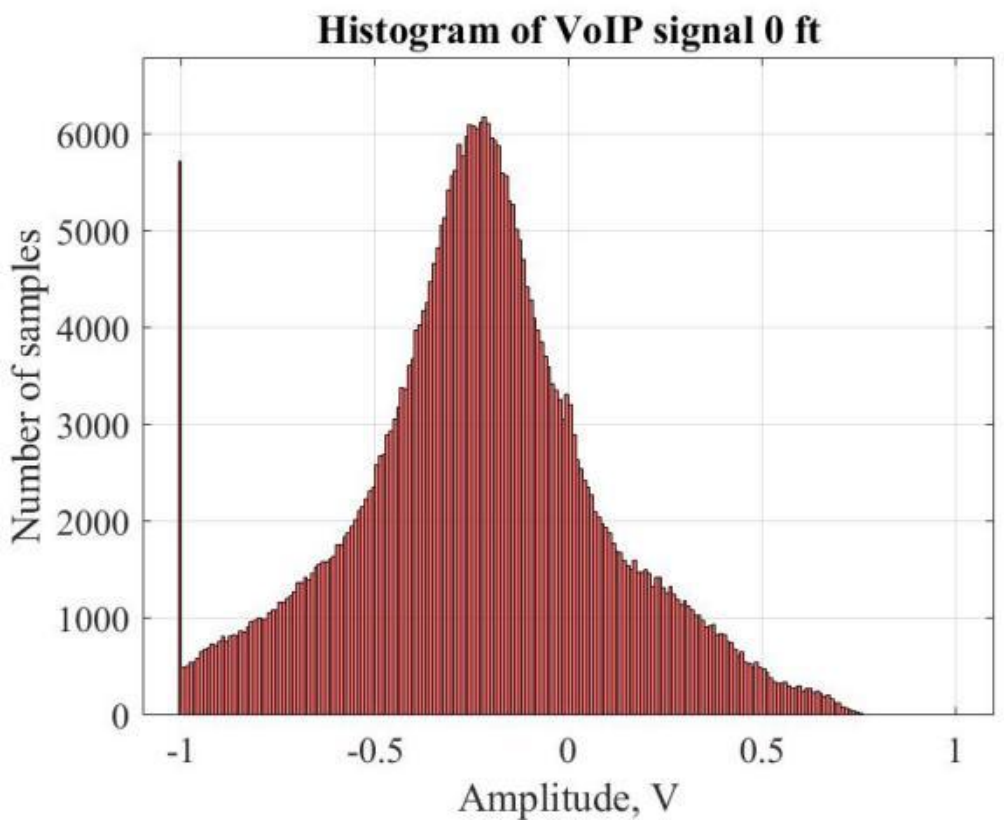

(b)

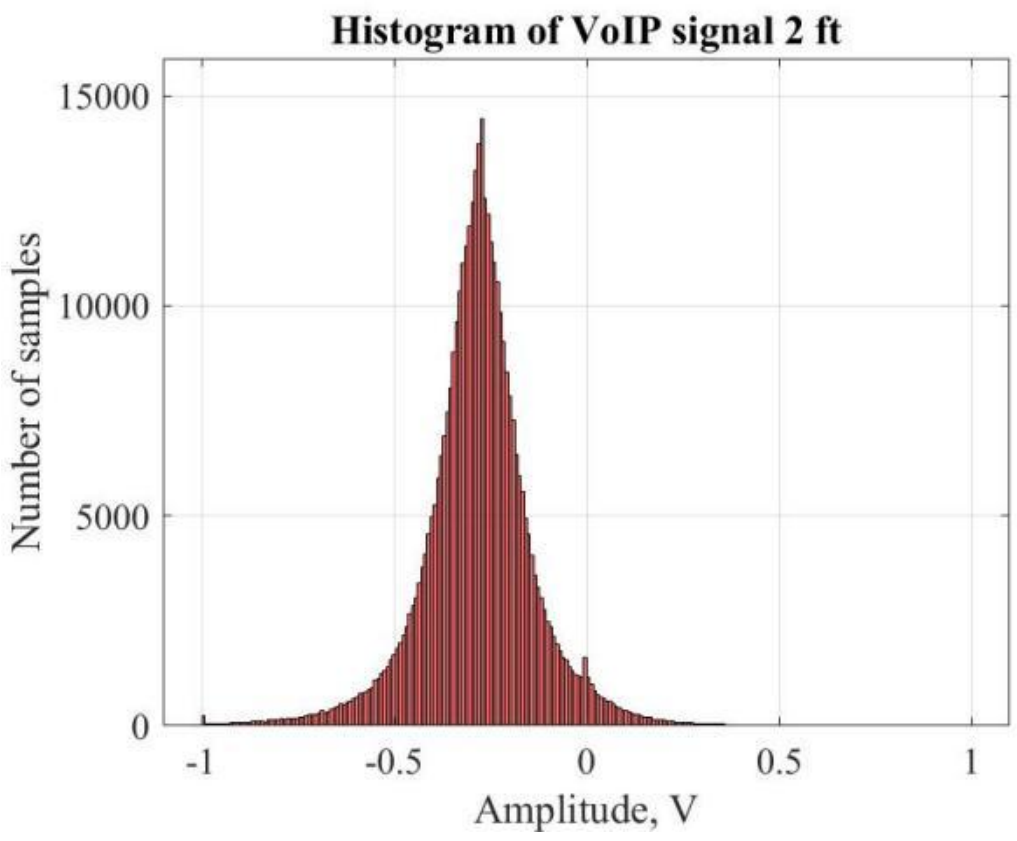

(c)

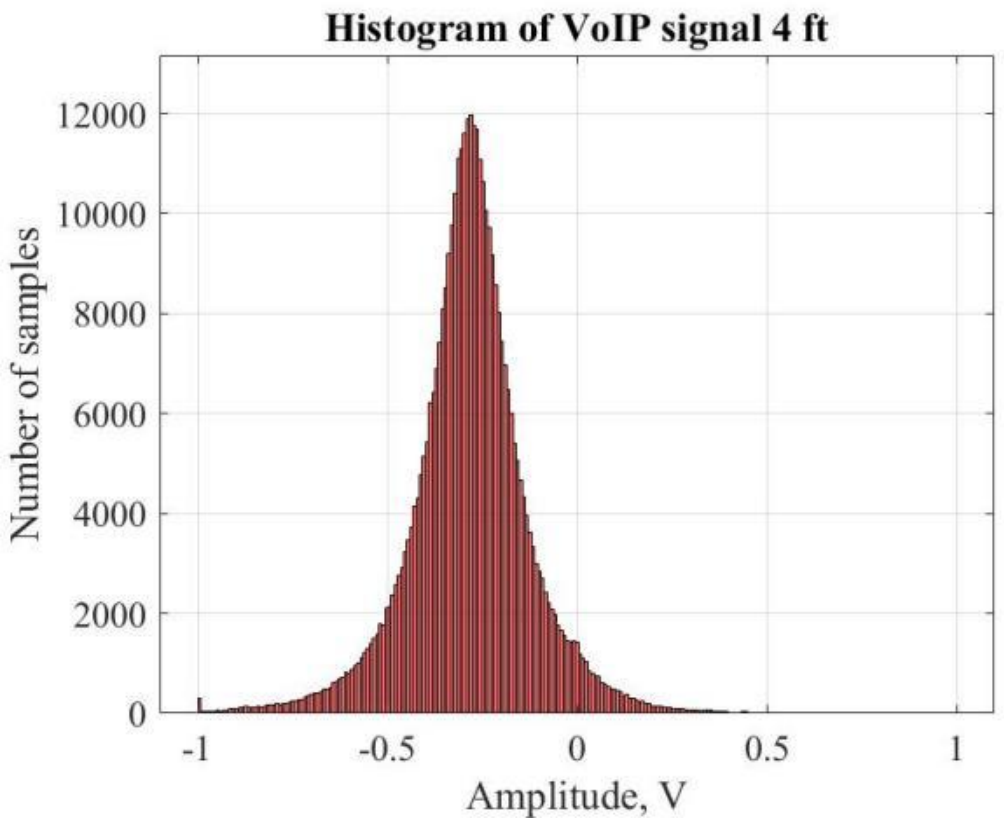

(d)

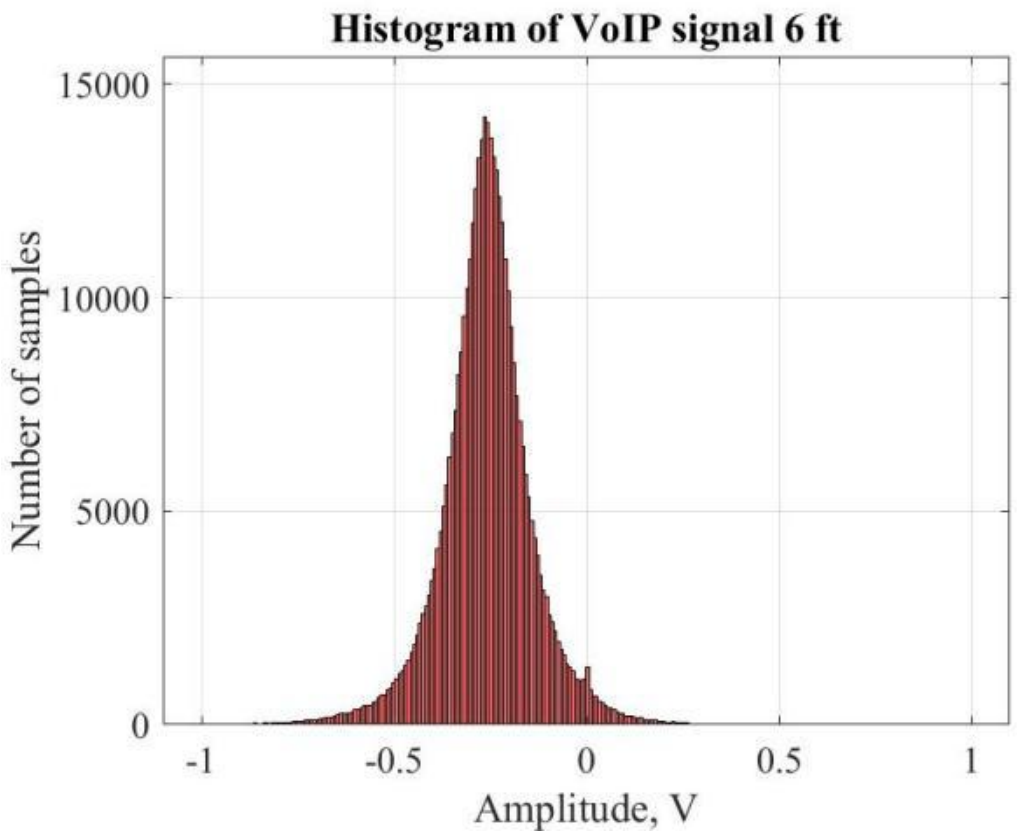

(e)

Figure 51 : Histrogram (a) main signal, (b) VoIP signal at 0 ft, (c) VoIP signal at 2 ft, (d) VoIP signal at 4 ft, (e) VoIP signal at 6 ft.

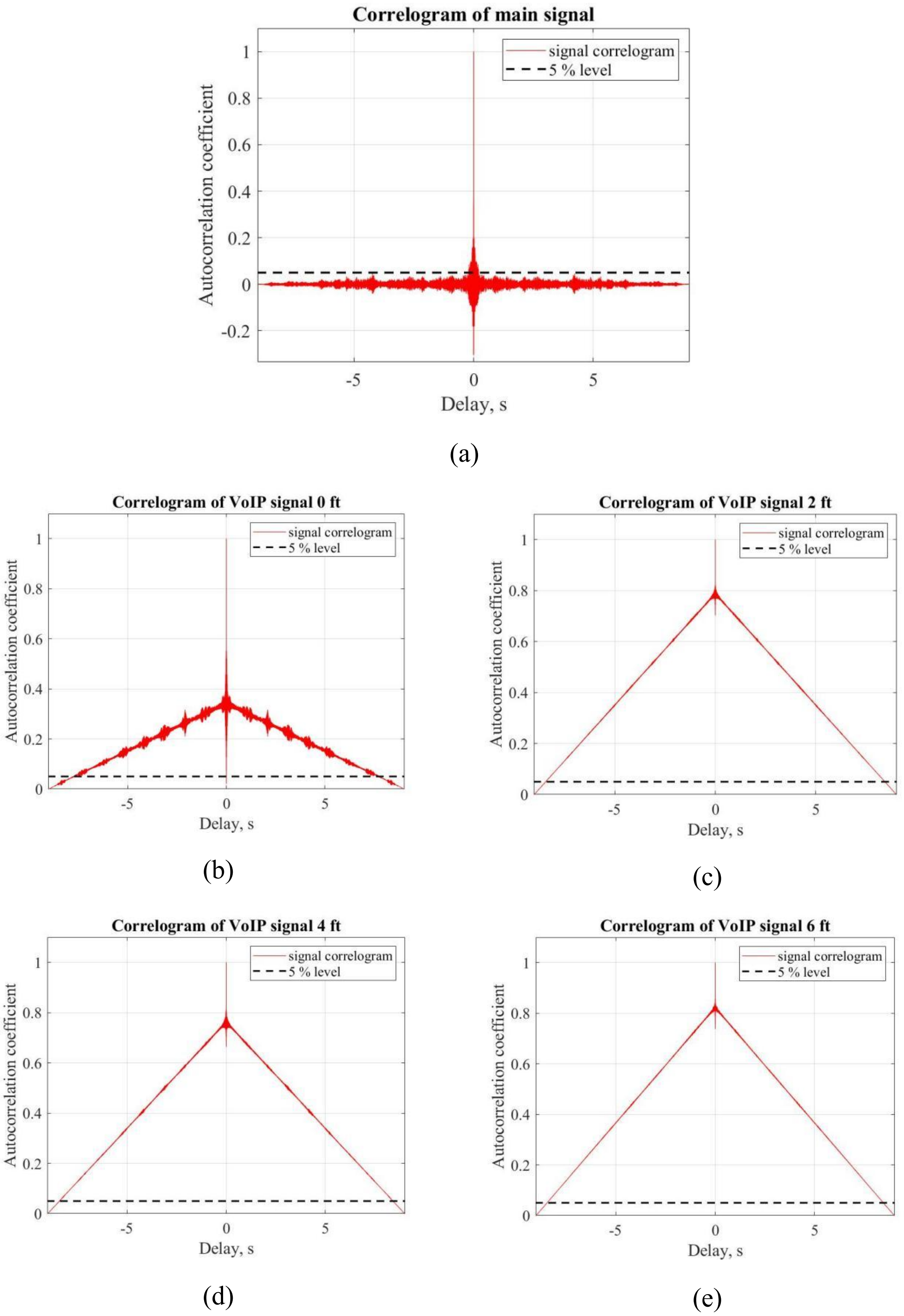

Figure 52 : Correlogram (a) main signal, (b) VoIP signal at 0 ft, (c) VoIP signal at 2 ft, (d) VoIP signal at 4 ft, (e) VoIP signal at 6 ft.

Table 7 : properties for VoIP signal for microphone set at different distance.

| Properties | Main signal | Voip Signal for 0 feet | Voip Signal for 2 feet | Voip Signal for 4 feet | Voip Signal for 6 feet |
|---|---|---|---|---|---|
| Max value | 0.59105 | 0.84014 | 0.62734 | 0.64458 | 0.52623 |
| Min value | -0.74599 | -1 | -1 | -1 | -1 |
| Mean value | 0.0012068 | -0.23353 | -0.28096 | -0.2828 | -0.25885 |
| RMS value | 0.098221 | 0.32333 | 0.14652 | 0.15851 | 0.12128 |
| Dynamic range DR | 87.7634 dB | 90.3087 dB | 90.3087 dB | 90.3087 dB | 90.3087 dB |
| Crest factor CF | 17.6106 dB | 9.807 dB | 16.682 dB | 15.9991 dB | 18.3242 dB |
| Signal duration | 9.0264 s | 9.0264 s | 9.0264 s | 9.0264 s | 9.0264 s |
| Autocorrelation time | 0.20587 s | 7.7759 s | 8.438 s | 8.4161 s | 8.4568 s |

## 4.6. Effectiveness of Security

The security used in this VoIP can not be easily tampered without the help of a good a computer or laptop.There are 3 security steps that an unauthorized person needs to a get through before they can use the transmitted data. First is accessing the WiFi router, second is finding the port of the TCP server, finally decryption of packet data.

### 4.6.1. Tampering of WPA

The WiFi router WPA/WPA2 system are very secured and do not send any data to unwanted device. It always send the data to the required device only. However the device that is connected to the WiFi is not always free from attack. Any device who wants to connect to the WiFi router needs to make a 4-way handshake with the WiFi router. Each time the device get disconnected it has to make the 4-way handshake.The 4-way handshake is the process of exchanging 4 messages between an access point (authenticator) and the client device (supplicant) to generate some encryption keys which can be used to encrypt actual data sent over Wireless medium.[80]

To start the penetration process the unauthorized device sends a signal to a device IP address which is connected to the WiFi and deauthenticates it thus disconnecting the device from the WiFi. Deauthentication attacks represent fraudulent requests that interfere with the communication between routers and devices.[80]

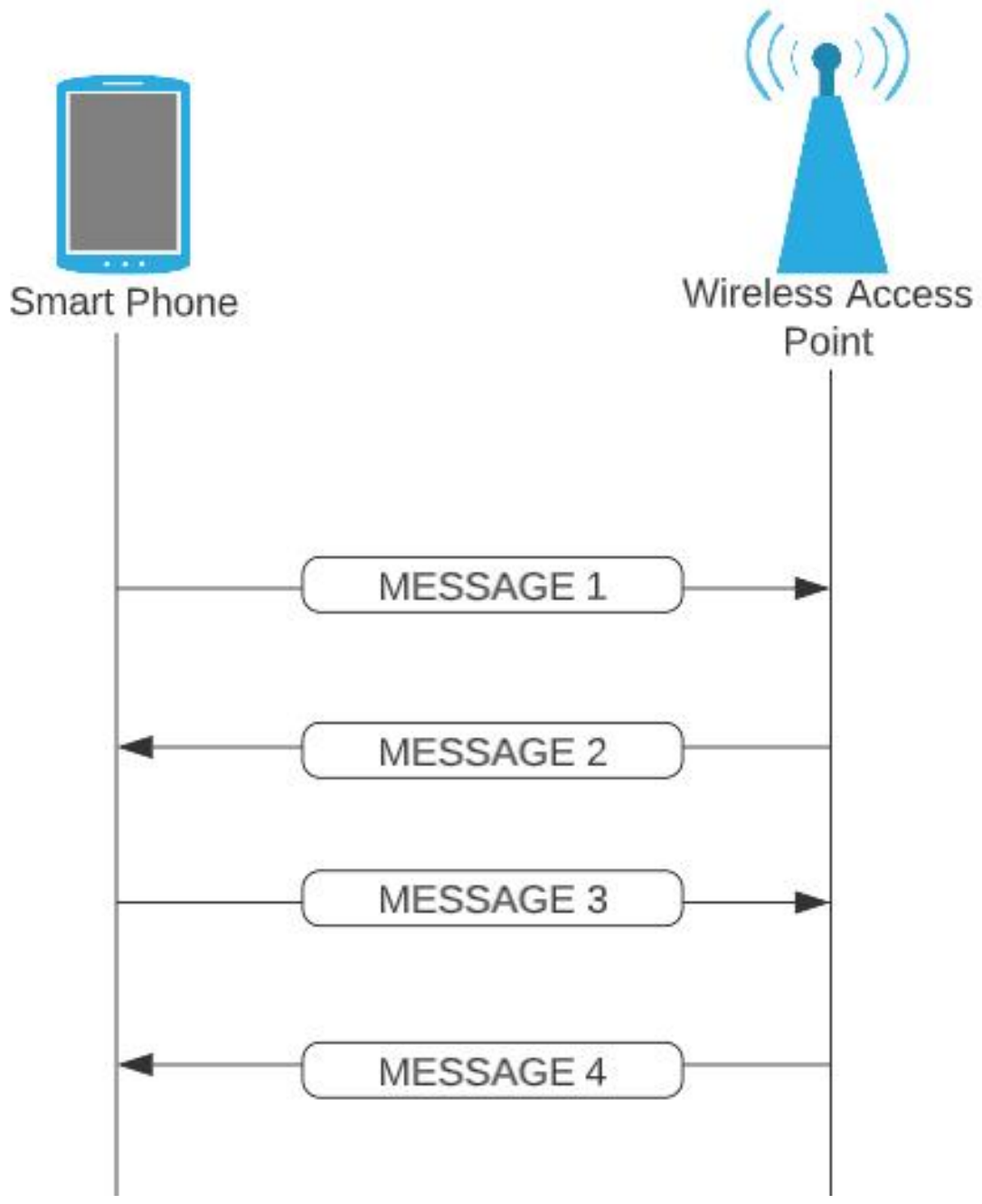

Figure 53 : 4-way handshake.

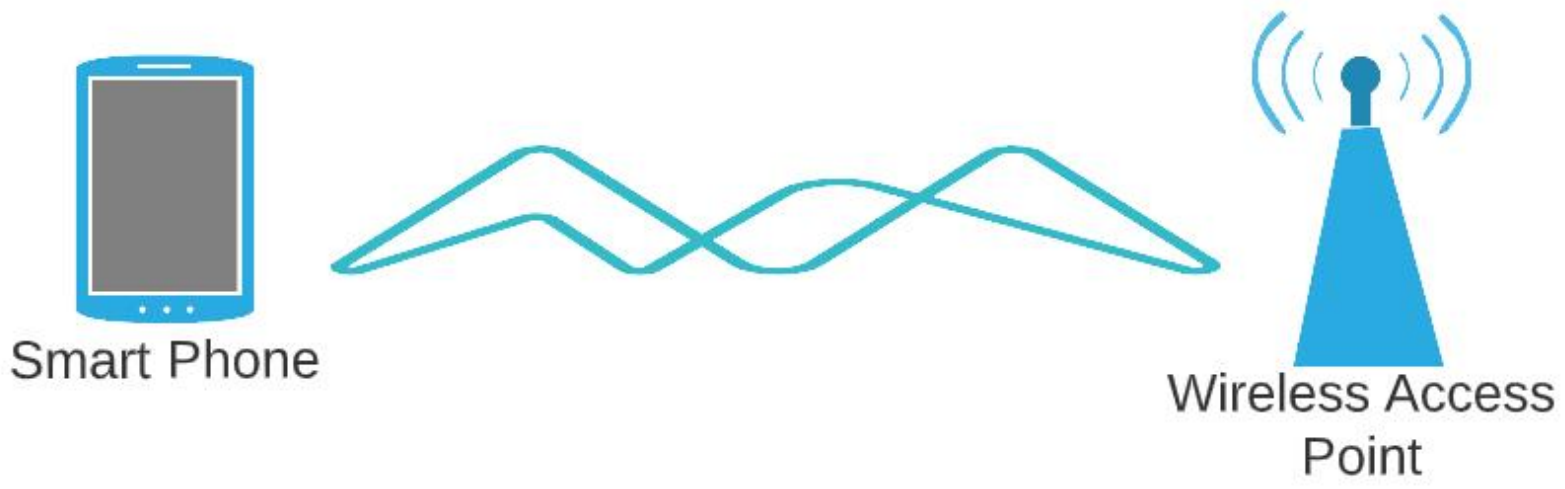

Figure 54 : A device connected to WiFI.

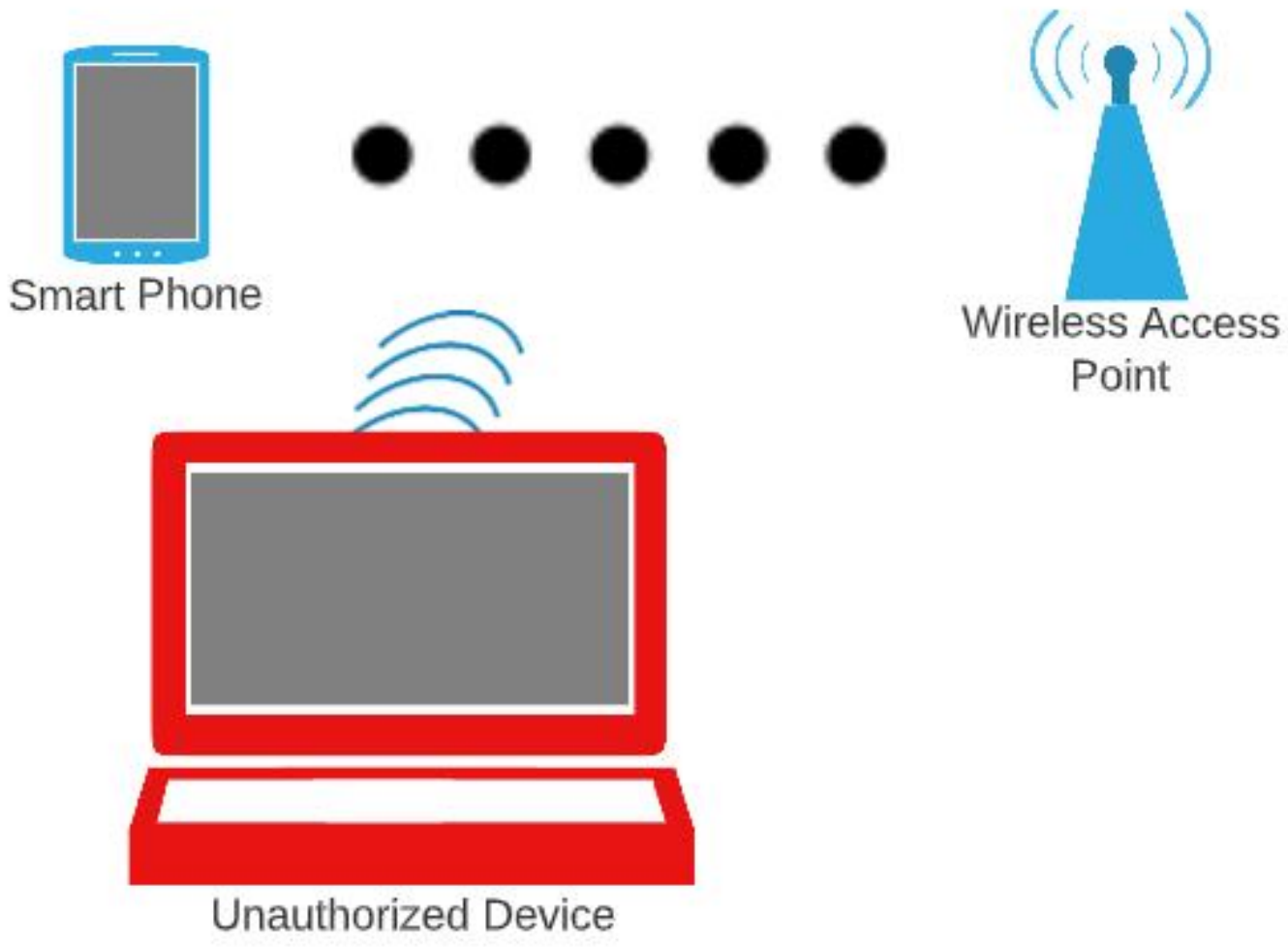

Figure 55 : Deauthentication of device.

Unauthorized device only deauthenticates the connected for a few seconds. As a result the device tried to connect to WiFi again, but this time the unauthorized devices catches all the messages between the device and WiFi router.[80]

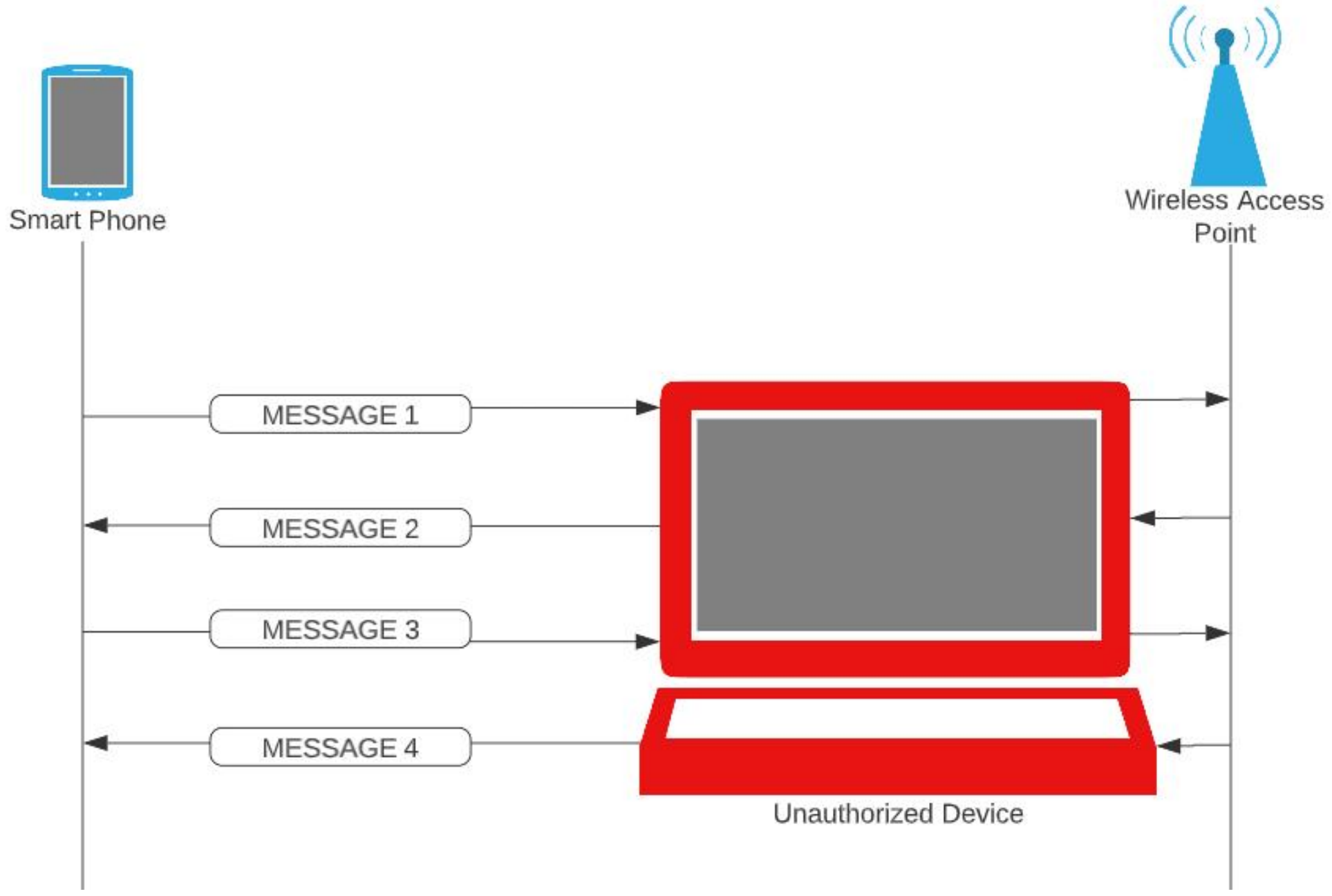

Figure 56 : Unauthorized device receiving the messages.

This messages are used to generate the WiFi password through using “Hash-cat”, which is a software to crack passwords. The time required for the software to generate the password depends upon password complexity and specifications of the device running Hash-cat. If the password is as simple as “1234567890” it would be able to regenerate it in less then 5 minutes if required device specification is fulfilled. More complex password would take hours to generate.[80]

After achieving the password of the WiFi it would be able to connect to the WiFi router. Even if an unauthorized device accesses the WiFi router, the router would notice the sudden increase in user/devices. If unknown user/devices are banned instantly no further attack can be conducted by the unauthorized device. Also if the password of WiFi is constantly changed it would be difficult for unknown user/devices to access it. This facility can be seen in WPA3, however used router did not use WPA3 security system.

### 4.6.2. TCP port scanning

If connected unauthorized device is not banned/removed from the WiFi router it would be then able to scan the available ports of different TCP server available inside the WiFi router with the software “Nmap”. The Process takes about 5 sec to 5 min depending on how many TCP/IP ports are available.[81]

After checking which port is sending data unauthorized device can then check packet data through a 3-way handshake. To get the packet data the correct IP and Port combination need to be used.

### 4.6.3. Decryption of data

The data received from the TCP server needs to be converted to Binary from Hexadecimal if hexadecimal packet data is used to create the audio data no audio file shall be created. Only Binary data can be combined to create the audio wav file.

### 4.6.4. Time Required for Infiltration

The most time is required during infiltration is while getting password needed to access the WiFi router. Thus the time required for getting different passwords are determined. The specifications of the device conducting the infiltration is Dell 15 5000 series, 16GB ram, 240 M.2 SSD, 4GB ATI Radeon GFX.

Table 8 : Time required to determine password of WiFi router.

| Password Given | Words Used | Time Required |
|---|---|---|
| 123456780 | 9 | 1 hour 39 min |
| 12345abcd | 9 | 1 hour 58 min |
| 12345AbCd | 9 | 2 hour 07 min |
| 1234567Tomal | 12 | 3 hour 55 min |
| 1234567#T& | 10 | Unable to determine |
| 1234567#T&l@ | 12 | Unable to determine |

As seen when non alphanumeric words are used it is very hard for the device to determine the password.

# CHAPTER 5

# CONCLUSION AND FUTURE RECOMMENDATION

## 8.1. Conclusions

Numerous tests were done to ensure that the fabricated VoIP system could work, all of which were successful. After running the JavaScript code on a computer, the VoIP system sends audio for a given duration, for signal analysis audio was recorded for 10 sec and another 2-3 seconds to upload the file to Google Drive. The time for uploading files can be reduced by using better internet. In this dissertation, The Baud Rate (sampling frequency) employed in the particle photon source system is 16 kHz. The output amplitude for the fabricated VoIP is higher in negative amplitude than the positive amplitude, rather than being equally distributed on both magnitudes. Also due to more frequent negative amplitude, the mean of the histogram is near -0.2 rather than 0. The reasons for a VoIP system signal to show DC offset or offsetting of a signal from zero can come from any number of issues from a bad analog-to-digital converter(ADC), broken hardware, or power issues. The simplest way to remove DC offset is to use a high-pass filter or process the audio file. A comparison of the Spectrum of power and spectrogram indicates VoIP system signal had developed slightly more magnitude than conventional recording for a given frequency period, this high magnitude is due to noise which is reduced using a 20pF ceramic disk capacitor as a low-pass filter. No Non-authorized person cannot access the audio for secure domain and encryption. The VoIP system's overall performance was practically adequate.

If WPA3 is used or WPA/WPA2 security with monitoring is maintained not easily unknown user/device would not be able to access the system. As tampering of the WPA/WPA2 system takes a lot of time then the other steps, it is favorable to continuously change the password of the WiFi router with a given time is also useful in stopping unknown user/device to access the WiFi. Also through monitoring the connected WiFi users/devices sudden accessed user/device can be banned/removed.

## 8.2. Recommendations

- There is a delay while uploading the audio to Google drive a fast, stable internet is required to upload the file with minimum delay.
- The electret microphone of the system is creating a DC offset which should be removed by using a high pass filter or processing it in “Audacity” software.
- A small lipo battery should be used to remove the need of adapter and help compact the device furthur.
- Microphone can be placed far from the controller by using long wire connectors.

# APPENDICES

## Appendix-I : Particle Photon Code-I for VoIP System

```
//Main Code
#include "SparkIntervalTimer.h"
#include "SimpleRingBuffer.h"
#include <math.h>

#define MICROPHONE_PIN DAC1
#define AUDIO_BUFFER_MAX 8192

#define SINGLE_PACKET_MIN 1024
#define SINGLE_PACKET_MAX 2048

#define TCP_SERVER_PORT 1234
#define SERIAL_DEBUG_ON true

// 1/8000th of a second is 125 microseconds
#define AUDIO_TIMING_VAL 62 /* 16kHz */

uint8_t txBuffer[SINGLE_PACKET_MAX + 1];

SimpleRingBuffer audio_buffer;

unsigned long lastRead = micros();
unsigned long lastSend = millis();
char myIpAddress[24];

TCPClient audioClient;
TCPClient checkClient;
TCPServer audioServer = TCPServer(TCP_SERVER_PORT);

IntervalTimer readMicTimer;

int _sendBufferLength = 0;
unsigned int lastPublished = 0;
```

```
bool _isRecording = false;

volatile int counter = 0;


void setup() {
    #if SERIAL_DEBUG_ON
    Serial.begin(115200);
    #endif

    setADCSampleTime(ADC_SampleTime_3Cycles);

    pinMode(MICROPHONE_PIN, INPUT);
    pinMode(D7, OUTPUT);

    int mySampleRate = AUDIO_TIMING_VAL;

    Particle.variable("ipAddress", myIpAddress, STRING);
    Particle.variable("sample_rate", &mySampleRate, INT);
    Particle.publish("sample_rate", " my sample rate is: " + String(AUDIO_TIMING_VAL));

    IPAddress myIp = WiFi.localIP();
    sprintf(myIpAddress, "%d.%d.%d.%d", myIp[0], myIp[1], myIp[2], myIp[3]);

    audio_buffer.init(AUDIO_BUFFER_MAX);

    lastRead = micros();

    readMicTimer.begin(readMic, AUDIO_TIMING_VAL, uSec);

}

unsigned int lastLog = 0;
unsigned int lastClientCheck = 0;

void loop() {
    unsigned int now = millis();
```

```
  if ((now - lastClientCheck) > 100) {
    lastClientCheck = now;

    checkClient = audioServer.available();
    if (checkClient.connected()) {
      audioClient = checkClient;
    }
  }


  #if SERIAL_DEBUG_ON
  if ((now - lastLog) > 1000) {
    lastLog = now;

    Serial.println("counter was " + String(counter));
     counter = 0;

  }

  #endif

  sendEvery(100);

}
void readMic(void) {
  uint16_t value = analogRead(MICROPHONE_PIN);
  value = map(value, 0, 4095, 0, 255);
  audio_buffer.put(value);
  counter++;
}

void sendEvery(int delay) {
  if ((millis() - lastSend) >= delay) {
    sendAudio();
    lastSend = millis();
  }
}
```

```
void sendAudio(void) {
   if (!audioClient.connected()) {
      return;
   }
   int count = 0;
   int storedSoundBytes = audio_buffer.getSize();
   while (count < storedSoundBytes) {
      if (audio_buffer.getSize() < SINGLE_PACKET_MIN) {
         break;
      }
      int size = min(audio_buffer.getSize(), SINGLE_PACKET_MAX);
      int c = 0;
      for(int c = 0; c < size; c++) {
         txBuffer[c] = audio_buffer.get();
      }
      count += size;
      audioClient.write(txBuffer, size);
   }
}
void write_socket(TCPClient socket, uint8_t *buffer, int count) {
   socket.write(buffer, count);
}
bool ledState = false;


void toggleLED() {
   ledState = !ledState;
   digitalWrite(D7, (ledState) ? HIGH : LOW);
}
```

## Appendix-II: Particle Photon Code-II for VoIP System

```
//SimpleRingBuff.cpp Code
#include "application.h"
#include "SimpleRingBuffer.h"

SimpleRingBuffer::SimpleRingBuffer() {
}
```

```
void SimpleRingBuffer::init(unsigned int size) {
    cur_idx = 0;
    cur_len = 0;

    _data = (uint8_t*)malloc(size * sizeof(uint8_t));
    data_size = size; //cap
}
bool SimpleRingBuffer::put(uint8_t value) {
    if (cur_len < data_size) {

        _data[(cur_idx + cur_len)%data_size] = value;
        cur_len++;
        return true;
    }
    return false;
}
uint8_t SimpleRingBuffer::get() {
    uint8_t val = 0;
    if (cur_len > 0) {
          val = _data[cur_idx];
        cur_idx = (cur_idx + 1) % data_size;
        cur_len--;
    }
    return val;
}


unsigned int SimpleRingBuffer::getSize() {
    return cur_len;
}

unsigned int SimpleRingBuffer::getCapacity() {
    return data_size;
}

void SimpleRingBuffer::clear() {
    cur_idx = 0;
    cur_len = 0;
```

```
}
void SimpleRingBuffer::destroy() {
   free(_data);
   _data = NULL;
}
```

## Appendix-III: Particle Photon Code-III for VoIP System

```
//SimpleRingBuff.h Code
#include "application.h"

#ifndef SIMPLERINGBUFFER_H
#define      SIMPLERINGBUFFER_H


class SimpleRingBuffer {
   protected:

      unsigned int data_size;
      unsigned int cur_idx;
      unsigned int cur_len;

   public:
      SimpleRingBuffer();

      void init(unsigned int size);

      bool put(uint8_t value);
      uint8_t get();

      unsigned int getSize();
      unsigned int getCapacity();

      void clear();
      void destroy();
};

#endif
```

## Appendix-IV : Node.js JavaScript for VoIP System

```
const { google } = require('googleapis');
const path = require('path');

const CLIENT_ID = '270425414783-47ugboaa4pq1m6108gjekmm97ccmldd6.apps.googleusercontent.com';
const CLIENT_SECRET = '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';
const REDIRECT_URI = 'https://developers.google.com/oauthplayground';
const REFRESH_TOKEN = '1//048clF1qEwzreCgYIARAAGAQSNwF-L9IrSvtvTuMUJ5CcvxPnUNEAgY3sCsnTxkLzrhR8PiFdRhApLmwsg2V_Ezo4Q0MflKorm3Q';
var settings = {
    ip: "192.168.43.126",
    port: 1234
};
//By 1602023
var fs = require("fs");
var samplesLength = 1000;
var sampleRate = 16000;
var bitsPerSample = 8;
var numChannels = 1;
var outStream = fs.createWriteStream("test.wav");
var writeHeader = function() {
    var b = new Buffer(1024);
    b.write('RIFF', 0);
    b.writeUInt32LE(32 + samplesLength * numChannels, 4);
    b.write('WAVE', 8);/
    b.write('fmt ', 12);
    b.writeUInt32LE(16, 16);
    b.writeUInt16LE(1, 20);
    b.writeUInt16LE(1, 22);
    b.writeUInt32LE(sampleRate, 24);
    b.writeUInt32LE(sampleRate * 1, 28);
    b.writeUInt16LE(numChannels * 1, 32);
    b.writeUInt16LE(bitsPerSample, 34);
    b.write('data', 36);
    b.writeUInt32LE(0, 40);
    outStream.write(b.slice(0, 50));
};
writeHeader(outStream);
```

```
var net = require('net');
console.log("connecting...");
client = net.connect(settings.port, settings.ip, function () {
        client.setNoDelay(true);
    client.on("data", function (data) {
        try {
                        outStream.write(data);
                        console.log("got chunk of " + data.length + " bytes ");
                        console.log("got chunk of " + data.toString('HEX'));
                }
        catch (ex) {
            console.error("Er!" + ex);
        }
    });
});

setTimeout(function() {
        console.log('recorded for 10 seconds');
const oauth2Client = new google.auth.OAuth2(
  CLIENT_ID,
  CLIENT_SECRET,
  REDIRECT_URI
);
oauth2Client.setCredentials({ refresh_token: REFRESH_TOKEN });
const drive = google.drive({
  version: 'v3',
  auth: oauth2Client,
});
const filePath = path.join(__dirname, 'test.wav');
async function uploadFile() {
  try {
    const response = await drive.files.create({
      requestBody: {
        name: 'test.wav',
        mimeType: 'audio/wav',
      },
      media: {
        mimeType: 'audio/wav',
        body: fs.createReadStream(filePath),
      },
```

```
        });
        console.log(response.data);
      } catch (error) {
        console.log(error.message);
      }
    }
    uploadFile();
            client.end();
            outStream.end();
    }, 10 * 1000);
```

## Appendix-V : Matlab Code-I for Acoustic Analysis

```
 clear, clc, close all

%% get a section of the sound file
[x, fs] = audioread('Voip audio.mp3');  % load an audio file
x = x(:, 1);                        % get the first channel
N = length(x);                      % signal length
to = (0:N-1)/fs;                    % time vector

%% detrend the signal
% organize question dialog menu about the detrending
quest = 'Do you want to detrend the signal?';
dlgtitle = 'Detrending';
btn1 = 'Yes, detrend the signal';
btn2 = 'No, do not detrend the signal';
defbtn = btn1;
answer = questdlg(quest, dlgtitle, btn1, btn2, defbtn);

% detrend the signal
switch answer
   case btn1
   % detrend the signal
   x = detrend(x);
   case btn2
   % do not detrend the signal
end
```

```
%% normalize the signal
% organize question dialog menu about the normalization
quest = 'What type of normalization do you want?';
dlgtitle = 'Normalization';
btn1 = 'Normalize the signal to unity peak';
btn2 = 'Normalize the signal to unity RMS-value';
btn3 = 'Do not normalize the signal';
defbtn = btn1;
answer = questdlg(quest, dlgtitle, btn1, btn2, btn3, defbtn);

% normalize the signal
switch answer
    case btn1
    % normalize to unity peak
    x = x/max(abs(x));
    case btn2
    % normalize to unity RMS-value
    x = x/std(x);
    case btn3
    % do not normalize the signal
end

%% plot the signal oscillogram
figure(1)
plot(to, x, 'r')
xlim([0 max(to)])
ylim([-1.1*max(abs(x)) 1.1*max(abs(x))])
grid on
set(gca, 'FontName', 'Times New Roman', 'FontSize', 14)
xlabel('Time, s')
ylabel('Amplitude, V')
title('Oscillogram of conventional recording signal')

%% plot the signal spectrum
% spectral analysis
winlen = N;
win = blackman(winlen, 'periodic');
```

```
nfft = round(2*winlen);
[PS, f] = periodogram(x, win, nfft, fs, 'power');
X = 10*log10(PS);

% plot the spectrum
figure(2)
semilogx(f, x, 'r')
xlim([0 max(f)])
grid on
set(gca, 'FontName', 'Times New Roman', 'FontSize', 14)
title('Spectrum of conventional recording signal')
xlabel('Frequency, Hz')
ylabel('Magnitude, dBV^2')

%% plot the signal spectrogram
% time-frequency analysis
winlen = 1024;
win = blackman(winlen, 'periodic');
hop = round(winlen/4);
nfft = round(2*winlen);
[~, F, T, STPS] = spectrogram(x, win, winlen-hop, nfft, fs, 'power');
STPS = 10*log10(STPS);

% plot the spectrogram
figure(3)
surf(T, F, STPS)
shading interp
axis tight
box on
view(0, 90)
set(gca, 'FontName', 'Times New Roman', 'FontSize', 14)
xlabel('Time, s')
ylabel('Frequency, Hz')
title('Spectrogram of conventional signal')

[~, cmax] = caxis;
caxis([max(-120, cmax-90), cmax])
```

```
hClbr = colorbar;
set(hClbr, 'FontName', 'Times New Roman', 'FontSize', 14)
ylabel(hClbr, 'Magnitude, dBV^2')

%% plot the cepstrogram
% cepstral analysis
[C, q, tc] = cepstrogram(x, win, hop, fs);          % calculate the cepstrogram
C = C(q >= 1e-3, :);                                % ignore all cepstrum coefficients for
                                                    % quefrencies bellow 1 ms
q = q(q >= 1e-3);                                   % ignore all quefrencies bellow 1 ms
q = q*1000;                                         % convert the quefrency to ms

% plot the cepstrogram
figure(4)
[T, Q] = meshgrid(tc, q);
surf(T, Q, C)
shading interp
axis tight
box on
view(0, 90)
set(gca, 'FontName', 'Times New Roman', 'FontSize', 14)
xlabel('Time, s')
ylabel('Quefrency, ms')
title('Cepstrogram of conventional recording signal')

[~, cmax] = caxis;
caxis([0 cmax])

%% plot the signal histogram
figure(5)
hHist = histogram(x, round(sqrt(N/10)), 'FaceColor', 'r');
xlim([-1.1*max(abs(x)) 1.1*max(abs(x))])
ylim([0 1.1*max(get(hHist, 'Values'))])
grid on
set(gca, 'FontName', 'Times New Roman', 'FontSize', 14)
xlabel('Amplitude, V')
ylabel('Number of samples')
title('Histogram of conventional recording signal')
```

```
%% autocorrelation function estimation
[Rx, lags] = xcorr(x, 'coeff');
tau = lags/fs;

% plot the signal autocorrelation function
figure(6)
plot(tau, Rx, 'r')
grid on
xlim([-max(tau) max(tau)])
ylim([1.1*min(Rx), 1.1])
set(gca, 'FontName', 'Times New Roman', 'FontSize', 14)
xlabel('Delay, s')
ylabel('Autocorrelation coefficient')
title('Correlogram of conventional recording signal')
line([-max(abs(tau)) max(abs(tau))], [0.05 0.05],...
    'Color', 'k', 'LineWidth', 1.5, 'LineStyle', '--')
legend('signal correlogram', '5 % level')

%% signal statistics
% compute and display the minimum and maximum values
maxval = max(x);
minval = min(x);
disp(['Max value = ' num2str(maxval)])
disp(['Min value = ' num2str(minval)])

% compute and display the the DC and RMS values
u = mean(x);
s = std(x);
disp(['Mean value = ' num2str(u)])
disp(['RMS value = ' num2str(s)])

% compute and display the dynamic range
DR = 20*log10(max(abs(x))/min(abs(nonzeros(x))));
disp(['Dynamic range DR = ' num2str(DR) ' dB'])

% compute and display the crest factor
CF = 20*log10(max(abs(x))/s);
```

```
disp(['Crest factor CF = ' num2str(CF) ' dB'])

% compute and display the autocorrelation time
ind = find(Rx>0.05, 1, 'last');
RT = (ind-N)/fs;
disp(['Signal duration = ' num2str(max(to)) ' s'])
disp(['Autocorrelation time = ' num2str(RT) ' s'])

commandwindow
```